\documentclass[pre,aps,eqsecnum,twocolumn]{revtex4}
\usepackage{graphicx}
\usepackage[dvips]{epsfig}
\usepackage{color}

\definecolor{red}{rgb}{1,0,0}
\definecolor{green}{rgb}{0.13,0.55,0.13}
\definecolor{blue}{rgb}{0,0,1}

\begin{document}

\title{Critical Casimir torques and forces \\ acting on needles in two spatial dimensions}

\author{O. A. Vasilyev$^{1,2}$, E. Eisenriegler$^3$, and S. Dietrich$^{1,2}$}

\affiliation{
$^1$
Max-Planck-Institut f{\"u}r Intelligente Systeme,
  Heisenbergstra\ss e~3, D-70569 Stuttgart, Germany\\
$^2$ IV. Institut f\"ur Theoretische Physik,
  Universit\"at Stuttgart, Pfaffenwaldring 57, D-70569 Stuttgart, Germany\\
$^3$Theoretical Soft Matter and Biophysics, Institute of Complex
Systems,
Forschungszentrum J\"ulich, D-52425 J\"ulich, Germany}

\date{\today}

\begin{abstract}
We investigate the universal orientation-dependent interactions
between non-spherical colloidal particles immersed in a critical
solvent by studying the instructive paradigm of a needle embedded
in bounded  two-dimensional Ising models at bulk criticality.
 For a needle in an Ising strip
the interaction on mesoscopic scales depends on the width of the
strip and the length, position, and orientation of the needle. By
lattice Monte Carlo simulations we evaluate the free energy
difference between needle configurations being parallel and
perpendicular to the strip. We concentrate on $\it small$ but
nonetheless mesoscopic needle lengths for which analytic
predictions are available for comparison. All combinations of
boundary conditions for the
 needles and boundaries are considered which belong
to either the ``normal'' or the ``ordinary'' surface universality
class, i.e., which induce local order or disorder, respectively.
We also derive exact results for needles of {\it arbitrary}
mesoscopic length, in particular for  needles embedded in a half plane
and oriented perpendicular to the corresponding
 boundary as well as for  needles embedded at the center line of a symmetric
strip with parallel orientation.
\end{abstract}

\maketitle

\section{INTRODUCTION}
\label{intro}
A solvent near its bulk critical point induces long-ranged forces
between immersed mesoscopic particles which are ``universal'',
i.e., independent of most microscopic details of the
system~\cite{FdG, Krech, BDT, Gambassi}. Apart from a few bulk
properties of the solvent and from its local interactions with the
surfaces of the immersed particles, these forces  depend only on
the {\it geometry of the confinement} of the critical fluctuations
of the  solvent imposed by the particle surfaces, that is on the
sizes, shapes, positions, and orientations of the particles. Such
forces have been observed experimentally in film geometry for
${^4}$He ~\cite{GCh1,Helium} and $^3$He/$^4$He
mixtures~\cite{GCh2} near the superfluid transition as well as for
classical binary liquid mixtures near their demixing
transition~\cite{FYP} and, directly, between a single spherical
particle immersed in a binary liquid mixture near the critical
demixing point and its confining planar
wall~\cite{directmeasurement,directmeasurement2}. Due to the
similarity of these forces with the Casimir forces arising in
quantum electrodynamics~\cite{KG,BMM} they are called critical
Casimir forces.

{\it Nonspherical} particles near other confinements, such as a
planar wall, experience orientation dependent forces giving rise
to critical Casimir {\it torques}~\cite{e,kondrat}.

In the following we concentrate on critical solvents belonging to
the Ising universality class such as demixing classical binary
liquid mixtures. In these systems the particle surfaces
generically prefer one of the two components of the mixture, i.e.,
in Ising language one of the two directions ($+$ or $-$) of the
order parameter is preferred. This preference is captured in terms
of surface fields the strength of which under renormalization
group flow attains $\pm \infty$, corresponding to fixed surface
spins and denoted by $\pm $ boundary conditions. However, using
suitable surface preparation one can suppress this preference,
being left with a weakened tendency to demix near the particle
surface~\cite{NHB}. These two types of surface universality
classes~\cite{hwd} are called ``normal'' ($+/-$) and ``ordinary''
(O), respectively.

Simple universal behavior arises in the scaling region where the
ranges of the interactions between the ordering degrees of freedom
other than short-ranged ones and the lengths characterizing
corrections to scaling and the crossover from less stable bulk and
surface universality classes, respectively, are much smaller than
the increasing correlation length of the bulk solvent upon
approaching criticality, than the distances between the particles,
and than the lengths characterizing the sizes and shapes of the
particles. In this region of a clear separation of ``microscopic''
and ``mesoscopic'' lengths, the forces and torques are given by
universal scaling functions which, apart from the surface
universality classes, only depend on ratios of the characteristic
mesoscopic lengths.

Here, our main interest is in the orientation dependent
interaction of nonspherical particles with boundaries of critical
systems. For the interaction between a prolate uniaxial ellipsoid
and a planar wall with $++$ or $+-$ boundary conditions at the
corresponding surfaces the complete universal scaling functions
for the critical Casimir forces, both for the disordered and
ordered bulk phases, have been obtained within mean field
theory~\cite{kondrat}. Beyond mean field theory such scaling
functions have been determined to a lesser degree of completeness.
For {\it nearby} particles, i.e., for closest surface to surface
distances much smaller than their size and radii of curvature, the
interaction can be expressed in terms of the Casimir force in film
geometry by using the Derjaguin approximation. The scaling
function of the latter is available for $++$ and $+-$ surfaces in
three~\cite{O1,O2,Hasenbusch} and two~\cite{evste} spatial
dimensions~\cite{fluct}. For a {\it small} spherical \cite{ber} or
nonspherical \cite{e,e'} particle, with a size much smaller than
the bulk correlation length and the distances to the surfaces of
other particles or to boundaries, the interactions can be obtained
from the so-called ``small particle expansion''~\cite{SPE}. This
is one of the field theoretic operator expansions for small
objects the most well known of which is the product of two nearby
operators first considered around 1970 by Wilson, Polyakov, and
Kadanoff~\cite{WK,opera}. These expansions are to a certain extent
reminiscent of the multipole expansion in electrostatics.

In view of the twofold asymptotic condition, i.e., the particle
being {\it small} on mesoscopic scales and {\it large} on the
microscopic scale \cite{SPE}, it is a nontrivial issue under which
circumstances the results of the ``small particle expansion'' can
be observed for a given actual system. As a first step to address
this issue, here we investigate whether the results of the
expansion can be observed in the two-dimensional Ising model on a
square lattice with (ferromagnetic) couplings ${\cal J} > 0$
between nearest neighbor spins only and right at the bulk critical
point.

\begin{figure}[h]
\includegraphics[width=0.49\textwidth]{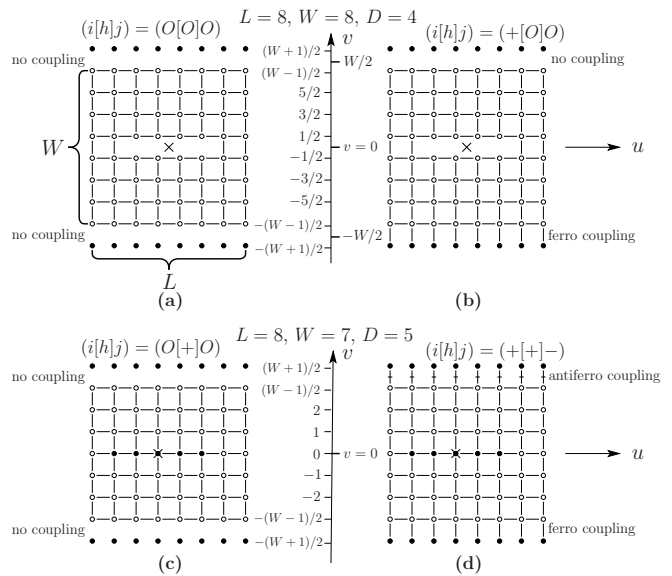}
\caption{
Needle, with their centers denoted by $\times$, embedded in a
strip of a ferromagnetic Ising model on a square lattice
comprising $W$ rows and $L$ columns~\cite{micromeso} of
fluctuating spins (empty circles) and periodic boundary condition
in the $u$ direction parallel to the rows. The two additional rows
of spins fixed in the $+$ direction (full circles) allow one to
induce positive or negative magnetic fields at the surfaces of the
strip by coupling them in a ferro- or antiferromagnetic way to the
bottom and top rows of fluctuating spins (see the main text). (a)
and (b): Needle of $D=4$ broken bonds in a strip with the number
$W$ of rows and the number $L$ of columns equal to $W=L=8$. (a)
shows the case $(O[O]O)$ of a strip with two free surfaces (no
coupling to the fixed spins) while (b) shows the case $(+[O]O)$ in
which a ferromagnetic coupling to the fixed spins leads to a
positive magnetic field at the lower surface. (c) and (d): Needle
of $D=5$ spins fixed in the $+$ direction (full circles) in a
strip with $W=7$ and $L=8$. (c) shows the case $(O[+]O)$ while (d)
shows the case $(+[+]-)$ with positive and negative magnetic
fields induced at the lower and upper surfaces by ferro- and
antiferromagnetic couplings, respectively. All the needles shown
are oriented in the $u$ direction and have their center $\times$
at the midline $v=0$ of the strip, half way between the strip
surfaces, i.e., $v_{N}=0$ for the vertical position of the center
of the needle. This requires to choose $W$ even (odd) for needles
of broken bonds (of fixed spins) so that spins reside at half odd
integer (integer) values of $v$. Needles oriented in the $v$
direction are shown in Figs.~\ref{FIG_IV_1} and~\ref{FIG_IV_2}
below.} \label{FIG_I_1}
\end{figure}

In our Monte Carlo simulations we consider a lattice of finite
size with the form of a strip (or rectangle) comprising $W$ rows
and $L$ columns with an embedded particle resembling a needle.
The directions along the rows and columns define, respectively,
the directions $u$ and $v$ parallel and perpendicular to the strip
of length \cite{micromeso} $L$ and width $W$. While in the $u$
direction we impose periodic boundary conditions by means of
couplings ${\cal J}$ between the first and last spin in each row,
we couple all of the spins in the lowest and uppermost row by
means of interactions 0 or $\pm {\cal J}$ to spins which are {\it
fixed} in the $+$ direction and are located in outside neighboring
rows (see Fig.~\ref{FIG_I_1}). Thus in the $v$ direction, the
strip is bounded by free surfaces or surfaces to which a magnetic
field $\propto \pm {\cal J}$ is applied, i.e., by surfaces of
``ordinary'' ($O$) or ``normal'' ($+/-$) character.

In order to generate an embedded needle with $O$ or $+/-$ boundary
conditions and orientation parallel to the strip we remove
couplings (break bonds) between two neighboring lattice rows or we
fix spins within a single lattice row, respectively (see Fig.~\ref{FIG_I_1}).
In the two cases we define the length $D$ of the
needle as the number \cite{micromeso} of broken bonds and of fixed
spins, respectively, which we choose to be even and odd,
respectively, in order that the needle centers $\times$ coincide
with the center of an elementary square and a vertex of the
lattice, respectively. This allows us to ``turn'' the needles
abruptly about their center by 90 degrees upon rearranging the
broken bonds and fixed spins, leading to broken bonds between
neighboring columns and fixed spins within a single column of the
lattice, respectively. In order to be able to position the centers
of $O$ and $+/-$ needles right at the midline of the strip (as
shown in Fig.~\ref{FIG_I_1}) we choose $W$ to be even and odd,
respectively.

In the simulation we calculate the free energy expense $\Delta F$
required to ``turn'' the needle about its center from an alignment
perpendicular to the strip to the parallel alignment. This is a
measure for the effective {\it torque} acting on the needle. In
line with the introductory remarks the free energy cost $\Delta F$
depends on the distances of the needle center from the strip
surfaces.

We put the origin of the $(u,v)$ coordinate system at the midline
of the lattice, i.e., for $O$ and $+/-$ needles at the center of
an elementary square and at a vertex so that the lattice spins are
located at half odd integer and integer values, respectively, of
the coordinates (see Fig.~\ref{FIG_I_1}). With this choice the
mirror symmetry of $\Delta F$ in a strip with equal boundary
conditions $i=j$ is described simply by its invariance if the
coordinate $v=v_{N}$ of the needle center changes sign. Here $i$
and $j$ denote the boundary conditions at the surfaces at $v=-W/2$
and $v=W/2$, respectively. Figure~\ref{FIG_I_1} shows the special
case in which the centers $\times$ of $O$ and $+/-$ needles
coincide with the origin, i.e., $v_{N}$ vanishes, while in the
general case $v_{N}$ takes integer values for both types of
needles.

It is useful to adopt a notation $(i[h]j)$ which characterizes the
boundary conditions $i$ and $j$ of the strip surfaces as well as
the boundary condition $h$ of the embedded needle. For example,
$(+[+]-)$ denotes the case of a needle of spins fixed in the $+$
direction which is embedded in a strip with outside couplings
${\cal J}$ and $-{\cal J}$ at the lower and upper surfaces
inducing strip surfaces $i=+$ and $j=-$, respectively, as shown in
Fig.~\ref{FIG_I_1}(d).

For completeness we consider also the case of a strip with
periodic boundary conditions in both $u$ and $v$ directions by
coupling, in both the $W$ rows and in the $L$ columns, the
corresponding end spins to each other with strength ${\cal J}$.
This is a strip without surfaces and is equivalent to a square
lattice on a torus.

\begin{figure}[b]
\includegraphics[width=0.49\textwidth]{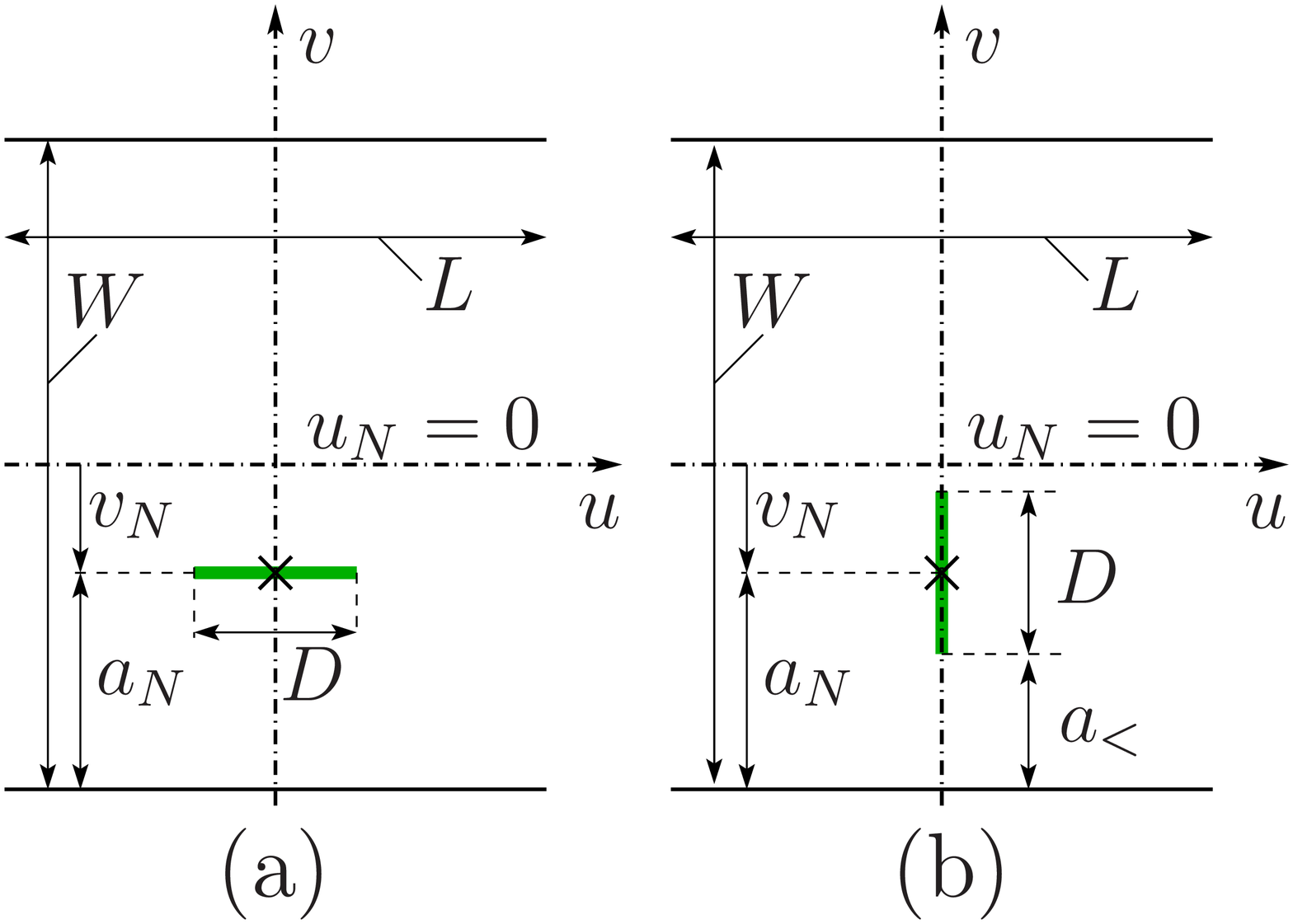}
\caption{Continuum \cite{micromeso} description of the geometry of
a needle in a strip. In the strip of length $L$ and width $W$ the
embedded needle of length $D$ is oriented parallel (a) and
perpendicular (b) to the strip boundaries at $v= \pm W/2$. The
needle center $\times$ is located at $u=u_{N}=0$ and $v=v_{N} \,
<0$ at a distance $a_{N} \equiv v_{N}+W/2 \, >0$ from the lower
boundary. In the perpendicular orientation the closer end of the
needle has a distance $a_{<} \equiv a_{N}-D/2 \, >0$ (and the
farther end a distance $a_{>}=a_{<}+D$) from the lower boundary.}
\label{FIG_I_2}
\end{figure}
If $D, \, W, \, L,$ and the closest distance between the needle
and the strip boundaries (corresponding to $a_{N}$ and $a_{<}$ in
Figs.~\ref{FIG_I_2}(a) and~\ref{FIG_I_2}(b), respectively)
are ``sufficiently'' large on the scale of the lattice constant,
the free energy cost $\Delta F$ in the lattice model is expected
to display universal scaling behavior and, as mentioned above,
$\Delta F/(k_B T)$ will depend only on the universality classes
$(i[h]j)$ of the boundaries and the needle and on three
independent ratios of $D, \, W, \, L$, and $v_{N}$ . In this case
one can adopt a mesoscopic continuum description with sharp strip
boundaries at $v= \pm W/2$ since lengths only differing by
approximately a lattice constant can be identified. Likewise,
different microscopic definitions of the length of the needles,
such as the number $D-1$ of bonds between the fixed spins of a $+$
needle (instead of its $D$ fixed spins), lead to the same
mesoscopic length $D$. Figure~\ref{FIG_I_2} shows the various
lengths characterizing the geometry of a needle in a strip within
the continuum description.

For small mesoscopic needles and a long strip, i.e., for $D \ll W
\ll L$, the universal behavior can be predicted by means of the
``small particle expansion'' and, as mentioned earlier, it is one
of our main goals to quantitatively investigate whether present
Monte Carlo simulations can access this regime.

For a needle perpendicular to the boundary of a half plane or
embedded at the midline of a symmetric ($i=j$) strip of infinite
length, we derive the analytic form of the universal scaling
behavior of the critical Casimir forces in the {\it complete}
range of mesoscopic needle lengths $D$, increasing from small to
large. This aspect is interesting in its own right and also allows
us to estimate the range of validity of the ``small needle
expansion''.

The predictions of the ``small particle expansion'' for $\Delta F$
in critical Ising strips and our analytic results for needles of
arbitrary length are presented in Secs.~\ref{SP}  and~\ref{ARB},
respectively. In Sec.~\ref{SIM} we describe the lattice model in
more detail and explain how one obtains via Monte Carlo
simulations the results for $\Delta F$ which are compared with the
analytic scaling predictions in Sec.~\ref{compare} . Our results
are summarized in Sec.~\ref{summary}. In
Appendices~\ref{appwithout} and~\ref{apparb} we present the input material
and the derivations which are necessary in order to obtain the
predictions and results of Secs.~\ref{SP} and~\ref{ARB},
respectively. For the convenience of the reader we present a
glossary of our symbols and notations in Appendix C.
%

Besides for critical systems \cite{e,kondrat} the orientation
dependent interaction between a wall and a nonspherical mesoscopic
object (such as an ellipsoid, a semi-infinite plate, or a
spherocylinder) has been studied also for quantum electrodynamics
\cite{emig2008,gies} and for purely entropic systems in which it
is induced by spherical so-called depletion agents with hard body
interaction only \cite{roth2002} or by free nonadsorbing polymer
chains \cite{e'}. As for correlation-induced forces we mention
also the attractive effective force generated by needles (rigid
rods) acting as depletion agents \cite{Helden2003} and the
repulsive force generated by a nonadsorbing polymer chain grafted
to the tip of a (model-) atomic force microscope near a repulsive
wall \cite{Magh}. Finally we note that the critical Casimir force
between inclusions in the {\it two}-dimensional Ising model has
been suggested as a possible mechanism for the presence of
long-ranged forces between membrane bound proteins
\cite{Machta2012}, which are typically non-circular.
%

\section{SMALL NEEDLE EXPANSION}
\label{SP}

Here we specialize the ``small particle expansion'' for
non-spherical particles (see~Ref.~[\onlinecite{e}]) to the present
case of a needle embedded in the two-dimensional Ising model and
apply it to the geometry of a needle in a strip.

We consider a needle of small mesoscopic length $D$, centered at
${\bf r}_N$, and pointing along the unit vector ${\bf n}$.
Inserting the needle into the $d=2$ Ising model at its critical
point \cite{bulkatcritpoint} changes the Boltzmann weight of the
corresponding field theory by a factor $\exp(-\delta {\cal H})$
which can be represented by the operator \cite{opera} expansion
\cite{e,e'}
\begin{eqnarray} \label{SPE}
\exp(-\delta {\cal H}) \propto 1+S_{I}+S_{A}
\end{eqnarray}
where
\begin{eqnarray} \label{SPE'}
&&S_{I}=\sum_{{\cal O}=\phi,\epsilon} {\cal A}_{\cal O}^{(h)}
\Biggl[ \Biggl( {D \over 2} \Biggr)^{x_{\cal O}} +  \nonumber \\
&& \quad + {1 \over 16 x_{\cal O}} \Biggl( {D \over 2}
\Biggr)^{2+x_{\cal O}} \Delta_{{\bf r}_N} \Biggr] {\cal O}({\bf
r}_N) + ...
\end{eqnarray}
with $\Delta_{{\bf r}_N}=\nabla^{2}_{{\bf r}_N}$ and

\begin{eqnarray} \label{SPE''}
&&S_{A}=\sum_{k,l=1,2} \Biggl (n_k n_l - {\delta_{kl} \over
2}\Biggr) \Biggl[ -{\pi \over
2} \Biggl( {D \over 2} \Biggr)^2 T_{kl}({\bf r}_N) + \nonumber \\
&&+ \sum_{{\cal O}=\phi,\epsilon} {\cal A}_{\cal O}^{(h)} {3 \over
8 (1+x_{\cal O})} \Biggl( {D \over 2} \Biggr)^{2+x_{\cal O}}
\partial_{(r_{N})_{k}}
\partial_{(r_{N})_{l}} {\cal O}({\bf r}_N) \Biggr] \nonumber \\
&& \qquad \qquad \qquad \qquad \qquad \qquad \qquad \qquad \qquad
+ \, ...
\end{eqnarray}
are the isotropic and anisotropic contributions, respectively.
Here ${\cal O}=\phi$ is the order-parameter-density operator and
${\cal O}=\epsilon$ is the difference of the energy-density
\cite{coarsegrain} operator $e$ and its average $\langle e
\rangle_{\rm bulk}$ in the unbounded plane (bulk) at bulk
criticality (so that their bulk averages $\left< \phi
\right>_{\mathrm{bulk}}$ and $\left< \epsilon
\right>_{\mathrm{bulk}}$ vanish at the bulk critical point). They
are normalized such that the bulk two-point correlation functions
\cite{bulkatcritpoint} are \cite{normzd}
\begin{eqnarray} \label{norm}
\langle {\cal O}({\bf r}) {\cal O}({\bf r}') \rangle_{\rm
bulk}=|{\bf r}-{\bf r}'|^{-2 x_{\cal O}} \, ,
\end{eqnarray}
where (in $d=2$) $x_{\phi}=1/8$ and $x_{\epsilon}=1$ are their
scaling dimensions \cite{Cardy}. The affiliation of the surface of
the needle to the ``ordinary'' ($h=O$) or to the ``normal'' ($h=+$
or $h=-$) surface universality class enters into
Eqs.~(\ref{SPE})-(\ref{SPE''}) via the universal half plane
amplitude ${\cal A}_{\cal O}^{(h)}$. The latter is the amplitude
of the profile \cite{bulkatcritpoint} of ${\cal O}$ in the half
plane, $\langle {\cal O} \rangle_{\rm half \, plane}={\cal
A}_{\cal O}^{(h)} a^{-x_{\cal O}}$, as function of the mesoscopic
distance $a$ from the boundary line of type $h$ and is given
by~\cite{Cardy}
\begin{eqnarray} \label{halfamp}
&&{\cal A}_{\phi}^{(O)}=0, \, {\cal A}_{\phi}^{(+)}=-{\cal
A}_{\phi}^{(-)}=2^{1/8}, \nonumber \\ &&{\cal
A}_{\epsilon}^{(O)}=-{\cal A}_{\epsilon}^{(+)}=-{\cal
A}_{\epsilon}^{(-)}=1/2 \, .
\end{eqnarray}
The properties ${\cal A}_{\phi}^{(O)}=0$ and ${\cal
A}_{\phi}^{(-)}=-{\cal A}_{\phi}^{(+)}$ of the amplitudes  of the
order parameter profile are obvious from the up-down symmetry of
the Ising spins \cite{bulkatcritpoint}. The energy density profile
increases (decreases) upon approaching an ``ordinary''
(``normal'') boundary where the Ising spins are more disordered
(more ordered) than in the bulk \cite{coarsegrain}. This implies
the positive (negative) sign of the corresponding amplitude ${\cal
A}_{\epsilon}^{(O)}$ (${\cal A}_{\epsilon}^{(+)}={\cal
A}_{\epsilon}^{(-)}$). The absolute values of ${\cal
A}_{\epsilon}^{(O)}$ and ${\cal A}_{\epsilon}^{(+)}={\cal
A}_{\epsilon}^{(-)}$ are the same, due to duality properties
\cite{duality,Afordlarger}. In Eq.~(\ref{SPE''})~the contribution
from the stress~\cite{Cardy} tensor $T_{kl}$ is the same
\cite{parthalf} for all  needle types $h=O,+,-$.

The ellipses in Eqs.~(\ref{SPE'})~and~(\ref{SPE''}) stand for
contributions of fourth order  and higher \cite{ellipses} in the
small mesoscopic length $D$. The common factor of proportionality
on the rhs of Eq.~(\ref{SPE}) is given \cite{bulkatcritpoint} by
$\langle {\rm exp}(-\delta {\cal H}) \rangle_{\rm bulk}$ because
the bulk averages of $S_I$ and $S_A$ vanish at the bulk critical
point.

For a two-dimensional Ising model with boundaries the insertion
free energy of the needle depends on its position and orientation
with respect to these boundaries. Removing the needle from the
bulk model and inserting it in the bounded one at a distance from
the boundaries much larger than $D$ changes the free energy by $
F({\bf r}_{N}, {\bf n})=-k_B T \ln \langle (1+S_{I} +S_{A})
\rangle_{\rm BM}$ where BM denotes the bounded model in the
absence of the needle. We shall concentrate on the geometry of a
needle in a strip as described in the Introduction (see
Fig.~\ref{FIG_I_2}) for which BM $\equiv$ ST is the strip with
boundaries $(i,j)$ in the absence of the needle and
correspondingly use the notation \cite{micromeso}
\begin{eqnarray} \label{uvnparnperp}
{\bf r}_N=(u_{N},v_{N}), \quad {\bf n}=(n_{||},n_{\perp})
\end{eqnarray}
for the center and direction vectors of the needle with the
components parallel and perpendicular to the strip. The density
averages $\langle {\cal O} \rangle_{\rm ST}$  at ${\bf r}_N$ which
enter $\langle S_{I}+S_{A} \rangle_{\rm ST}$ are, within this
model, independent of $u_{N}$  and in the scaling region given by
\begin{eqnarray} \label{densscale}
\langle {\cal O}({\bf r}_N) \rangle_{\rm ST}=W^{-x_{\cal O}}
f_{\cal O}^{(i,j)}(v_{N}/W,W/L),
\end{eqnarray}
where $f$ are universal scaling functions and $f_{\cal O}^{(j,i)}$
follows from $f_{\cal O}^{(i,j)}$ upon replacing $v_{N}/W$ by
$-v_{N}/W$.

Due to the continuity equation of the stress tensor \cite{Cardy}
its average $\langle T_{kl}({\bf r}_{N}) \rangle_{\rm ST}^{(i,j)}$
in the strip is independent of both $u_{N}$ and $v_{N}$ and
follows from the universal, scale-free, and shape-dependent
contribution~\cite{shapefreeenergy} $\Phi_{\rm ST}^{(i,j)}(L/W)$
to the free energy $F_{\rm ST}$ per $k_B T$ of the strip ST
without the needle:
\begin{eqnarray} \label{stressscale}
\langle [T_{\perp \perp}, T_{\parallel \parallel}] \rangle_{\rm
ST}^{(i,j)}&=& -[L^{-1} \partial_W, W^{-1} \partial_{L}] \Phi_{\rm
ST}^{(i,j)}(L/W) \nonumber \\&=&[1,-1]W^{-2}\Delta_{i,j}(W/L) \, .
\end{eqnarray}
Like in Eq.~(\ref{uvnparnperp}), here $\parallel$ and $\perp$
denote directions parallel and perpendicular to the $u$ axis of
the strip,
\begin{eqnarray} \label{stressscale'}
\Delta_{i,j}(W/L)=(d/d \delta)\Phi_{\rm ST}^{(i,j)}(\delta), \quad
\delta=L/W \, ,
\end{eqnarray}
and the off-diagonal components $\langle T_{\perp
\parallel} \rangle_{\rm ST}=\langle T_{\parallel \perp}
\rangle_{\rm ST}$ vanish by symmetry. Cardy \cite{Cardybc} has
obtained an explicit form for all functions $\Phi_{\rm
ST}^{(i,j)}(\delta)$. While an extensive discussion of
$\Delta_{i,j}(W/L)$ is deferred to Appendix~\ref{appstripbound},
here we mention a few basic properties. Obviously $\Phi_{\rm
ST}^{(i,j)}$ and $\Delta_{i,j}$ are symmetric in $i$ and $j$. For
a long strip the leading behavior $\Phi_{\rm ST}^{(i,j)}(L/W \to
\infty) \,=\, \Delta_{i,j}(0) \times \delta$ of $\Phi_{\rm
ST}^{(i,j)}$ is linear in $\delta$ with $\Delta_{i,j}(0) \equiv
\Delta_{i,j}$ given \cite{Cardy} by $\Delta_{i,j}=(\pi /48) [-1,
-1, 23, 2]$ for $(i,j)$\,=\,$[(O,O), (+,+), (+,-), (+,O)]$. While
due to the ($+ \leftrightarrow -$) symmetry the equalities
$\Delta_{+,+}(1/\delta)=\Delta_{-,-}(1/\delta)$ and
$\Delta_{+,O}(1/\delta)=\Delta_{-,O}(1/\delta)$ hold for arbitrary
$1/\delta$, the equality $\Delta_{O,O}=\Delta_{+,+}$ holds only
for infinitely long strips~\cite{dual}, i.e., $1/\delta=0$.

For the free energies $F_{\parallel}$ and $F_{\perp}$ associated
with removing the needle from the bulk system and inserting it in
the strip with its center at ${\bf r}_{N}$ and with its
orientation ${\bf n}$ parallel and perpendicular, respectively, to
the $u$ axis, Eqs.~(\ref{SPE})-(\ref{SPE''}) and
(\ref{uvnparnperp}) yield
\begin{eqnarray} \label{Fparal}
F_{\parallel} \equiv F({\bf r}_N, {\bf n}=(1,0))=-k_{B}T \ln
Z_{\parallel}
\end{eqnarray}
and
\begin{eqnarray} \label{Fperp}
F_{\perp} \equiv F({\bf r}_N, {\bf n}=(0,1))=-k_{B}T \ln Z_{\perp}
\end{eqnarray}
with
\begin{eqnarray} \label{Z}
Z_{\parallel}=1+\zeta_{I}-\zeta_{A}, \quad
Z_{\perp}=1+\zeta_{I}+\zeta_{A}
\end{eqnarray}
where
\begin{eqnarray} \label{zetaI}
&&\zeta_{I}=\sum_{{\cal O}=\phi,\epsilon} {\cal A}_{\cal O}^{(h)}
\Biggl[ \Biggl( {D \over 2W} \Biggr)^{x_{\cal O}} + {1 \over 16
x_{\cal O}} \Biggl( {D \over 2W} \Biggr)^{2+x_{\cal O}} \times
\nonumber \\
&& \qquad \qquad \times  \frac{\partial^2} {\partial \left({v_{N}/W}\right)^{2}} \Biggr]
f_{\cal O}^{(i,j)}\left(\frac{v_{N}}{W}, \frac{W}{L}\right) + ...
\end{eqnarray}
and
\begin{eqnarray} \label{zetaA}
&&\zeta_{A}= -{\pi \over 2} \Biggl( {D \over 2W} \Biggr)^2
\Delta_{i,j}\left(\frac{W}{L} \right)
+ {1 \over 2} \sum_{{\cal O}=\phi,\epsilon} \frac{3{\cal A}_{\cal
O}^{(h)} }{ 8 (1+x_{\cal O})} \times \nonumber \\
&&\times \Biggl( {D \over 2W} \Biggr)^{2+x_{\cal O}}
\frac{\partial^2} {\partial \left(v_{N}/W\right)^{2}} f_{\cal
O}^{(i,j)}\left(\frac{v_{N}}{W}, \frac{W}{L}\right) + ...
\end{eqnarray}
follow from $S_I$ and $S_A$ in Eqs.~(\ref{SPE'})
and~(\ref{SPE''}), respectively. This implies the expression
\begin{eqnarray} \label{DeltaF'}
\Delta F = - k_{B} T \ln (Z_{\parallel}/Z_{\perp})
\end{eqnarray}
for the free energy required to turn the needle about its center
from its orientation perpendicular to the $u$ axis of the strip to
the parallel orientation.

In the expansions in Eqs.~(\ref{zetaI}) and~(\ref{zetaA}) a term
$\propto D^{\cal X}$ is of the  order of $(D/W)^{\cal X}$ near the
strip center, where $|v_{N}| \ll W$, and of the order of
$(D/a_{N})^{\cal X}$ near the strip boundaries, where $a_{N}
\equiv (W/2)-|v_{N}| \ll W$. These expansions for the {\it
partition functions} $Z_{\parallel}$ and  $ Z_{\perp}$ are useful
if $D/W$ and $D/a_{N}$ are much smaller than 1. However, expanding
the {\it free energies} $F_{\parallel}$, $F_{\perp}$, and $\Delta
F$ in powers of $D$ is not always useful for the comparison with
our simulation data. While $D/W \ll 1$ and $D/a_{N} \ll 1$ can
readily be achieved in the MC simulations for a needle with a
mesoscopic length $D$ of many lattice constants, at present it
seems to be unrealistic to achieve $(D/W)^{1/8} \ll 1$ and
$(D/a_{N})^{1/8} \ll 1$. Thus we expand the logarithm in
Eq.~(\ref{DeltaF'}) in terms of powers of $D$ only in those cases
in which the power $D^{x_{\phi}} \equiv D^{1/8}$ does not appear
\cite{useless} in Eq.~(\ref{zetaI}). These are the cases with an
``ordinary'' needle $h=O$, for which ${\cal A}_{\phi}^{(h)}$
vanishes, and with a ``normal'' needle in a strip with two
``ordinary'' boundaries $i=j=O$, for which the density profile in
Eq.~(\ref{densscale}) vanishes for the order parameter ${\cal
O}=\phi$. Expanding the logarithm in these cases yields
\begin{eqnarray} \label{expand}
\Delta F=\Delta F_{l}+\Delta F_{nl}+...
\end{eqnarray}
with the leading contribution
\begin{eqnarray} \label{expand'}
{\Delta F_l \over k_B T} = -\pi \Biggl( {D \over 2W} \Biggr)^{2}
\Delta_{i,j}(W/L)
\end{eqnarray}
and the next-to-leading contribution
\begin{eqnarray} \label{expand''}
{\Delta F_{nl} \over k_B T} &=& \Biggl( {D \over 2W }\Biggr)^{3}
{\cal A}_{\epsilon}^{(h)} \Biggl[ \pi \Delta_{i,j}(W/L) +
\nonumber \\
&+&{3 \over 16} \frac{\partial^2}{\partial \left(v_{N}/W
\right)^{2}} \Biggr]f_{\epsilon}^{(i,j)}\left(\frac{v_{N}}{W},
 \frac{W}{L}\right) \, .
\end{eqnarray}
The ellipses in Eq.~(\ref{expand}) are of order $D^4$.

For completeness we  also consider a double periodic rectangle or
strip with periodic boundary conditions in both the $u$ and
$v$-directions, i.e., the surface of a torus. In this
boundary-free case at $T_{c}$ the  average of $\phi$ vanishes in
the strip, that of $\epsilon$ is independent of both $u$ and $v$,
and for the average of the stress tensor Eq.~(\ref{stressscale})
again applies with $\Delta_{i,j}(W/L)$ replaced by a function
$\Delta_{P}(W/L)=(d/d\delta)\Phi_{\rm ST}^{(P)}(\delta)$ with the
infinite strip limit $\Delta_{P}(0) \equiv \Delta_{P}=-\pi/12$
(see Subsec.~\ref{comperper} and Appendix~\ref{appperper} for more
details). For the double periodic strip
Eqs.~(\ref{expand})-(\ref{expand''}) also apply if
$\Delta_{i,j}(W/L)$ and $f_{\epsilon}^{(i,j)}(v_{N}/W, W/L)$ are
replaced by the corresponding stress amplitude $\Delta_{P}(W/L)$
and the energy density scaling function $f_{\epsilon}^{(P)}(W/L)$
which is independent of $v_{N}/W$.

Explicit expressions for the functions $\Delta(W/L)$ are
known~\cite{Cardybc} for all types of strips considered here.
Concerning the scaling functions $f_{\cal O}$ the dependence on
the aspect ratio $W/L$ is known~\cite{FF,FSZ} for
$f_{\epsilon}^{(P)}$ while in the presence of boundaries the
functions $f_{\cal O}^{(i,j)}(v_{N}/W,W/L)$ are, to the best of
our knowledge, known only~\cite{symmfinL} for $W/L=0$, i.e., for
strips of infinite~\cite{BX} length $L=\infty$. For the
convenience of the reader we collect these expressions in
Appendix~\ref{appwithout}.

For the special case in which the distance $a_{N}=v_{N}+W/2$ of
the center of the small needle of type $h$ from the boundary of
type $i$ at $v=-W/2$ (see Fig.~\ref{FIG_I_2}) is much smaller than
the width of the infinitely long strip, i.e., for the case $D \ll
a_{N}$ and $W/a_{N}, \, L/a_{N} \, \to \infty$, the above
expressions for the needle in the strip with free energies $F
\equiv F^{(i[h]j)}$ reduce to those for the needle in the {\it
half plane} $v+W/2 \equiv a>0$ with free energies $F \equiv
F^{(i[h])}$ where Eqs.~(\ref{zetaI}) and~(\ref{zetaA}) are
replaced by
\begin{eqnarray} \label{halfzeta}
&&\zeta_{I}=\sum_{{\cal O}=\phi,\epsilon} {\cal A}_{\cal O}^{(h)}
{\cal A}_{\cal O}^{(i)} \Biggl[ \vartheta^{x_{\cal
O}} + {1+x_{\cal O} \over 16} \vartheta^{2+x_{\cal O}} \Biggr] + ... \, , \nonumber \\
&&\zeta_{A}= \sum_{{\cal O}=\phi,\epsilon} {\cal A}_{\cal O}^{(h)}
{\cal A}_{\cal O}^{(i)} {3x_{\cal O} \over 16}
\vartheta^{2+x_{\cal O}} + ... \, ,
\end{eqnarray}
and Eqs.~(\ref{expand'}) and~(\ref{expand''}) by
\begin{eqnarray} \label{halfDelF}
{\Delta F_{l} \over k_B T} = 0 \, , \quad
{\Delta F_{nl} \over k_B T} = {\cal A}_{\epsilon}^{(h)} {\cal
A}_{\epsilon}^{(i)} {3 \over 8} \vartheta^{3} \,
\end{eqnarray}
where
\begin{eqnarray} \label{vartheta}
\vartheta={D \over 2 a_{N}} \, .
\end{eqnarray}
As mentioned above these relations only apply if $0 \leq \vartheta
\ll 1$.
%
%

\section{NEEDLES OF ARBITRARY LENGTH}
\label{ARB}

The ``small needle expansion'' is valid if the mesoscopic length
$D$ of the needle is ``small'' compared to the other mesoscopic
lengths of the system, i.e., within the present model much smaller
than the width and length of the strip and the distance of the
needle from the two boundaries. Given the limited set of operators
appearing in Eqs.~(\ref{SPE'}) and~(\ref{SPE''}), it is an open
issue what in this context ``small'' means quantitatively. In
order to get a clue,  in this section we study a few situations in
which results for an {\it arbitrary} mesoscopic needle length $D$
can be obtained. These results, which we derive in
Appendix~\ref{apparb}, are interesting also in their own right.

(i) {\it Needle in half-plane}

We consider a needle of universality class $h$ embedded in the
half plane, oriented {\it perpendicular} to the boundary line of
surface universality  class $i$, and with its center located at a
distance $a_{N}$ from the boundary. Here we also introduce the
distance $a_{<}=a_{N}-(D/2)$ of the closer end of the needle from
the boundary (compare Fig.~\ref{FIG_I_2}(b)) so that one has
\begin{eqnarray} \label{asmaller}
\vartheta \equiv {D \over 2 a_{N}} = {1 \over 1+(2a_{<} / D)}
\end{eqnarray}
for the length ratio $\vartheta$ defined in Eq.~(\ref{vartheta}).
Note that $\vartheta$ tends to 0 and 1 for a small (or distant)
needle with $D \ll a_{N}$ and a long (or close) needle with $a_{<}
\ll D$, respectively. Both $D$ and $a_{<}$ are assumed to be
mesoscopic lengths, i.e., both are large on the microscopic scale.
We are interested in the free energy $F_{\perp} \equiv
F_{\perp}^{(i[h])} = k_{B}T f_{i[h]}^{\perp}(\vartheta)$ required
to insert the needle from the bulk into the half plane, and in the
corresponding Casimir force $-(\partial/\partial
a_{N})F_{\perp}^{(i[h])} \equiv -(\partial/\partial
a_{<})F_{\perp}^{(i[h])} =
(k_{B}T/a_{N})g_{i[h]}^{\perp}(\vartheta)$ with \cite{integration}
universal scaling functions $f_{i[h]}$ and $g_{i[h]}$. The above
partial derivatives are taken at fixed needle length $D$. The
force pushes the needle away from (towards) the boundary if
$g_{i[h]}$ is positive (negative).

Due to  symmetries the identities
$F_{\perp}^{(+[+])}=F_{\perp}^{(-[-])}$,
$F_{\perp}^{(+[O])}=F_{\perp}^{(-[O])}$, and $F_{\perp}^{(h[i])} =
F_{\perp}^{(i[h])}$ hold so that exchanging the surface
universality classes of the needle and of the boundary leaves the
free energy unchanged. These follow from the $(+ \leftrightarrow
-)$ and the duality~\cite{duality'} symmetries of the Ising model
and are consistent with the symmetries of the corresponding small
needle expression which follows from Eqs.~(\ref{Fperp}),
(\ref{Z}), and~(\ref{halfzeta}).

(ia) For $i=h=O$ the effective interaction has the form (see the
paragraph containing Eqs.~(\ref{Thalfperp})-(\ref{shifthalfperp}))
\begin{eqnarray} \label{arbperp}
{F_{\perp}^{(O[O])} \over k_{B}T} \, =
f^{\perp}_{O[O]}(\vartheta)=
 \, - 2^{-5} \ln
{(1+\vartheta)^{5} \over (1-\vartheta)^{3}} \,
\end{eqnarray}
and is attractive within the entire range $0 \leq \vartheta\leq
1$. For $\vartheta \ll 1$, Eq.~(\ref{arbperp}) is in agreement
with the corresponding result
$F_{\perp}^{(O[O])}/(k_{B}T)=-\vartheta/4 + \vartheta^{2}/2^{5} -
\vartheta^{3}/12 + {\cal O}(\vartheta^{4})$ of the small needle
expansion which follows from Eqs.~(\ref{Fperp}),~(\ref{Z}),
and~(\ref{halfzeta}). For $\vartheta \nearrow 1$,
$F_{\perp}^{(O[O])}/k_{B}T \to - 2^{-5} \times 3 \ln
[(2^{2/3}D)/a_{<}]$ shows a logarithmic dependence. The
logarithmic divergence of the free energy for $D/a_{<} \to \infty$
is related to the long-ranged behavior $\propto 1/a_{<}$ of the
Casimir force for a needle of {\it infinite} length $D= \infty$
which is addressed in Eq.~(\ref{long}) below; its integral
diverges for $a_{<} \to \infty$ (while for a needle of {\it
finite} length $D$, the force decays more rapidly than $1/a_{<}$
as $a_{<}$ increases beyond $D$ so that the integral is finite).

(ib) For $i=+, \, h=O$ the interaction is always repulsive and has
the form
\begin{eqnarray} \label{arbperp+O}
{F_{\perp}^{(+[O])} \over k_{B}T} \, =  \, 2^{-5} \ln
{(1+\vartheta)^{3} \over (1-\vartheta)^{5}} \, ,
\end{eqnarray}
i.e., the same form as Eq.~(\ref{arbperp}) but with $\vartheta$
replaced by $-\vartheta$ (see Appendix~\ref{apparbord}).

(ic) For the Casimir forces, in Appendix~\ref{apparbbreak} we find
expressions which apply to {\it arbitrary} combinations $(i,h)$ of
the universality classes. For later convenience we present them
here in two forms

\begin{eqnarray} \label{iJ}
&& -{\partial \over \partial a_{N}} {F_{\perp}^{(i[h])}
\over k_{B} T} \nonumber \\
&& \qquad= -{\partial \over \partial a_{N}} {F_{\perp}^{(O[O])}
\over k_{B} T} - {1 \over 2 a_{N}} {1 \over
1-\vartheta^{2}} \tilde{\tau}_{i,h}(\vartheta) \nonumber \\
&& \qquad \equiv - {1 \over 2 a_{N}} {1 \over 1-\vartheta^{2}}
\Biggl( {1 \over 12} - {\vartheta^{2} \over 24} -
\rho_{i,h}(\vartheta) \Biggr)
\end{eqnarray}
which are equivalent due to Eq.~(\ref{arbperp}) and the definition
\begin{eqnarray} \label{sigmatilde}
\tilde{\tau}_{i,h}(\vartheta) \equiv {1 \over 12}
(1-6\vartheta+\vartheta^{2}) - \rho_{i,h}(\vartheta) \, .
\end{eqnarray}
The dependence of $\rho_{i,h}$ on $\vartheta$ is given by
\begin{eqnarray} \label{sigma}
\rho_{i,h}(\vartheta) \, = \, {4 \pi \Delta_{i,h}(1/\delta) \over
[K^{*}(\vartheta)]^{2}} \, = \, {\pi \delta^{2}
\Delta_{i,h}(1/\delta) \over 4 \, [K(\vartheta)]^{2}} \,
\end{eqnarray}
with the variable $1/\delta$ replaced by the function
\begin{eqnarray} \label{K+K^{*}}
1/\delta \, = \, K^{*}(\vartheta) / [4 K(\vartheta)]
\end{eqnarray}
of $\vartheta$. Here
%
\begin{eqnarray} \label{theta'}
K^{*}(\vartheta) \equiv K(\bar \vartheta) \, , \quad \bar
\vartheta \equiv \sqrt{1-\vartheta^{2}} \, ,
\end{eqnarray}
and $K$ is the complete elliptic integral function  (see
Eqs.~(8.113.1) and (8.113.3) in Ref. [\onlinecite{GR}]). The
quantity $\Delta_{i,h}(1/\delta)$ in Eq.~(\ref{sigma}) is the
well-studied \cite{Cardybc} Casimir (or stress tensor) amplitude
for a strip ST without needle, with boundaries $(i,h)$, and finite
aspect ratio $W/L = 1/\delta$ which has been introduced in
Eqs.~(\ref{stressscale}) and~(\ref{stressscale'}) and is given
explicitly in Appendix~\ref{appstripbound}.

The symmetries of $F_{\perp}^{(i[h])}$ addressed above
Eq.~(\ref{arbperp}) are reflected in
Eqs.~(\ref{iJ})-(\ref{theta'}) by the symmetries of
$\Delta_{i,j}(1/\delta)$ discussed below Eq.~(\ref{stressscale'}).

For the special cases $(i,h)=(O,O)$ and $(+,O)$ we have checked
that Eqs.~(\ref{iJ})-(\ref{theta'}) are consistent with the simple
expressions given in Eqs.~(\ref{arbperp}) and~(\ref{arbperp+O}).
In particular, $\tilde{\tau}_{O,O}(\vartheta)$ vanishes for all
$\vartheta$.

The present results for a needle embedded in a half plane involve,
via $\Delta_{i,h}(1/\delta)$, knowledge about the strip ST without
needle but with finite aspect ratio because the two geometries are
related by a conformal mapping as explained in Appendix
\ref{apparbbreak}.

Varying $\vartheta$ from 0 to 1,  for all combinations $(i,h)$
Eqs.~(\ref{iJ})-(\ref{theta'}) provide  the complete crossover of
the force $-(\partial/\partial a_{N})F_{\perp}^{(i[h])}/k_{B}T$ as
the length $D$ and the position (determined by the distances
$a_{N}$ or $a_{<}$) of the needle in the half plane change from
small and remote from the boundary to large and close to the
boundary. Equation (\ref{K+K^{*}}) tells us that the corresponding
change in the geometry of the strip ST is from remote ($\delta
\searrow 0$) to close ($\delta \nearrow \infty$) strip boundaries,
as expected.

In the small needle limit $\vartheta \, \searrow \, 0$ the
quantity $\tilde{\tau}_{i,h}$ vanishes (compare
Eq.~(\ref{Deldist})) and the expansion for small $\vartheta$ is
provided by Eqs.~(\ref{Deltaijdistant}), (\ref{Deltadistant'}),
and (\ref{Deltadistant''}) and the form of $\sigma$ in Eq.
(\ref{phi+gamma}). We check in the paragraph containing Eq.
(\ref{+-expand}) that this expansion is consistent with the small
needle expansion in Eqs.~(\ref{Fperp}),~(\ref{Z}),
and~(\ref{halfzeta}).

In the long needle limit $\vartheta \nearrow 1$ one has
$\tilde{\tau}_{i,h} \to 16[\Delta_{i,i}(0)-\Delta_{i,h}(0)] /\pi$
with $\Delta_{i,h}(0)$ given below Eq.~(\ref{stressscale'}) so
that \cite{Maghrebi}
\begin{eqnarray} \label{long}
&&-(\partial/\partial a_{N})F_{\perp}^{(i[h])}/k_{B}T
 \, \to \, [-3, \, -3, \, 61, \, 5]/(32 a_{<}) \, , \nonumber \\
&& \qquad (i[h])=[(O[O]),(+[+]),(-[+]),(+[O])] \, .
\end{eqnarray}
The behavior $\propto a_{<}^{-1}$ of the force per $k_{B}T$ acting
on the needle of infinite length $D=\infty$ follows easily
\cite{Maghrebi} from comparing its scaling dimension with that of
the only remaining mesoscopic variable $a_{<}$. Unlike the
expansion of the force for a small needle in which {\it
fractional} powers such as $a_{N}^{-1} (D/ a_{N})^{1/8}$ may
occur, the expansion around the long needle limit involves only
contributions for which Eq.~(\ref{long}) is multiplied with
positive {\it integer} powers of $a_{<}/D \equiv \alpha$ (compare
Eqs.~(\ref{Deltaijclose}) and~(\ref{Deltaclose''}), the relation
between $\kappa$ and $\bar \vartheta$ in Eq.~(\ref{phi+gamma}),
and the relation $\bar
\vartheta^{2}=4\alpha(1+\alpha)/(1+2\alpha)^{2}$ which follows
from Eq.~(\ref{asmaller})).

In the limit $\vartheta \nearrow 1$ the free energy per $k_{B}T$ tends to
\begin{eqnarray} \label{long'}
&&F_{\perp}^{(i[h])}/ (k_{B}T) \equiv f_{i[h]}^{\perp} \nonumber
\\
&& \to ([-3,-3,61,5]/32) \ln (C_{i,h}/(1-\vartheta)) \, , \nonumber \\
&& (i[h])=[(O[O]),(+[+]),(-[+]),(+[O])] \, ,
\end{eqnarray}
where $1/(1-\vartheta) \to D/(2 a_{<})$ in terms of $a_{<}$. Here
$C_{i,h}$ are numbers, in particular $C_{O,O}=2^{5/3}$ and
$C_{+,O}=2^{3/5}$ (see Eqs.~(\ref{arbperp}) and
(\ref{arbperp+O})). This logarithmic behavior of
$F_{\perp}^{(i[h])}/(k_{B}T)$ for the long {\it perpendicular}
needle should be compared with the power law behavior
$F_{\parallel}^{(i[h])}/(k_{B}T) \to \Delta_{i,h}D/a_{<}$ of the
free energy for a long  needle aligned at a small distance $a_{<}$
{\it parallel} to the boundary which is obtained by turning the
perpendicular needle about that end which is closer to the
boundary. For a particle of {\it circular} shape \cite{ber} with a
radius $R$ much larger than the distance $a_{<}$ between the
closest points of the circle and the boundary, the interaction
free energy $F^{(i[h])}/(k_{B}T) \to  \pi \Delta_{i,h}
(2R/a_{<})^{1/2}$ exhibits, as expected intuitively, a power law
exponent the value of which lies in between those corresponding to
perpendicular and parallel needles. Unlike the long perpendicular
needle, for the long parallel needle and the large circle the
exact results of the interaction free energies quoted above are
found already within the Derjaguin approximation and exhibit an
amplitude proportional to $\Delta_{i,h}$ which depends on $(i,h)$.
For all three particles (the perpendicular and parallel needles as
well as the circle) the particle-boundary interaction is
attractive (repulsive) for $i=h$ ($i \neq h$).

\begin{figure*}
\includegraphics[width=0.49\textwidth]{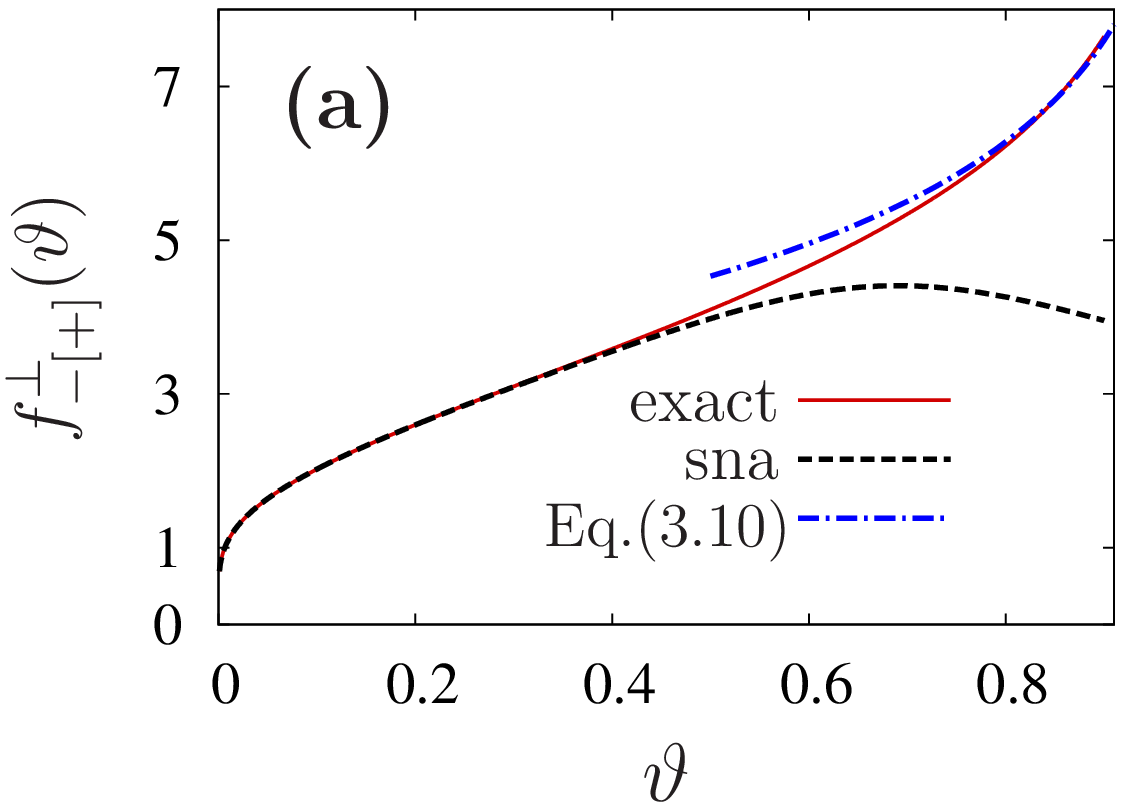}
\includegraphics[width=0.49\textwidth]{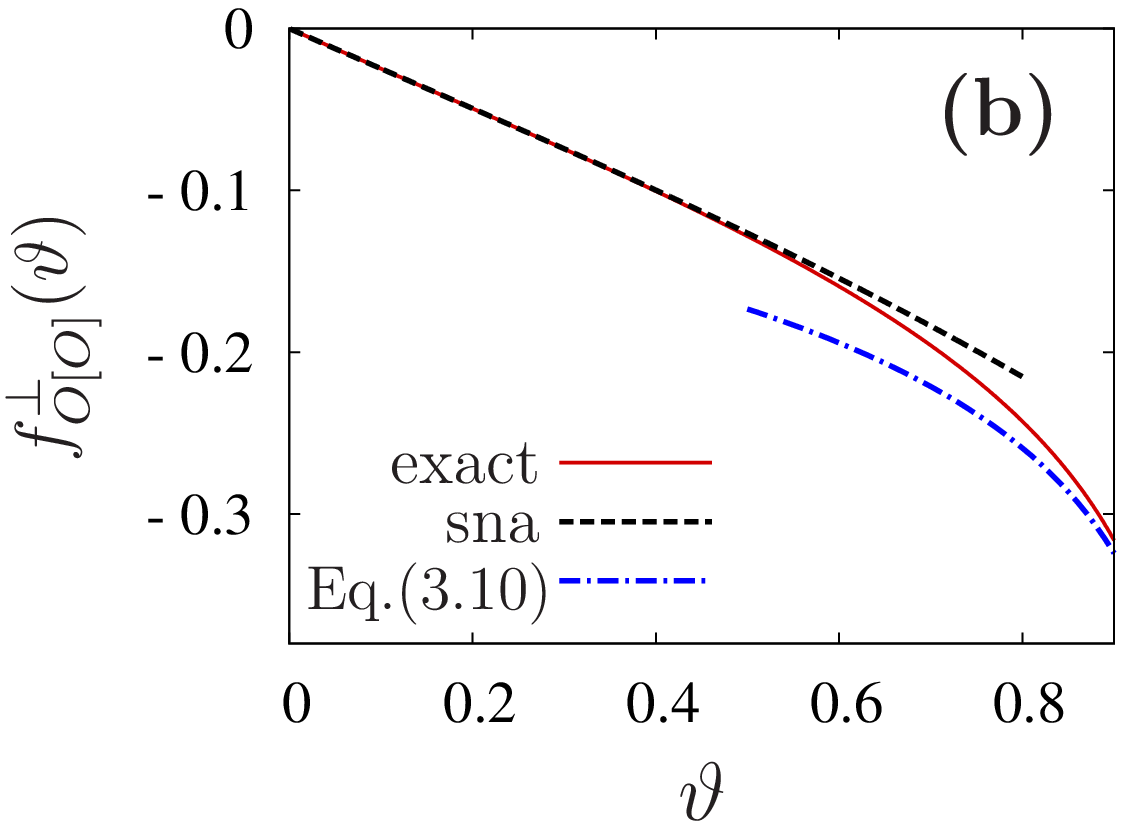}
\includegraphics[width=0.49\textwidth]{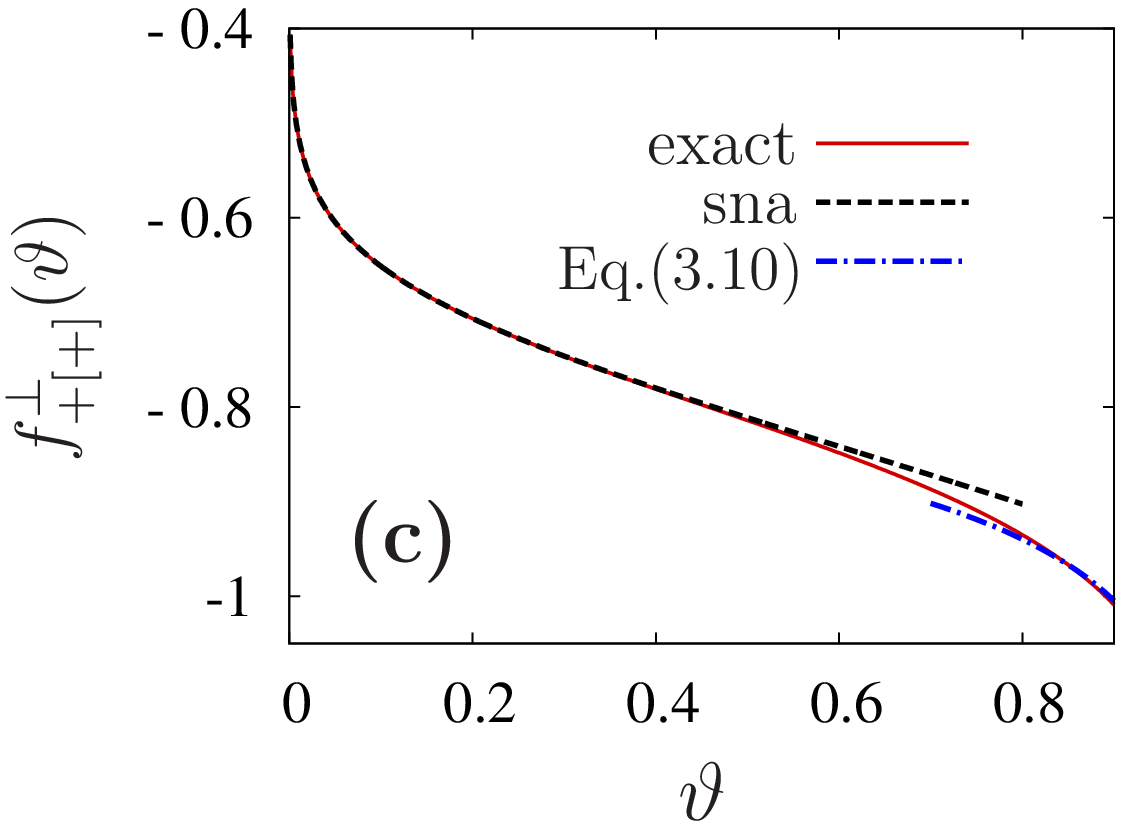}
\includegraphics[width=0.49\textwidth]{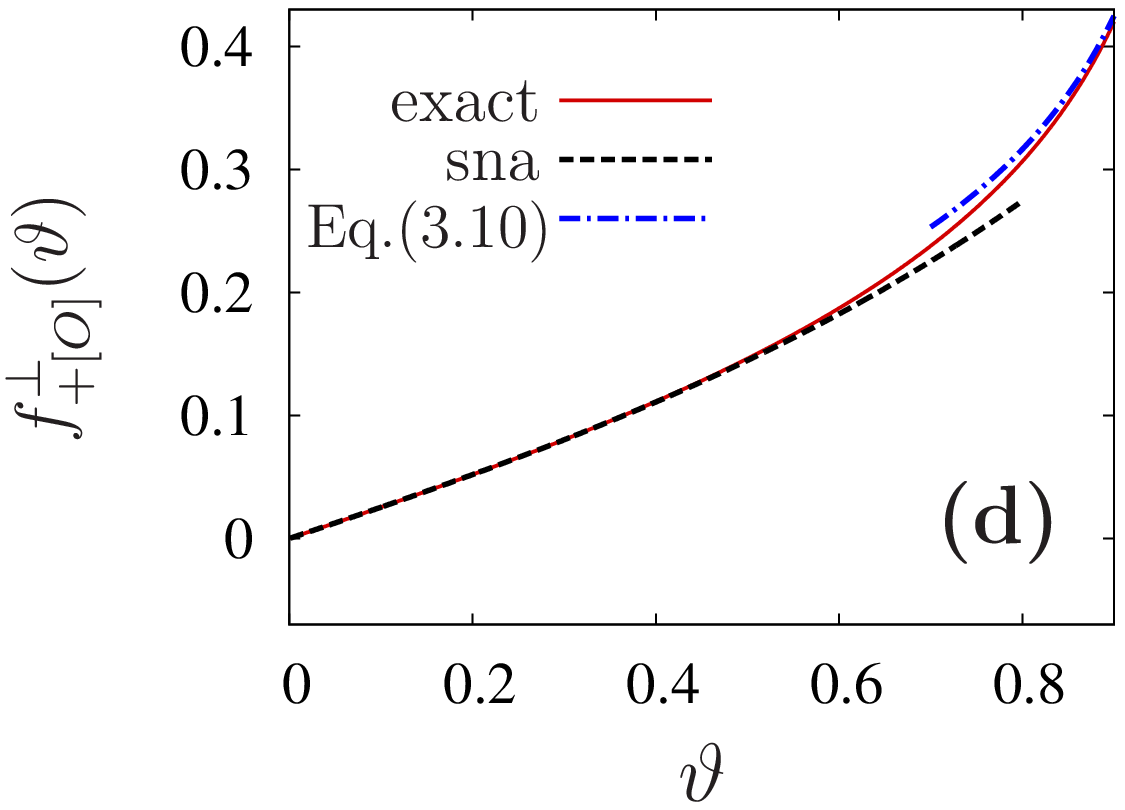}
\caption{Free energy cost per $k_{B}T$, $f_{i[h]}^{\perp}$, to
transfer a needle of length $D$ and universality class $h$ from
the unbounded plane (bulk) to the half plane, with an orientation
which is perpendicular to the boundary of the half plane of
universality class $i$. The dependence on $\vartheta=D/(2 a_{N})$,
with $a_{N}$ as the distance of the needle center from the
boundary, is shown for various combinations $(i[h])$. The exact
results for arbitrary $D/2 \leq a_{N}$ are compared with the
``small needle approximation'' (``sna'', Eq.~(\ref{SNE})), which
is  valid for small $\vartheta$, and with the long needle behavior
for $\vartheta \to 1$ as given by Eq.~(\ref{long'}). The
interaction between the boundary and the needle is attractive
(repulsive) for equal (different) universality classes $i=h$ ($i
\neq h$) shown in (b) and (c) ((a) and (d)).} \label{FIG_III_1}
\end{figure*}
Figure~\ref{FIG_III_1}(a) shows  a comparison between the exact
result for $F_{\perp}^{(-[+])}/(k_{B}T) \equiv
f_{-[+]}^{\perp}(\vartheta)$ which follows from Eq.~(\ref{iJ})
upon integration \cite{integration} and its ``small needle
approximation'' (``sna'') (compare
Eqs.~(\ref{+-predict})~and~(\ref{SNE})). The approximation
reproduces the exact result very well for
$\vartheta<\vartheta_{0}$ with $0.3 \lesssim \vartheta_{0}
\lesssim 0.4$ while it fails  for  $\vartheta \gtrsim 0.6$ where
it shows an unphysical maximum. For the {\it strip} situation
$(-[+]j)$ with the center of the perpendicular needle at $v_{N}$,
the question arises down to which proximity of the $+$ needle to
the lower $-$ boundary of the strip one can expect the ``small
needle approximation'' to be reasonably valid. Identifying the
distance $a_{N}=D/(2\vartheta)$ in the half plane with the
distance $W/2-|v_{N}| \equiv W/2 + v_{N}$ in the strip suggests
the approximation  for $F_{\perp}^{(-[+]j)}$ to be valid if
$|v_{N}|/W$ is smaller than $[1-D/(W\vartheta_{0})]/2$. For $0.3
\lesssim \vartheta_{0} \lesssim 0.4$ and the value $D/W=21/101$
used in our MC simulations (see, c.f., Subsec.~\ref{compnorm})
this rough argument predicts that $|v_{N}|/W$ should be kept
smaller than about $1/6$ or $1/4$, i.e., $v_{N}$ should be kept
rather close to the midline $v=0$. For the anisotropy $\Delta
F^{(-[+]j)}$ of the free energy of the needle in the strip there
is even more uncertainty concerning the validity of the ``sna''
because the corresponding comparison for $F_{\parallel}^{(-[+])}$
is lacking, i.e., the counterpart of Fig.~\ref{FIG_III_1}(a) for a
needle parallel to the boundary of the half space is missing.

Figures~\ref{FIG_III_1}(b)-(d) show the corresponding results for
the free energies $k_{B}T f_{i[h]}^{\perp}$ with $(i,h)=(O,O), \,
(+,+)$, and $(+,O)$. For $i=h$ and $i \neq h$, $f_{i[h]}^{\perp}$
is negative and positive, respectively.

(ii) {\it Needle in infinitely long symmetric strip}

Here we consider a needle of universality class $h$ and with
parallel orientation $||$ at the center line $v=0$ of an
infinitely long $(i,i)$ strip of width $W$. We determine the free
energy cost $F_{\parallel} \equiv F_{\parallel}^{(i[h]i)} = k_{B}T
f_{i[h]i}^{\parallel}(\theta)$ and the corresponding disjoining
force $-(\partial/\partial W)F_{\parallel}^{(i[h]i)} =
(k_{B}T/W)g_{i[h]i}^{\parallel}(\theta)$ between the two boundary
lines of the strip upon inserting the needle from the bulk with
parallel orientation. Here, the partial derivative is taken with
the needle length $D$ kept fixed. The disjoining force increases
(decreases) if $g_{i[h]i}$ is positive (negative). The $(i,h)$
symmetries of $F_{\parallel}^{(i[h]i)}$ are those of
$\Delta_{i,h}$ as can be inferred from Eq.~(\ref{arbiJ}) below.
The results depend on the length ratio \cite{simp}
\begin{eqnarray} \label{theta}
\theta={\pi D \over W}
\end{eqnarray}
and provide the crossover from the small needle behavior for
$\theta \ll 1$, consistent with the ``small needle expansion'', to
the long needle behavior for $\theta \gg 1$. In the latter case
one has
\begin{eqnarray} \label{longiJ}
&&{F_{\parallel}^{(i[h]i)} \over k_{B} T} \equiv
f_{i[h]i}^{\parallel} \, \to \, D \Biggl[ 2 {\Delta_{i,h}(0)
\over (W/2)} - {\Delta_{i,i}(0) \over W} \Biggr] \nonumber \\
&& \qquad \qquad \qquad \qquad = \, \theta[-1, \, -1, \, 31, \, 3]/16 \, ,  \nonumber \\
&& \quad (i,h) = [(O,O),(+,+),(-,+),(+,O)]
\end{eqnarray}
because upon inserting a long parallel needle with its center at
$u=v=0$ transforms, within the interval $-D/2 < u < D/2$, the
$(i,i)$ strip of width $W$ into two independent adjacent $(i,h)$
strips of width $W/2$ each.

(iia) For $i=h=O$ the free energy to insert the needle from the
bulk reads
\begin{eqnarray} \label{arbord}
{F_{\parallel}^{(O[O]O)} \over k_{B}T} \, = \, - {\theta \over 8}
+ {1 \over 16} \ln {\sinh \theta \over \theta} \, .
\end{eqnarray}

(iib) For $i=+, \, h=O$ we find
\begin{eqnarray} \label{arbord+O}
{F_{\parallel}^{(+[O]+)} \over k_{B}T} \, = \, {\theta \over 8} +
{1 \over 16} \ln {\sinh \theta \over \theta}
\end{eqnarray}
which is Eq.~(\ref{arbord}) with $\theta$ replaced by $-\theta$.
Equations~(\ref{arbord}) and~(\ref{arbord+O}) are derived in
Appendix~\ref{apparbord}.

(iic) For {\it arbitrary} $(i,h)$ we find (see Appendix
\ref{apparbbreak})
\begin{eqnarray} \label{arbiJ}
&&-{\partial \over \partial W} {F_{\parallel}^{(i[h]i)}
 \over k_{B} T} = -{\partial \over \partial W} {F_{\parallel}^{(O[O]O)}
 \over k_{B} T} - {1 \over 4 W} {\theta
\over t} \tilde{\tau}_{i,h}(t) \nonumber \\
&& \qquad \qquad \equiv  {1 \over 4W} \Biggl\{- {1 \over 4} +
\theta \Bigl[ {1 \over 24} \Bigl( {1 \over t} + t \Bigr) + {1
\over t}
\rho_{i,h}(t) \Bigr] \Biggr\} , \nonumber \\
&& \quad \quad \quad t \, \equiv \, \tanh (\theta/2)\, ,
\end{eqnarray}
where $\tilde{\tau}_{i,h}(t)$ and $\rho_{i,h}(t)$ are the
functions $\tilde{\tau}_{i,h}(\vartheta)$ and
$\rho_{i,h}(\vartheta)$ from Eqs.~(\ref{sigmatilde}) and
(\ref{sigma}), respectively, evaluated at $\vartheta=t$.
Equations~(\ref{arbord}) and~(\ref{arbord+O}) as well as Eq.
(\ref{arbiJ}) are consistent with the ``small needle expansion''
for small $\theta$ (see Eqs.~(\ref{expandpar})) and with
Eq.~(\ref{longiJ}) for large $\theta$.

\begin{figure}[h]
\includegraphics[width=0.49\textwidth]{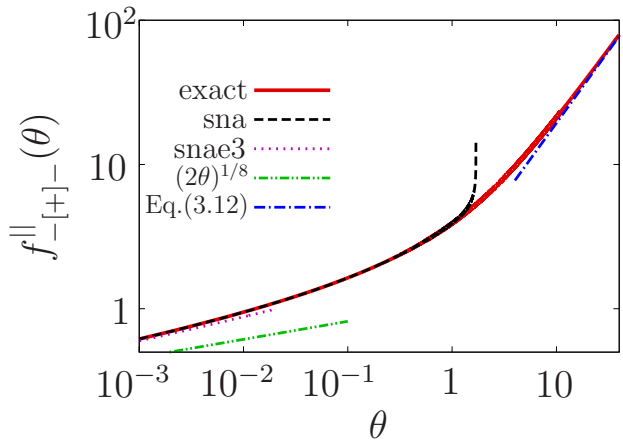}
\caption{Free energy cost per $k_{B}T$, $f_{-[+]-}^{\parallel}$,
to remove a needle of universality class $+$ and length $D$ from
the bulk and insert it along the midline of an infinitely long
strip with $(-,-)$ boundary conditions and width $W$ as a function
of $\theta=\pi D/W$ (see Ref. [\onlinecite{integration}] and Eq.
(\ref{arbiJ})). The exact result for arbitrary length ratio
$\theta$ is compared with the ``small needle approximation''
(``sna'', Eq.~(\ref{barfpar})), valid for small $\theta$, and with
the long needle behavior according to Eq.~(\ref{longiJ}). Also
shown are the first term, $(2\theta)^{1/8}$, and the sum of the
first three terms
$(2\theta)^{1/8}+\theta^{1/4}/2^{3/4}+(2\theta)^{3/8}/3$, denoted
as ``snae3'', in the expansion of $f_{-[+]-}^{\parallel}$ in terms
of powers of $\theta$ for $\theta \ll 1$ (see Eqs.
(\ref{expandpar}) and~(\ref{since'})). Expanding
$f_{-[+]-}^{\parallel}$ proves to be less successful than the
``sna'' within which $\exp (-f_{-[+]-}^{\parallel})$ is expanded
and which agrees with the exact result up to $\theta \approx 1$.
The first term (dash-double-dotted green line) approaches the
exact result (full red line) for $\theta \to 0$ only very slowly.}
\label{FIG_III_2}
\end{figure}
Figure~\ref{FIG_III_2} shows a comparison between the exact result
for $F_{\parallel}^{(-[+]-)}/(k_{B}T) \equiv
f_{-[+]-}^{\parallel}(\theta)$ which follows from
Eq.~(\ref{arbiJ}) via integration \cite{integration} and its
``small needle approximation'' (compare Eqs.~(\ref{full}),
(\ref{approxpar}),~(\ref{since'}), and~(\ref{barfpar})). For the
value $\theta=21 \pi / 101=0.653$ as used in our MC simulations
(see Subsec.~\ref{compnorm}) there is good agreement. However, we
have no such comparison for $F_{\perp}^{(i[h]i)}$.

\begin{figure}[h]
\includegraphics[width=0.49\textwidth]{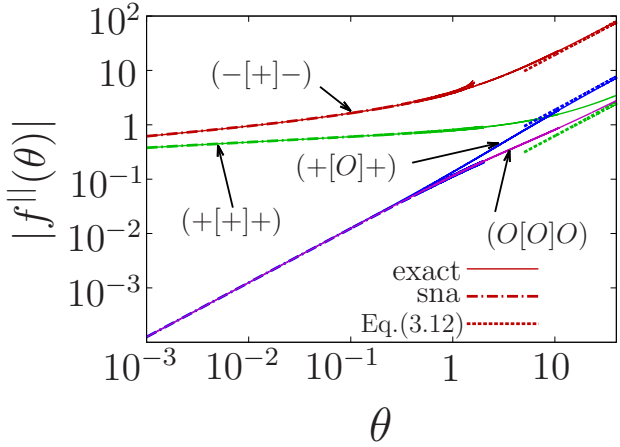}
\caption{Same as Fig.~\ref{FIG_III_2} for all four independent
combinations $(i[h]i)$ of a  needle of universality class $h$
placed along the midline of an $(i,i)$ strip as function of
$\theta=\pi D/W$. Full, dash-dotted, and dotted lines correspond
to the exact results, the ``sna'', and Eq.~(\ref{longiJ}),
respectively. For $h=i$ ($h \neq i$) $f_{i[h]i}^{\parallel}$ is
negative (positive), decreasing (increasing) with increasing
$\theta$, and implies a negative (positive) contribution
$-\partial_{W} F_{\parallel}^{(i[h]i)}=(k_{B}T/W)\theta
\partial_{\theta}f_{i[h]i}^{\parallel}$ to the disjoining force.
Thus the attractive Casimir interaction (see Eq.
(\ref{stressscale})) with the negative universal disjoining force
$-\partial_{W}k_{B}T \Phi_{\rm ST}^{(i,i)}=\Delta_{i,i} k_{B}T
L/W^{2}=-(\pi/48)k_{B}T L/W^{2}$ between the $(i,i)$ boundaries of
the long strip without the needle becomes even more attractive
(less attractive) due to the presence of the needle with $h=i$ ($h
\neq i$).} \label{FIG_III_3}
\end{figure}
All four possible independent free energies $k_{B}T
f_{i[h]i}^{\parallel}(\theta)$ for needle insertion are collected
in Fig.~\ref{FIG_III_3}. They are negative (positive) for $i=h$
($i \neq h$). In this double logarithmic plot the power laws as
obtained for small $\theta$ from the ``small needle expansion''
and for large $\theta$ from the long needle expression
(Eq.~(\ref{longiJ})) appear as straight lines.

(iii) {\it Periodic strip}

Finally we consider an infinitely long strip of width $W$ with
{\it periodic} boundary condition in $v$ direction. In this case
the free energy cost to insert a needle parallel to the strip
reads (see Appendix~\ref{apparbord})
\begin{eqnarray} \label{arbperiod}
{F_{\parallel} \over k_{B}T} \, = \, {1 \over 16} \ln {\sinh
\theta \over \theta} \, .
\end{eqnarray}
This result holds for any needle universality class $h=O,+,-$ and
it is valid for an arbitrary ratio $\theta=\pi D / W$.
Equation~(\ref{arbperiod}) should be compared with Eq.
(\ref{arbord}). For $\theta \gg 1$ the rhs of
Eq.~(\ref{arbperiod}) tends to $\theta/16=D
\times[\Delta_{h,h}(0)-\Delta_{P}(0)]/W$. This is consistent with
the periodic strip of width $W$ being transformed, by inserting a
parallel needle,   to a $(h,h)$ strip of width $W$ within a
$u$-interval of length $D$. Consistent with the ``small needle
expansion'', for $\theta \ll 1$ the rhs of Eq.~(\ref{arbperiod})
tends to $\theta^2 /96+{\cal O}(\theta^4)$.

\section{FREE ENERGY ANISOTROPY FROM MONTE CARLO SIMULATIONS} \label{SIM}
The anisotropic shape of a bounded critical system induces
orientation dependent properties for embedded non-spherical
particles. In Sec.~\ref{SP} the operator expansion has provided
the asymptotic scaling properties for a ``small'' but
``mesoscopic'' particle in a ``large'' system. In the present
section we address the issue to which extent these asymptotic
properties capture the actual behavior in specific critical model
systems. Concerning the needles studied here, we want to check
whether the asymptotic predictions of Sec.~\ref{SP} for $\Delta F$
can already be observed within a lattice model with numerous, but
not too many rows and columns so that the model is amenable to
simulations. In this section we describe how to set up the
corresponding Monte Carlo simulations and to calculate $\Delta F$.
In Sec.~\ref{compare} we compare these simulation data with the
corresponding analytic predictions. This allows us to judge both
the achievements and the limitations of the ``small particle
approximation''.

\subsection{Model} \label{simA}

For the simulation we use the lattice version of the Ising strip
described in the Introduction and shown in Fig. 1. The
implementation of the double periodic boundary conditions is
obvious. Beyond that, here we describe in more detail strips with
boundaries $(i,j)$. In this case the lattice Hamiltonian ${\cal
H}_{\rm ST}$ for strips without a needle reads (see Sec.
\ref{intro}):
\begin{eqnarray} \label{eq:hini}
&&{\mathcal H}_{\mathrm ST} /{\cal J}= - \sum \limits_{\langle
u,v;u',v' \rangle} \, J_{u,v;u',v'} \, s_{u,v} \, s_{u',v'}
\nonumber \\
&& \qquad \,  -\Lambda_{i}^{(1)}\sum_{u}s_{u,-(W+1)/2} \,
s_{u,-(W-1)/2} \nonumber \\
&& \qquad \,  -\Lambda_{j}^{(1)}\sum_{u}s_{u,(W+1)/2} \,
s_{u,(W-1)/2} \,
\end{eqnarray}
with ${\cal J} > 0$ and $J_{u,v;u',v'} =1$ for nearest neighbors
(denoted by $\langle u,v;u',v' \rangle$) and zero otherwise. The
fluctuating Ising spins $s= \pm 1$ reside on the vertices of the
square lattice consisting of $W$ rows and $L$ columns with
periodic boundary conditions in the $u$ direction. The two last
terms generate ``ordinary'' or ``normal'' ($+/-$) boundaries near
$v=-W/2$ and $v=W/2$ by means of fixed values
$s_{u,-(W+1)/2}=s_{u,(W+1)/2}=1$ for the spins in the two
additional outside rows and by choosing independently
$\Lambda_{i}^{(1)}$ and $\Lambda_{j}^{(1)}$ equal to 0
(``ordinary'') or $\pm 1$ ($+/-$ ``normal''). This serves as a
microscopic realization of all the pairs $(i,j)$ of strip boundary
types considered in Sec.~\ref{SP} which exhaust all possible
surface universality classes in $d=2$ (up to the ``extraordinary''
boundary type \cite{duality,dual} corresponding to infinitely
strong couplings between surface spins).

For reasons given in the Introduction and shown in Fig. 1, for
``ordinary'' (``normal'') needles the number $W$ of rows is taken
to be even 2, 4, 6, ... (odd 1, 3, 5, ...) and the components of
the lattice vertices $u, \, v$ are half odd integers $\pm 1/2, \,
\pm 3/2, \, \pm 5/2, ...$ (integers $0, \, \pm 1, \, \pm 2, \,
...$). The components $u=u_{N}$ and $v=v_{N}$ of the needle
centers $\times$ are integers for both types of needles.

In order to be able to compare the Monte Carlo data with the
results of Sec. II we consider the system at its bulk critical
point $T_c$ with ${\cal J}/(k_BT_{c})=\frac{1}{2}\ln(\sqrt{2}+1)$
where ${\cal J}$ is the coupling constant scaled out of ${\cal
H}_{\rm ST}$ (Eq.~(\ref{eq:hini})).

Inserting a needle of class $h=O$ or $h=\pm$  into strips amounts
to appropriately removing bonds or fixing spins in accordance with
Fig.~\ref{FIG_I_1}  via additional terms in the Hamiltonian. For
given surface universality classes $(i,j)$ of the strip,
universality class $h$, needle length $D$, and the coordinate
$v_{N}$ of the center of the needle we introduce the notation
\begin{eqnarray} \label{Hperp}
{\cal H}={\cal H}_{0} \equiv {\cal H}_{\rm ST}+{\cal
H}_{\perp}^{(h)}
\end{eqnarray}
and
\begin{eqnarray} \label{Hparallel}
{\cal H}={\cal H}_{1} \equiv {\cal H}_{\rm ST}+{\cal
H}_{\parallel}^{(h)}
\end{eqnarray}
for the total lattice Hamiltonian ${\cal H}$ corresponding to the
needle being oriented perpendicular and parallel, respectively, to
the $u$ axis. The explicit forms of ${\cal H}_{\perp}^{(h)}$ and
${\cal H}_{\parallel}^{(h)}$ will be given in
Eqs.~(\ref{eq:HOperp}),~(\ref{eq:HOparallel}), and
(\ref{addH+perp})-(\ref{ext3}) below.
\subsection{Numerical algorithm} \label{simB}
\subsubsection{Coupling parameter approach} \label{simB1}
We consider two systems with the same configurational space
$\mathcal C$ (i.e., number and spatial connectivity of spins)
 and with Hamiltonians
${\mathcal H}_{0}$ and ${\mathcal H}_{1}$ as given in
Eqs.~(\ref{Hperp}) and~(\ref{Hparallel}). The corresponding free
energies are $F_{0,1}=-\frac{1}{\beta} \ln \sum_{\mathcal C}
\exp(-\beta {\mathcal H}_{0,1})$; $\beta=1/(k_{B}T)$ is the
inverse thermal energy. We are interested in the free energy
difference $\Delta F=F_{1}-F_{0}$.

For the computation of this free energy difference we use the
coupling parameter approach~\cite{Bennet}. To this end we
introduce the {\it cr}ossover Hamiltonian ${\mathcal H}_{\rm
cr}(\lambda)$ which depends on the coupling parameter $\lambda$,
\begin{equation}
\label{eq:hcr} {\mathcal H}_{\rm cr}(\lambda)={\mathcal
H}_{0}+\lambda \left( {\mathcal H}_{1}-{\mathcal H}_{0} \right)=
{\mathcal H}_{0}+\lambda \Delta {\mathcal H},
\end{equation}
with the Hamiltonian difference $\Delta {\mathcal H}= {\mathcal
H}_{1}-{\mathcal H}_{0} $ which in the present context is ${\cal
H}_{\parallel}^{(h)}-{\cal H}_{\perp}^{(h)}$. The derivative of
the corresponding {\it cr}ossover free energy $F_{\rm
cr}(\lambda)=-\frac{1}{\beta} \ln \sum_{\mathcal C} \exp(-\beta
{\mathcal H}_{\rm cr}(\lambda))$ with respect to the coupling
parameter reads
\begin{equation}
\label{eq:df} \frac{{\rm d} F_{\rm cr}(\lambda)}{\rm d \lambda}
\equiv F_{\rm cr}'(\lambda) = \frac{\sum_{\mathcal C} \Delta
{\mathcal H}
 \rm e^{-\beta {\mathcal H}_{\rm cr}(\lambda)}}{\sum_{\mathcal C}
 \rm e^{-\beta {\mathcal H}_{\rm cr}(\lambda)} }
=\langle\Delta {\mathcal H}\rangle_{\rm cr}(\lambda),
\end{equation}
where $\langle \Delta {\mathcal H} \rangle_{\rm cr}(\lambda) $ is
the Hamiltonian difference $\Delta {\mathcal H}$ averaged with
respect to ${\mathcal H}_{\rm cr}(\lambda)$. Therefore one can
compute the free energy difference by integrating $\left< \Delta
{\cal H } \right>_{\rm cr}$ with respect to $\lambda$:
\begin{equation}
\label{eq:alg} \Delta F=F_{1}-F_{0} = \int \limits_{0}^{1}
 F'_{\rm cr}(\lambda) {\rm d} \lambda
 =\int \limits_{0}^{1} \left< \Delta {\cal H } \right>_{\rm cr}(\lambda)
 {\rm d} \lambda.
\end{equation}
For the  forms of ${\cal H}_{0}$ and ${\cal H}_{1}$ given by Eqs.
(\ref{Hperp}) and~(\ref{Hparallel}), respectively, one has
\begin{eqnarray} \label{eq:hcr'}
&&\Delta {\cal H}={\cal H}_{\parallel}^{(h)}-{\cal
H}_{\perp}^{(h)} \, ,
\nonumber \\
&&{\cal H}_{\rm cr} \equiv {\cal H}_{\rm cr}^{(h)}={\cal H}_{\rm
ST} + (1-\lambda){\cal H}_{\perp}^{(h)}+\lambda {\cal
H}_{\parallel}^{(h)} \, .
\end{eqnarray}

\subsubsection{Needle of broken bonds} \label{simB2}

\begin{figure}[h]
\includegraphics[width=0.49\textwidth]{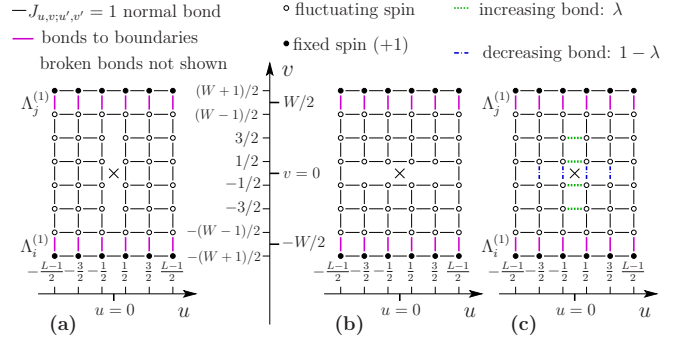}
\caption{Bond arrangements for a needle ($O$) of $D=4$ {\it broken
bonds} with center ($\times$) at $(u_{N},v_{N})=(0,0)$ at the
midline $v=0$ of a strip with $L=W=6$: (a) for the perpendicular
orientation of the needle with Hamiltonian ${\cal H}_{0}$; (b) for
the parallel orientation of the needle with Hamiltonian ${\cal
H}_{1}$; (c) for the crossover Hamiltonian ${\mathcal H}_{\rm
cr}^{(O)}(\lambda)$ which interpolates between (a) for $\lambda=0$
and (b) for $\lambda=1$. Bonds of strength $J=1$ are indicated by
thin black lines. Broken bonds ($J=0$) are not shown. The bonds
indicated by green dashed and blue dot-dashed lines have strengths
$\lambda$ and $1-\lambda$ which increase and decrease,
respectively, as $\lambda$ increases. As in Fig.~\ref{FIG_I_1} the
fluctuating spins $s_{u,v}= \pm 1$ with $|v| \leq (W-1)/2$ are
indicated by empty circles while the fixed spins $s_{u, \pm
(W+1)/2}=1$ in the two additional outside rows are indicated by
full circles. The nearest neighbor bonds between fixed and
fluctuating spins are indicated by thick magenta lines. For the
lower and upper boundary $i$ and $j$ they have the strengths
$\Lambda_{i}^{(1)}$ and $\Lambda_{j}^{(1)}$, respectively, where
$\Lambda^{(1)}$  equals 0 for an $O$ boundary and $\pm 1$ for a
$+/-$ boundary.} \label{FIG_IV_1}
\end{figure}

Here we consider the combination $(i[O]j)$ corresponding to an
``ordinary'' needle which consists of an even number $D$ of broken
bonds in a strip with an even number $W$ of rows. Figure~\ref{FIG_IV_1}
shows the example of a needle with $D=4$ broken
bonds in the center of an $L \times W=6 \times 6$ strip. The
Hamiltonians ${\mathcal H}_{0}$ and ${\mathcal H}_{1}$ for
perpendicular and parallel needle orientation have the same form
as the right hand side of Eq.~(\ref{eq:hini}) but with reduced
interaction constants $J_{u,v;u',v'}$ which depend suitably on the
coordinates $(u,v)$ and $(u',v')$ of nearest neighbor spins.

Choosing without restriction $u_{N}=0$, the needle with center at
$(0,v_{N})$ is inserted in perpendicular orientation by
``breaking'', i.e., removing the $D$ lattice bonds which at $v=
v_{N} \pm 1/2, \, v_{N} \pm 3/2, \, ..., \, v_{N} \pm (D-1)/2$
extend from $u=-1/2$ to $u=1/2$. This is accomplished by adding
\begin{eqnarray} \label{eq:HOperp}
{\mathcal H}^{(O)}_{\perp}/{\cal J}&=& \sum \limits_{k=1}^{D}
s_{-1/2,v_{N}-(D+1)/2+k} \, s_{1/2,v_{N}-(D+1)/2+k} \nonumber \\
&\equiv& \sum \limits_{\langle {\rm inc.}\rangle }
\end{eqnarray}
to the Hamiltonian ${\cal H}_{\rm ST}/{\cal J}$ without needle
(see Eq.~(\ref{Hperp})). Similarly for the parallel orientation of
the needle one has to add
\begin{eqnarray} \label{eq:HOparallel}
{\mathcal H}^{(O)}_{||}/{\cal J}&=&  \sum \limits_{k=1}^{D}
s_{-(D+1)/2+k,v_{N}-1/2} \, s_{-(D+1)/2+k,v_{N}+1/2} \nonumber \\
&\equiv& \sum \limits_{\langle {\rm decr.} \rangle} \,
\end{eqnarray}
so that $\Delta {\cal H}={\cal H}_{\parallel}^{(O)}-{\cal
H}_{\perp}^{(O)}$. Figures~\ref{FIG_IV_1}(a) and (b) illustrate
these configurations for the special case $v_{N}=0$.

The sums in Eqs.~(\ref{eq:HOperp}) and~(\ref{eq:HOparallel}) have
been characterized by subscripts $\langle {\rm inc.}\rangle$ and
$\langle {\rm decr.}\rangle$ because in the crossover Hamiltonian
following from Eq.~(\ref{eq:hcr'}),
\begin{equation} \label{eq:HOcr}
{\mathcal H}_{\rm cr}^{(O)}(\lambda)=\tilde{\cal H}^{(O)}-\lambda
{\cal J}\sum \limits_{\langle {\rm inc.}\rangle} -(1-\lambda)
{\cal J} \sum \limits_{\langle {\rm decr.} \rangle} \, ,
\end{equation}
they appear with a prefactor $ -\lambda$ and $ -(1-\lambda)$,
respectively, representing sums of products of spins coupled by
nearest neighbor bonds with strengths $\lambda {\cal J}$ and
$(1-\lambda){\cal J}$ which increase and decrease, respectively,
as $\lambda$ increases. Here $\tilde{\cal H}^{(O)} \equiv {\cal
H}_{\rm ST} +{\cal J}\sum \limits_{\langle {\rm inc.} \rangle}
+{\cal J}\sum \limits_{\langle {\rm decr.} \rangle}$ equals ${\cal
H}_{\rm ST}$ in Eq.~(\ref{eq:hini}) but with both types of nearest
neighbor bonds missing which are broken in the perpendicular or
the parallel orientation of the needle. This corresponds to
Fig.~\ref{FIG_IV_1}(c) with both dashed and dash-dotted bonds removed
(so that in $\tilde{\cal H}^{(O)}$ only those bonds of ${\cal
H}_{\rm ST}$ remain which are outside a hole with the shape of a
cross). In ${\mathcal H}_{\rm cr}^{(O)}(\lambda)$, however, the
two types of missing bonds are replaced by the bonds of increasing
and decreasing strength as illustrated in Fig.~\ref{FIG_IV_1}(c)
by the green dashed and blue dash-dotted lines. This obviously
leads to the crossover from the perpendicular to the parallel
needle orientation as $\lambda$ increases from 0 to 1.

On this basis, following the steps described by Eqs.
(\ref{eq:hcr})-(\ref{eq:hcr'}) allows us to calculate
$F_{\parallel}-F_{\perp} \equiv F_{1}-F_{0}=\Delta F$ for the
combination $(i[O]j)$.

\subsubsection{Needle of fixed spins} \label{simB3}

In this subsection we consider the lattice version of the case
$(i[+]j)$ in which a needle consisting of an odd number $D$ of
spins fixed in the $+$ direction is embedded in a strip with an
odd number $W$ of rows.
\begin{figure}[h]
\includegraphics[width=0.49\textwidth]{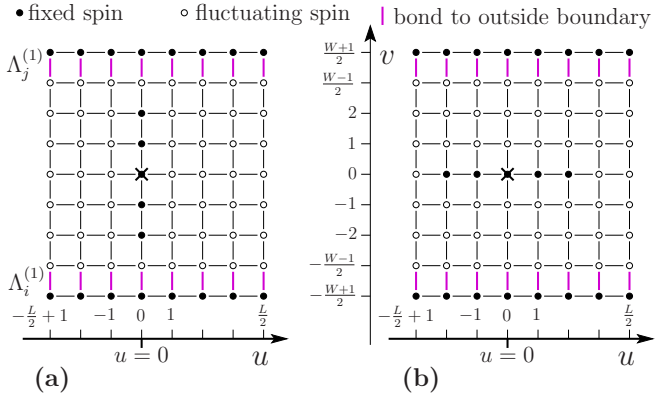}
\caption{
Bond arrangements for a needle of $D=5$ {\it spins fixed} in the
$+$ direction with the needle center ($\times$) at
$(u_{N},v_{N})=(0,0)$ at the midline $v=0$ of a strip with $L=8$
and $W=7$ for the perpendicular direction of the needle (a) and
for the parallel direction of the needle (b). The strengths
$\Lambda^{(1)}$ of the bonds near the strip boundaries are as
explained in Fig.~\ref{FIG_IV_1}.}
\label{FIG_IV_2}
\end{figure}
Figures~\ref{FIG_IV_2}(a) and (b) show the
example of a needle with $D=5$ in the center of an $L \times W=8
\times 7$ strip. The partition functions and free energies of
these lattice models remain unchanged if the bonds between the
fixed needle spins and their, in all, $2D+2$ fluctuating nearest
neighbors are removed and these $2D+2$ neighbors are coupled
instead with the bulk strength ${\cal J}$ to a {\it single}
exterior spin $s_{0}=+1$ which is kept fixed in the $+$ direction.
The coupling to this single spin or to the $D$ fixed spins has the
same effect on the fluctuating spins, namely that of a magnetic
field acting on the $2D+2$ neighboring spins. Once this coupling
to $s_{0}$ is in place, for the following it is convenient to
replace each of the $D$ fixed needle spins by a freely fluctuating
spin, i.e., free of any couplings. This changes the free energy
per $k_{B}T$ only by $D \ln 2$, independent of the orientation of
the needle and thus drops out of $\Delta F$. For the above example
discussed in Fig.~\ref{FIG_IV_2} this alternative model is
illustrated in Figs.~\ref{FIG_IV_3}(a) and (b), with the couplings
to the external spin denoted by north-east arrows.
\begin{figure}
\includegraphics[width=0.49\textwidth]{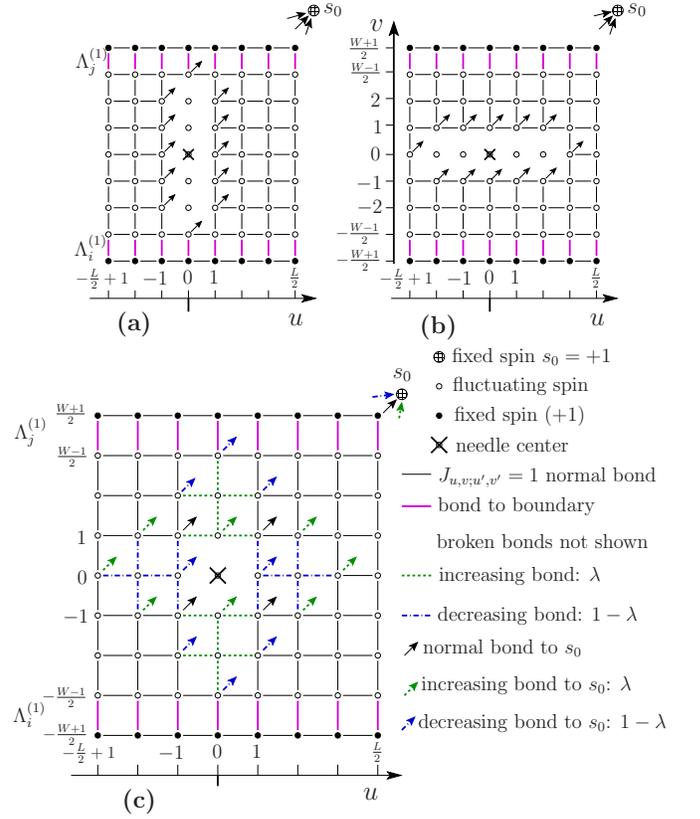}
\caption{
(a) and (b) describe how to mimic the configurations discussed in
Fig.~\ref{FIG_IV_2} by means of couplings (north-east arrows) to a
fixed external spin $s_{0}=+1$; (c) shows the arrangement of bonds
in the crossover Hamiltonian ${\mathcal H}_{\rm
cr}^{(+)}(\lambda)$ (see Eq.~(\ref{+cross})) with their strengths
indicated in the right margin. The arrangement reduces to that of
(a) and (b) for $\lambda=0$ and $\lambda=1$, respectively. For
further explanations see the main text. } \label{FIG_IV_3}
\end{figure}

The corresponding additional terms in the Hamiltonian can be
written as
\begin{equation} \label{addH+perp}
{\mathcal H}^{(\mathrm +)}_{\perp}= \sum \limits_{\langle {\rm
zero} \rangle}+\widehat {\sum \limits_{\langle {\rm inc.}
\rangle}} - \sum^{(+)} \limits_{\langle {\rm one} \rangle} -
\sum^{(+)}_{\langle {\rm decr.} \rangle}
\end{equation}
and
\begin{equation} \label{addH+parallel}
{\mathcal H}^{(\mathrm +)}_{||}= \, \sum \limits_{\langle {\rm
zero} \rangle} + \widehat {\sum \limits_{\langle {\rm decr.}
\rangle}}-\sum^{(+)} \limits_{\langle {\rm one}\rangle}-\sum^{(+)}
\limits_{\langle {\rm inc.} \rangle}
\end{equation}
(so that $\Delta {\cal H}={\cal H}_{\parallel}^{(+)}-{\cal
H}_{\perp}^{(+)}$) where the sums
\begin{equation} \label{nn1}
\sum \limits_{\langle {\rm zero} \rangle}=s_{0,v_{N}}
(s_{-1,v_{N}}+s_{1,v_{N}}+s_{0,v_{N}+1}+s_{0,v_{N}-1})  ,
\end{equation}
\begin{eqnarray} \label{nn2}
&&\widehat {\sum \limits_{\langle {\rm inc.} \rangle}}
=\sum \limits_{k=1}^{(D-1)/2}
\Bigl[s_{0,v_{N}+k} (s_{-1,v_{N}+k}+s_{1,v_{N}+k}) \\
&& \qquad \qquad \qquad + s_{0,v_{N}-k} (s_{-1,v_{N}-k}+s_{1,v_{N}-k}) \nonumber \\
&& \qquad \qquad \qquad + s_{0,v_{N}+k} s_{0,v_{N}+k+1} +
s_{0,v_{N}-k} s_{0,v_{N}-k-1}\Bigr] , \nonumber
\end{eqnarray}
and
\begin{eqnarray} \label{nn3}
&&\widehat {\sum \limits_{\langle {\rm decr.} \rangle}}
=\sum \limits_{k=1}^{(D-1)/2}
\Bigl[s_{k,v_{N}} (s_{k,v_{N}+1}+s_{k,v_{N}-1}) \\
&& \qquad \qquad \qquad +s_{-k,v_{N}} (s_{-k,v_{N}+1}+s_{-k,v_{N}-1}) \nonumber \\
&& \qquad \qquad \qquad +s_{k,v_{N}} s_{k+1,v_{N}} +
s_{-k,v_{N}} s_{-k-1,v_{N}}\Bigr] \nonumber
\end{eqnarray}
contain products of nearest neighbor lattice spins while the sums
\begin{equation} \label{ext1}
\sum \limits_{\langle {\rm one} \rangle}^{(+)}
=s_{-1,v_{N}+1}+s_{1,v_{N}+1}+s_{-1,v_{N}-1}+s_{1,v_{N}-1} ,
\end{equation}
\begin{eqnarray} \label{ext2}
&&\sum \limits_{\langle {\rm decr.} \rangle}^{(+)}
= s_{0,v_{N}+(D-1)/2+1} + s_{0,v_{N}-(D-1)/2-1} \nonumber \\
&& \; + s_{-1,v_{N}}+s_{1,v_{N}} \\
&& \; + \sum \limits_{k=2}^{(D-1)/2}
\Bigl[s_{-1,v_{N}+k} + s_{1,v_{N}+k}
+ s_{-1,v_{N}-k} + s_{1,v_{N}-k}\Bigr] \, , \nonumber
\end{eqnarray}
and
\begin{eqnarray} \label{ext3}
&&\sum \limits_{\langle {\rm inc.} \rangle}^{(+)}
= s_{(D-1)/2+1,v_{N}} + s_{-(D-1)/2-1,v_{N}} \nonumber \\
&& \; + s_{0,v_{N}+1} + s_{0,v_{N}-1} \\
&& \; + \sum \limits_{k=2}^{(D-1)/2}
\Bigl[s_{k,v_{N}+1} + s_{k,v_{N}-1}
+ s_{-k,v_{N}+1} + s_{-k,v_{N}-1}\Bigr] \nonumber
\end{eqnarray}
contain products of a lattice spin and the fixed external spin
$s_{0}$ (which has the value $s_{0}=+1$ and thus does not appear
in Eqs.~(\ref{ext1})-(\ref{ext3})). Here $\sum\limits_{\langle
{\rm zero} \rangle}$ contains the products of the center
($\times$) spin of the needle with its four nearest neighbor spins
which correspond to lattice bonds broken in both the perpendicular
and parallel needle orientation (compare Figs.~\ref{FIG_IV_3}(a)
and (b)). The products in $\widehat{ \sum\limits_{\langle {\rm
inc.} \rangle}}$ and $\widehat{\sum\limits_{\langle {\rm decr.}
\rangle}}$ correspond to the remaining nearest neighbor bonds
which are broken in the perpendicular and parallel needle
orientation, respectively. The four terms in $\sum\limits_{\langle
{\rm one} \rangle}^{(+)}$ correspond to the bonds between the
external spin and those four lattice spins which are coupled to it
in both the perpendicular and the parallel needle configuration
(compare Figs.~\ref{FIG_IV_3}(a) and~(b)). The sums
$\sum\limits_{\langle {\rm decr.} \rangle}^{(+)}$ and
$\sum\limits_{\langle {\rm inc.} \rangle}^{(+)}$ contain the terms
which correspond to the rest of the bonds to the external spin in
the perpendicular and the parallel needle orientation,
respectively.

As in the previous subsection the notation for the various sums
reflects the modulus of their corresponding prefactors in the
crossover Hamiltonian
\begin{eqnarray} \label{+cross}
&&{\mathcal H}_{\rm cr}^{(+)}(\lambda) = \tilde{\cal H}^{(+)} -
\lambda {\cal J} \widehat {\sum \limits_{\langle {\rm inc.}
\rangle}} -(1-\lambda) {\cal J} \widehat { \sum \limits_{\langle
{\rm decr.} \rangle}} \nonumber \\
&& \quad \quad - {\cal J} \sum^{(+)} \limits_{\langle {\rm
one}\rangle}-\lambda {\cal J} \sum^{(+)} \limits_{\langle {\rm
inc.} \rangle}-(1-\lambda) {\cal J}\sum^{(+)} \limits_{\langle
{\rm decr.} \rangle} .
\end{eqnarray}
The first term $\tilde{\cal H}^{(+)} \equiv {\cal H}_{\rm ST} +
{\cal J}\sum \limits_{\langle {\rm zero} \rangle}+ {\cal J}
\widehat {\sum \limits_{\langle {\rm inc.} \rangle}} + {\cal J}
\widehat {\sum \limits_{\langle {\rm decr.} \rangle}}$ corresponds
to a strip with a cross-shaped hole where the bonds belonging to
$\sum\limits_{\langle {\rm zero} \rangle}$, $\widehat {\sum
\limits_{\langle {\rm inc.} \rangle}}$, and $\widehat {\sum
\limits_{\langle {\rm decr.} \rangle}}$ are missing (i.e., in
Fig.~\ref{FIG_IV_3}(c) this means that the dashed and dash-dotted
bonds and all the arrows are removed). In ${\mathcal H}_{\rm
cr}^{(+)}(\lambda)$ the contributions due to the two last types of
bonds missing in $\tilde{\cal H}^{(+)}$, i.e., $\widehat {\sum
\limits_{\langle {\rm inc.} \rangle}}$ and $\widehat {\sum
\limits_{\langle {\rm decr.} \rangle}}$ carry the prefactors
$-\lambda {\cal J}$ and $-(1-\lambda){\cal J}$ of increasing and
decreasing strengths , respectively. Moreover, bond contributions
are added which couple the lattice spins contained in $\sum^{(+)}
\limits_{\langle {\rm one}\rangle}$, $\sum^{(+)} \limits_{\langle
{\rm inc.} \rangle}$, and $\sum^{(+)} \limits_{\langle {\rm decr.}
\rangle}$ with strengths 1, $\lambda$, and $1-\lambda$,
respectively, to the external spin $s_{0}=1$.
For $L=8$, $W=7$, $D=5$, and the needle center $\times$ at
$v_{N}=0$ at the midline of the strip, the various bond strengths
in ${\mathcal H}_{\rm cr}^{(+)}(\lambda)$ are shown in
Fig.~\ref{FIG_IV_3}(c) which clearly illustrates the crossover from the
perpendicular to the parallel needle orientation considered in
Figs.~\ref{FIG_IV_3}(a) and~\ref{FIG_IV_3}(b) as $\lambda$
increases from 0 to 1.

\subsubsection{Details of the numerical implementation}
For the sequential generation of system configurations we have
used the hybrid Monte Carlo method~\cite{LB}. One step consists of
updating a Wolff cluster~\cite{Wolff} followed by $L \times W/4$
attempts of Metropolis updates~\cite{Metr} of randomly chosen
spins and of additional $(D+3)^2$ updates of randomly chosen spins
in the square of size $(D+3) \times (D+3)$ with the center at
position $(0,v_{N})$.

In order to determine the dependence of the free energy on $v_{N}$
we have used system sizes $1000 \times 100$ and $1000\times 101$
for $(i[O]j)$ and $(i[+]j)$ needles, respectively. For
thermalization we have used $1.5 \times 10^{7}$ MC steps, followed
by the  computation of the thermal average using
 $8 \times 10^{7}$ MC steps. These latter MC steps have been split
 into 16 intervals which facilitates to estimate the numerical inaccuracy.

In order to determine the aspect ratio dependence of the free
energies we have used various numbers of MC steps (split into 8
intervals for estimating again the numerical inaccuracy) for
various system sizes, varying from $6 \times 10^{6}$ MC steps for
$L=4000$ to $2.4 \times 10^{9}$ for $L=200$. We have used one
fifth of these MC steps in order to achieve thermalization.

Concerning the  numerical integration over the crossover variable
$\lambda$, for every selected set of parameters (i.e., type of
needle and boundary conditions $(i[h]j)$, $L,W,D$, and $v_{N}$) we
have performed computations for 32 points
$\lambda_{k}=\frac{k}{31},\;k=0,1,\dots,31$ and then we have
carried out the numerical integration by using the extended
version of Simpson's rule.

\section{COMPARISON OF ANALYTIC RESULTS WITH SIMULATION DATA} \label{compare}

Here we compare the ``small needle'' predictions from Sec.
\ref{SP} for the quasi-torque $\Delta F$ with corresponding
results obtained by the Monte Carlo simulations described in Sec.
\ref{SIM}. On one hand this allows us to assess the performance of
the truncated form of the ``small needle approximation'', i.e., to
determine the {\it smallness} of the mesoscopic needle length $D$
needed in order to be able to neglect higher order terms in this
expansion. On the other hand, good agreement signals that all the
mesoscopic distances and lengths, including $D$, chosen in the
simulations turn out to be {\it large} enough for the lattice
system to lie within the universal scaling region.

\subsection{Aspect ratio dependence for the double periodic strip}
\label{comperper}

For a needle in a double periodic strip, one can study the full
dependence of $\Delta F$ (Eq.~(\ref{DeltaF'})) on the aspect ratio
$W/L$. Due to the symmetry of these boundary conditions, $\Delta
F$ keeps its modulus but changes its sign as the values of $W$ and
$L$ are exchanged. This implies that $\Delta F$ vanishes for
$W=L$. Within the small needle approximation, its quantitative
behavior follows from the remarks below Eq.~(\ref{expand''})
yielding (see Eqs.~(\ref{expand})-(\ref{expand''}))
\begin{eqnarray} \label{expandP}
&&{\Delta F \over k_{B}T} \, \simeq \, {\Delta F_{l}+\Delta F_{nl}
\over k_{B}T} \, = \, -\pi \Biggl( {D \over 2W} \Biggr)^{2}
\Delta_{P}(W/L) \times \nonumber \\
&& \qquad \qquad \times \Biggl[ 1 + \left\{-{1 \over 2}, \, {1
\over 2} \right\}{D \over 2W} f_{\epsilon}^{(P)}(W/L) \Biggr] \,
\end{eqnarray}
for an \{``ordinary'', ``normal''\} needle with ${\cal
A}_{\epsilon}^{(h)}=\{1/2, \, -1/2 \}$  (see Eq.~(\ref{halfamp})).
Equation~(\ref{expandP}) is consistent with the sign change
mentioned above because both $\Phi_{\rm ST}^{(P)}$ (see
Eqs.~(\ref{stressscale}) and~(\ref{stressscale'})) and $\langle
\epsilon \rangle_{\rm ST}^{(P)}=W^{-1}f_{\epsilon}^{(P)}(W/L)$
(see Eq.~(\ref{densscale}) with $x_{\epsilon}=1$)
 remain unchanged upon exchanging
the values of $W$ and $L$, implying
$\Delta_{P}(W/L)/W^{2}=-\Delta_{P}(L/W)/L^{2}$ and
$f_{\epsilon}^{(P)}(W/L)/W=f_{\epsilon}^{(P)}(L/W)/L$. Since, due
to Appendix~\ref{appperper}, $-{\rm sgn}(L-W) \times
\Delta_{P}(W/L)$ is positive and $f_{\epsilon}^{(P)}$ is negative,
Eq.~(\ref{expandP}) implies the following:
\begin{itemize}
\item[(i)] Both ``ordinary'' and ``normal'' needles prefer to
align {\it perpendicular} to the longer axis of the double
periodic strip. (Concerning related effects see Ref.
[\onlinecite{dimer}].)
\item[(ii)] While in leading order $\propto \, D^2$ the strength
of this preference is independent of the needle type $h$, it is
strengthened (weakened) for an ``ordinary'' (``normal'') needle by
the correction of order $D^3$.
\end{itemize}

\begin{figure}
\includegraphics[width=0.49\textwidth]{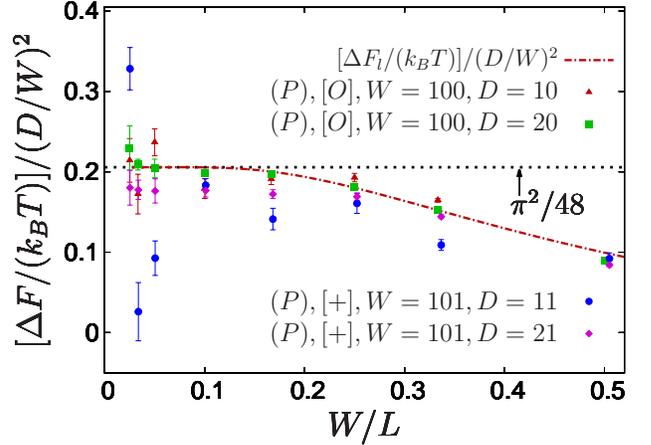}
\caption{
Normalized quasi-torque $\Delta F=F_{||}-F_{\perp}$
(Eq.~(\ref{DeltaF'})) acting on a small mesoscopic needle of
length \cite{micromeso} $D$ in a double periodic strip $P$ as a
function of the aspect ratio $W/L$ of the strip. $\Delta F>0$
implies that there is a preference for the perpendicular
orientation. $\Delta F $ vanishes at $W/L=1$ and is antisymmetric
around this point. The black dotted line denotes its limit for
$W/L=0$. For ``ordinary'' $[O]$ and ``normal'' $[+]$ needles with
$D=20$ and $D=21$, respectively, the simulation data (green
squares and magenta diamonds) agree quite well with the analytic
prediction (red dash-dotted line) according to the first term on
the rhs of Eq.~(\ref{expandP}). The fact that the (green) squares
lie above the (magenta) diamonds agrees with the tendency arising
from the second term $\Delta F_{nl}$ in Eq.~(\ref{expandP}) due to
$f_{\epsilon}^{(P)} < 0$. However, the splitting between $[O]$ and
$[+]$ due to $\Delta F_{nl}$ is quite small. (Considering $D=20$
and $W=100$, the explicit form of $f_{\epsilon}^{(P)}$ given in
Eqs.~(\ref{epsP})-(\ref{epsP''}) leads at, e.g., $W/L=0.2$ and
$W/L=0.5$ to the values $1 \pm 0.003$ and $1 \pm 0.027$,
respectively, of the square bracket in Eq.~(\ref{expandP})). A quantitatively
reliable check of this small splitting effect requires accurate
data from lattice systems with larger values of $D, \, W$, and $L$
which are closer to the asymptotic scaling limit than the ones
presently available. For more details see the main text.
} \label{FIG_V_1}
\end{figure}
A quantitative comparison of the aspect ratio dependence predicted
in Eq.~(\ref{expandP}) with our simulation data is provided in
Fig.~\ref{FIG_V_1}. This figure shows data~\cite{errorbars} for
the normalized quantity $[\Delta F/(k_{B}T)]/(D/W)^2$ within the
range $0<D/W<1/2$. For the ``ordinary'' needle of ``broken bonds''
the data were obtained from systems with $W=100$ and $D=10$ and
20. The data for the ``normal'' needle of ``fixed spins'' stem
from systems with $W=101$ and $D=11$ as well as $D=21$. The data
for $D=20$ and $D=21$ are in rather good agreement with the
expression $[\Delta F_{l}/(k_{B}T)]/(D/W)^2$, as predicted
according to Eq.~(\ref{expandP}) and Appendix~\ref{appperper} for
the leading order contribution $\propto \, D^2$, shown as the red
dash-dotted line. The data for the ``ordinary'' needle are indeed
slightly larger than those for the ``normal'' needle, as predicted
by the next-to-leading order contribution $\propto \, D^3$ in
Eq.~(\ref{expandP}). The stronger deviations of the data for
$D=10$ and $D= 11$ from the analytic approximation presumably
indicate that these smaller needle lengths lie outside the
mesoscopic scaling region required \cite{SPE} for the validity of
Eq.~(\ref{expandP}).

\subsection{Dependence of the free energy anisotropy on the
spatial position of needles in strips.} \label{compord}

In strips with boundaries $\Delta F$ depends on the position
$v_{N}$ of the needle in the strip. Here we consider strips of
infinite length $L$, either with arbitrary boundaries $(i,j)$ and
containing an ``ordinary'' needle or with boundaries $(O,O)$ and
containing a ``normal'' needle. In all these cases $\Delta F \,
\simeq \, \Delta F_{l}+\Delta F_{nl}$ is predicted to have the
form given by Eqs.~(\ref{expand})-(\ref{expand''}).

{\it Ordinary needles}

While $\Delta F_{l}$ does not depend on  $v_{N}$ and is given by
Eq.~(\ref{expand'}) with $\Delta_{i,j}$ given below
Eq.~(\ref{stressscale'}), in the  cases $(i[O]j)$ one has
\begin{eqnarray} \label{nlexplicit}
{\Delta F_{nl} \over k_{B}T}={1 \over 256} \Biggl( {\pi D \over W}
\Biggr)^{3} g_{i,j}
\end{eqnarray}
which depends on $v_{N}$ via the following simple expressions for
$g_{i,j}$:
\begin{eqnarray} \label{nlexplicit'}
&&g_{O,O}=3(\cos V)^{-3} - (5/3)(\cos V)^{-1} \nonumber \\
&&g_{+,+}=-g_{O,O} \nonumber \\
&&g_{+,O}=-g_{O,+}=(\tan V) [3(\cos V)^{-2} + (1/3)] \nonumber \\
&&g_{+,-}=g_{-,+}=-[3(\cos V)^{-3} + \nonumber \\
&&\qquad \qquad \qquad + (7/3)((\cos V )^{-1} - 4\cos V)]
\end{eqnarray}
with $V=\pi v_{N}/W$. They follow upon inserting
$f_{\epsilon}^{(i,j)}$ from Eq.~(\ref{f}) and ${\cal
A}_{\epsilon}^{(h)} \equiv {\cal A}_{\epsilon}^{(O)} = 1/2$ (see
Eq.~(\ref{halfamp})) into Eq.~(\ref{expand''}).

\begin{figure}
\includegraphics[width=0.49\textwidth]{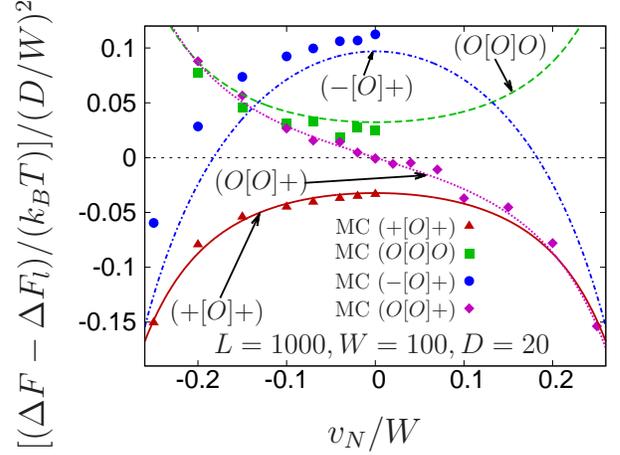}
\caption{Quasi-torque $\Delta F$ acting on a small mesoscopic
``ordinary'' needle $O$ in a long strip with boundaries $(i,j)$
corresponding to the cases $(i[O]j)$. This plot shows its
dependence on the position $v_{N}$ of the needle in the strip and
on the boundary conditions $(i,j)$. The lines are suitably
normalized expressions $c_{ij}$ (see Eqs.~(\ref{nlexplicit}) and
(\ref{nlexplicit'}) and the text below the latter one) for $\Delta
F_{nl}$ which, according to Eq.~(\ref{nlexplicit'}), are symmetric
around $v_{N}/W=0$ for $(i,j)=(O,O)$, $(+,+)$, and $(-,+)$ but
antisymmetric for $(i,j)=(O,+)$. In order to obtain the numerical
data we have computed $\Delta F$ by means of the coupling
parameter approach (Sec.~\ref{SIM}) and then have subtracted the
analytic expression~(\ref{expand'}) for $\Delta F_{l}.$ There is
very favorable agreement of the simulation data, which correspond
to $W/L=0.1$, with the analytic predictions for $W/L=0$. The case
$(-[O]+)$ is exceptional in that for it the agreement is only
fair, i.e., in this case there are sizeable corrections to $\Delta
F=\Delta F_{l}+\Delta F_{nl}$ beyond $\Delta F_{nl}$. The
statistical error bars are comparable with the symbol sizes and
therefore they are omitted.} \label{FIG_V_2}
\end{figure}
Figure~\ref{FIG_V_2} compares our simulation data for the cases
$(i[O]j)$ with the corresponding analytic predictions for various
positions $v_{N}$ of the needle and for various boundary
conditions $(i,j)$ of the strip. The plots show the data for
$[(\Delta F - \Delta F_{l})/(k_{B}T)]/(D/W)^2$ with $\Delta F$
obtained from a system with $L=1000, \, W=100$, and $D=20$. The
comparison with $[\Delta F_{nl}/(k_{B}T)]/(D/W)^2 \equiv c_{i,j}$
following from Eqs.~(\ref{nlexplicit}) and~(\ref{nlexplicit'}) and
shown by lines is very favorable (i.e., $\Delta F$ is captured
well by $\Delta F_{l}+\Delta F_{nl}$), except for the case
$(-[O]+)$ in which it is only fair. According to
Eqs.~(\ref{nlexplicit}) and~(\ref{nlexplicit'}) one has
$c_{i,j}=(\pi^{3}/256)(D/W)g_{i,j}$.

\begin{figure}
\includegraphics[width=0.49\textwidth]{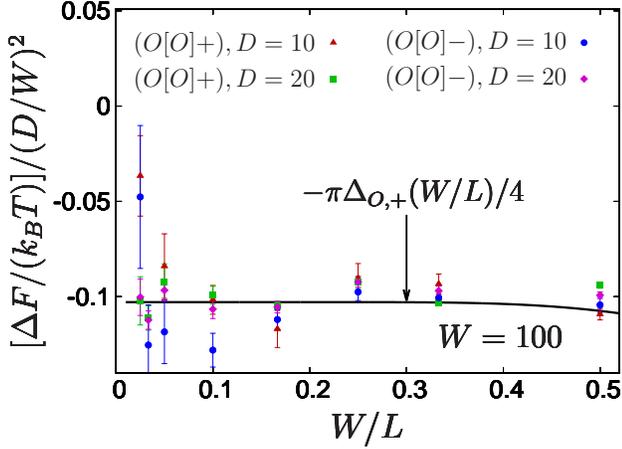}
\caption{Dependence of the quasi-torque $\Delta F$ on the aspect
ratio $W/L$ of the strip for $(O[O]+)$ and $(O[O]-)$ boundary
conditions with the needle at the strip center $v_{N}=0$. In these
cases the next-to-leading contribution $\Delta F_{nl}$
vanishes~\cite{symmfinL} for arbitrary $W/L$ and the simulation
data for $D=20$ (squares and diamonds) are indeed close to the
values given by $-\pi \Delta_{O,+}(W/L)/4$ (full black line)
corresponding to the leading contribution $\Delta F_{l}$ which are
$-\pi^{2}/96$ for $W/L=0$ and about 5 percent smaller for
$W/L=1/2$ (see Eqs.~(\ref{Deltaijrect'}),~(\ref{Deltaijclose}),
and~(\ref{Deltaclose'})). This decrease is too weak to be
reflected by the present simulation data. Due to the inherent ($+
\leftrightarrow - $) symmetry, one has $\Delta F^{(O[O]+)}=\Delta
F^{(O[O]-)}$ for {\it arbitrary} values of $D$, $v_{N}$, $W$, and
$L$. This exact identity is embodied in the form of the
corresponding lattice Hamiltonians $\Delta {\cal H}$ and ${\cal
H}_{\rm cr}(\lambda)$ in Sec.~\ref{SIM}. However, the results for
the thermal average $\langle \Delta {\cal H} \rangle_{\rm
cr}(\lambda)$ and its integral $\Delta F$ in Eq.~(\ref{eq:alg}),
calculated by means of the statistical Monte Carlo method, violate
this identity within the numerical inaccuracy. Note that the
ensuing deviations in the above data are of tolerable size, at
least for the larger value of $D$.} \label{FIG_V_3}
\end{figure}
\begin{figure}
\includegraphics[width=0.49\textwidth]{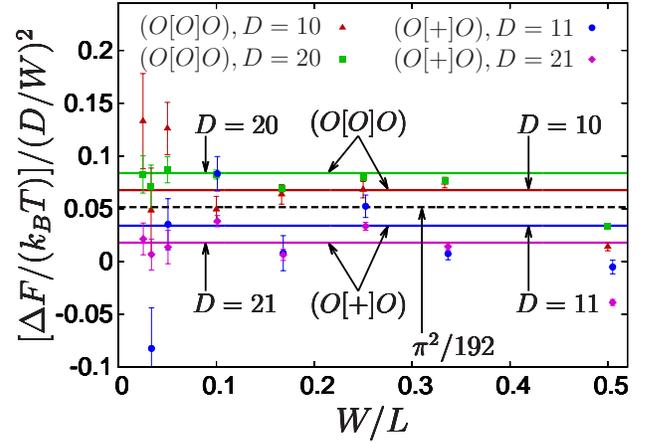}
\caption{Similar as Fig.~\ref{FIG_V_3} but for the cases $(O[O]O)$
and $(O[+]O)$ in which, for $L=\infty$, $[(\Delta F_{l}+\Delta
F_{nl})/(k_{B}T)]/(D/W)^{2}$ equals $(\pi^{2}/192)[1+\pi D/W]$ and
$(\pi^{2}/192)[1-\pi D/W]$, respectively, so that the leading
contributions (dashed line) are the same. For small $W/L$ the
simulation data for $D=20, W=100$ and $D=21, W=101$ (green squares
and magenta diamonds) are indeed close to the corresponding
uppermost and lowest horizontal line, respectively. }
\label{FIG_V_4}
\end{figure}
In order to visualize the  approach of the limit of infinite strip
length $L=\infty$, Figs.~\ref{FIG_V_3} and~\ref{FIG_V_4} show
\cite{errorbars} the dependence of the quasi-torque  on the aspect
ratio $W/L$ for needles of length $D=10$ or $D=20$ with their
center $v_{N}=0$ at the midpoint of a strip of width $W=100$. The
cases $(O[O]+)$ and $(O[O]-)$ should lead to the same $\Delta F$,
due to the $(+ \leftrightarrow -)$ symmetry \cite{bulkatcritpoint}
of the Ising model, and $\Delta F_{nl}$ should vanish for
$v_{N}=0$, due to Eq.~(\ref{expand''}) and Ref.
[\onlinecite{symmfinL}]. These properties are reflected rather
well by the data in Fig.~\ref{FIG_V_3} for $D=20$ (shown as
squares and diamonds) which are close to the values of $[\Delta
F_{l}/(k_{B}T)]/(D/W)^2$, equal to $-\pi^{2} /96$ for $W/L=0$
(shown as the horizontal straight line) and about 5$\%$ smaller
for $W/L=1/2$. Here we discard the data for $W/L \leq 0.1$ which
carry large error bars. This should be compared with the case
$(O[O]O)$ discussed in Fig.~\ref{FIG_V_4} where in a strip of
width $W=100$ the data for the needle of length $D=20$ (shown by
squares) attain much closer the asymptotic value
$(\pi^{2}/192)(1+20\pi/100)$ of $[(\Delta F_{l}+\Delta
F_{nl})/(k_{B}T)]/(D/W)^2$ predicted for $v_{N}=0$ and $L=\infty$
(shown as the uppermost straight line) as $L$ increases, i.e., as
$W/L$ becomes smaller.

{\it Normal needles in strips without broken symmetry}

\begin{figure}
\includegraphics[width=0.49\textwidth]{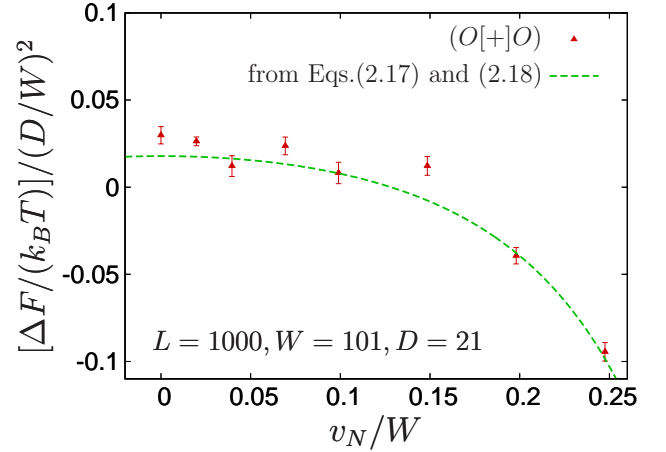}
\caption{Dependence of the quasi-torque $\Delta F$ on the position
$v_{N}$ of the needle in the $(O[+]O)$ case. The simulation data
(triangles) agree well with the analytic prediction (dashed line)
given by $\Delta F_{l}+\Delta F_{nl}$ (see the main text).}
\label{FIG_V_5}
\end{figure}
Now we consider the (equivalent) cases $(O[+]O)$ and $(O[-]O)$ of
a ``normal'' needle with $h=+$ or $-$ in an $(O,O)$ strip.
Figure~\ref{FIG_V_5} shows data \cite{errorbars} for $[\Delta
F/(k_{B}T)]/(D/W)^2$ for such a needle of length $D=21$ at various
positions $v_{N}$ in a strip with $L=1000$ and $W=101$. They
compare favorably with the corresponding analytic expression
$-(\pi/4)\Delta_{O,O}+(\pi^{3}/256)(D/W)(-g_{O,O})$ for $[(\Delta
F_{l}+ \Delta F_{nl})/{(k_{B}T)}]/(D/W)^2$ with $L= \infty$, which
is shown as dashed line. This expression follows from
Eqs.~~(\ref{expand})-(\ref{expand''}) by repeating analogously the
line of arguments leading to
Eqs.~(\ref{nlexplicit})~and~(\ref{nlexplicit'}). Here the sign in
front of $g_{O,O}$ differs from that in Eq.~(\ref{nlexplicit}) for
the ``ordinary'' needle because ${\cal A}_{\epsilon}^{(+)}={\cal
A}_{\epsilon}^{(-)}=-{\cal A}_{\epsilon}^{(O)}$
(Eq.~(\ref{halfamp})). In  strips of infinite length $L$ the
pseudo-torques $\Delta F(v_{N})$ for the present case $(O[+]O)$
and for the case $(+[O]+)$ considered above are {\it identical}.
This follows from a duality argument similar to the one in Ref.
[\onlinecite{dual}], is in agreement with the expression $\Delta
F_{l}+ \Delta F_{nl}$ in Eqs.~(\ref{expand})-(\ref{expand''}), and
is quite well reflected by the simulation data in
Figs.~\ref{FIG_V_2} and~\ref{FIG_V_5}. Concerning the aspect ratio
dependence, for $W/L \le 0.1$ the data for the case $(O[+]O)$ of a
``normal'' needle with $v_{N}=0$ and $D=21$ (shown by diamonds in
Fig.~\ref{FIG_V_4}) have the tendency to approach the predicted
limiting value $(\pi^{2}/192)(1-21 \pi/101)$ (indicated in
Fig.~\ref{FIG_V_4} by the lowest horizontal straight line).

\subsection{``Normal'' needles in strips with broken symmetry} \label{compnorm}

In this subsection we consider a ``normal'' needle with $h=+$
embedded in strips with at least one ``normal'' boundary. These
are the boundary conditions $(i,j)=[(+,+), (-,-), (+, -), (+,O),
(-,O)]$ for which the order parameter profiles $\langle \phi
\rangle_{\rm ST}=W^{-1/8} f_{\phi}^{(i,j)}$ in
Eq.~(\ref{densscale}) are nonvanishing and $\zeta_{I}$ in
Eq.~(\ref{zetaI}) contains a contribution $\propto D^{1/8}$. In
these cases we do not compare our simulation data with the
expanded analytic form of $\Delta F$, because confining this
expansion to the two leading powers of $D$, as in
Eq.~(\ref{expand}), is expected to be insufficient for reaching
agreement with presently accessible simulation data. As explained
there, we rather set out to compare the simulation data with the
full expression ``sna'' of the ``small needle approximation''
following from Eqs.~(\ref{Fparal})-(\ref{DeltaF'}). While we
disregard the terms of order $D^{4}$ denoted by the
ellipses~\cite{ellipses} in Eqs.~(\ref{zetaI}) and~(\ref{zetaA}),
we leave the logarithm in Eq.~(\ref{DeltaF'}) unexpanded.

\begin{figure}
\includegraphics[width=0.49\textwidth]{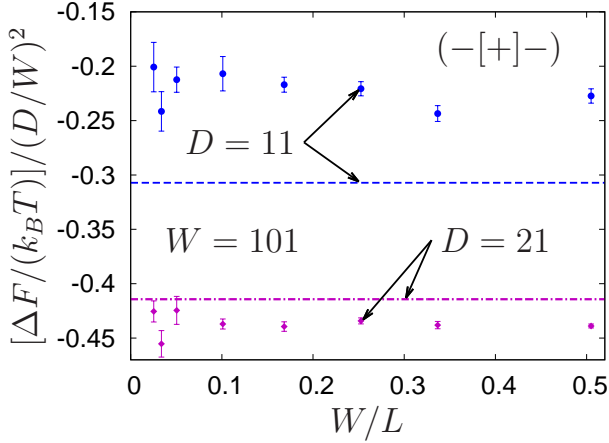}
\caption{Quasi-torque $\Delta F$ acting on a $+$ needle in the
center of a $(-,-)$ strip. We show  simulation data for needles of
length $D=11$ and 21 in strips of width $W=101$ and of various
lengths $L$. The data for $D=21$ (diamonds) agree quite well with
the ``sna'' prediction $-0.4143$ (dash-dotted line) for a strip of
infinite length as explained in the first two paragraphs of
Subsec.~\ref{compnorm}. For the smaller needle of length $D=11$
there is a significant discrepancy between the simulation data and
the ``sna'' prediction -0.3072 (dashed line) for a strip of
infinite length.} \label{FIG_V_6}
\end{figure}
We illustrate this point for the case $(-[+]-)$, i.e., a $+$
needle embedded in a $(-,-)$ strip. Figure~\ref{FIG_V_6}
shows\cite{errorbars} that $[\Delta F/(k_{B}T)]/(D/W)^2$ as
obtained from simulations for a needle (with $D=21$ and located in
the center of a strip with $W=101$) agrees, for smaller aspect
ratios $W/L$, quite well with the prediction $-0.4143$ for
$L=\infty$ (see the diamonds and the dash-dotted bottom line in
Fig.~\ref{FIG_V_6}). The prediction follows by inserting the
present value $\theta =21 \pi/101=0.653$ of $\theta \equiv \pi D/W
$ into the ``sna'' expression $\Delta F/(k_{B}T)=- \ln
(Z_{\parallel}/Z_{\perp})$ where
\begin{eqnarray} \label{full}
Z_{\parallel, \perp}=
&1&-(2\theta)^{1/8}-2^{-6+1/8}\theta^{17/8}+2^{-3}\theta+2^{-9}\theta^{3}
\nonumber \\
&+&(1,-1)\times [-2^{-7}(1/3)\theta^{2}\nonumber \\
&& \qquad +2^{-6+1/8}(1/3)\theta^{17/8}- 2^{-10} 3\theta^{3}] \, ,
\end{eqnarray}
see Eqs.~(\ref{halfamp}) and~(\ref{Z})-(\ref{zetaA}) with the
value $\Delta_{-,-}=-\pi/48$ taken from below Eq.
(\ref{stressscale'}) as well as Eqs.~(\ref{f}) and~(\ref{CS}). For
this value of $\theta$ the validity of the ``sna'' is confirmed by
Fig.~\ref{FIG_III_2}, as far as the contribution $F_{\parallel}$
to $\Delta F$ is concerned (compare the discussion in the
paragraph below Eq.~(\ref{arbiJ})). This should be contrasted with
the prediction $-0.0017$ (as compared with $-0.4143$, see above)
which would follow from $\Delta F$ being truncated after the
leading and the next-to-leading order in $D$, i.e., from $(\Delta
F_{l}+\Delta F_{nl})/k_{B}T=[(2\theta)^{2}-(2\theta)^{17/8}]/(3
\times 2^8)$ (by using again Eq.~(\ref{full})) and which would
disagree with the simulation data by a factor of about 250.

The reasonable agreement between the simulation data for $D=21$
and the ``sna'' persists for the dependence on the needle position
$v_{N}$ shown as the squares and the green dashed line in
Fig.~\ref{FIG_V_7}, provided $|v_{N}|/W$ remains small.

However, for $|v_{N}|/W \gtrsim 0.15$ an even qualitative
deviation develops in that the analytic approximation predicts a
point of inflection and a minimum which is not supported by the
simulation data. We attribute this failure to inadequacies of this
approximation near the strip boundary which have  been discussed
in detail for $F_{\perp}$ in the half plane (see the paragraph in
Sec.~\ref{ARB} addressing Fig.~\ref{FIG_III_1}(a)). In particular
the unphysical minimum of the green dashed line in
Fig.~\ref{FIG_V_7} is related to the maximum of the ``sna'' result
shown in Fig.~\ref{FIG_III_1}(a).

This dependence on $v_{N}/W$ is shown in Fig.~\ref{FIG_V_7} also
for the other cases $(-[+]O), (+[+]O), (+[+]-)$, and $(+[+]+)$ of
a + needle embedded in strips with at least one ``normal''
boundary. Also in these cases the dependences of the data on
$v_{N}/W$ are reproduced reasonably well by the ``sna''.

\section{SUMMARY AND CONCLUDING REMARKS} \label{summary}

A critical solvent such as a binary liquid mixture at its
continuous demixing transition induces a long-ranged, so-called
critical Casimir interaction between immersed particles which is
universal on mesoscopic length scales. For {\it nonspherical}
particles the interaction depends not only on their sizes and
distances among each other but also on their shapes and mutual
orientations. We have studied  a critical system belonging to the
Ising bulk universality class  in two spatial dimensions and
particles of needle shape.

As described in the Introduction we have  considered  Ising strips
at the bulk critical point with an embedded needle which is
oriented either parallel or perpendicular to the symmetry axis of
the strips. For such systems our analysis benefits from the wealth
of knowledge accumulated for the two-dimensional Ising model at
criticality and the comparative ease to implement a needle along a
row or column of the square lattice. The corresponding effective
interaction between the needle and the two strip boundaries is
probably the simplest example to check by Monte Carlo simulations
the range of validity of asymptotic analytic predictions for the
orientation dependence of critical Casimir interactions between
nonspherical mesoscopic particles which go beyond the Derjaguin
approximation.

In Sec.~\ref{SIM} and Figs.~\ref{FIG_I_1},~\ref{FIG_IV_1},~\ref{FIG_IV_2},
and~\ref{FIG_IV_3} we explain how to implement
in the lattice model boundary properties at the two confining
surfaces of the strip and for the needle, which locally induce one
of the two demixing bulk phases (``normal'' surface universality
classes $+$ (or $-$)) or induce disorder and suppress demixing
(``ordinary'' surface class $O$). The geometrical features of the
corresponding continuum description are explained in Fig.~\ref{FIG_I_2}.

Primarily, we analyze the free energy $\Delta
F=F_{\parallel}-F_{\perp}$ required at bulk criticality to turn
the needle from an alignment perpendicular to the strip (with free
energy $F_{\perp}$) to parallel alignment (with free energy
$F_{\parallel}$). Thus one has $\Delta F > 0$ and $\Delta F < 0$
if the needle prefers the perpendicular and parallel alignment,
respectively. $\Delta F$ depends on the length $D$ of the needle,
the width $W$ and length $L$ of the strip, the distance $v_{N}$ of
the needle center from the midline of the strip, and the surface
universality classes $h$ of the needle and $(i,j)$ of the two
boundaries of the strip. Here we consider needles of {\it small}
mesoscopic length $D$ for which predictions are available from the
so-called ``small needle expansion'' explained in Sec.~\ref{SP}.

Before presenting below an itemized summary of the quantitative
comparison with the simulations we point out a few qualitative
observations. For the needle it is advantageous to reside in a
spatial region and take an orientation which suits its boundary
condition.

(i) First, consider a needle of universality class $h$ in a {\it
half plane} with the boundary belonging to surface universality
class $i$. The needle will prefer the vicinity of this boundary if
$i=h$. Since in the case $i=+$ ($i=O$) the boundary-induced order
(disorder) increases upon approaching the boundary - as described,
according to Eq.~(\ref{halfamp}), by the concomitant increasing
density profiles of the order parameter (of the energy) in the
half plane without needle - this leads to an attractive force
between needle and boundary. Since both increases are stronger
than linear, for a fixed needle center in both cases the needle
will adopt an orientation perpendicular to the boundary of the
same class. Likewise, for different universality classes $i \neq
h$ of the needle and the boundary the force will be repulsive and
for a fixed center the needle will orient parallel to the
boundary~\cite{e}. These orientational preferences depend on the
needle position, increasing and decreasing, respectively, with
decreasing and increasing distance $a_{N}$ of the needle center
from the boundary.

(ii) In a {\it strip} there is an additional qualitative effect in
that there is a contribution to the orientational preference of
the needle which is $\it independent$ of its position $v_{N}$ in
the strip but depends on the combination $(i,j)$ of surface
universality classes of the two strip boundaries. In a long strip
this contribution favors an alignment of the needle perpendicular
(parallel) to the boundaries if they belong to the same
(different) universality class $i=j$ ($i \neq j$). For the double
periodic strip it favors an alignment perpendicular to the longer
axis. This is reminiscent of - and related to - a well known
corresponding anisotropy \cite{dimer} of the two-point averages
$\langle {\cal O}({\bf r}-{\bf s}/2) {\cal O}({\bf r}+{\bf s}/2)
\rangle_{\rm ST}$ in the strip of the densities ${\cal O}=\phi$ of
the order parameter and ${\cal O}=\epsilon$ of the energy. For
small mesoscopic distances $|{\bf s}|$ the anisotropy causes the
two-point averages to be larger for ${\bf s}$ perpendicular
(parallel) to the two strip boundaries if $i=j$ ($i \neq j$) and
for ${\bf s}$ perpendicular to the longer axis of the double
periodic strip.

All of these expectations are confirmed and further refined by the
quantitative predictions of the ``small needle expansion''. It
predicts, in particular, that the anisotropy $\Delta F$ due to
\cite{ii} the aforementioned effect (ii) depends, apart from the
boundary classes ($i,j$) of the strip, only on the length and
not~\cite{parthalf} on the universality class ($h=+, \, -$, or
$O$) of the small mesoscopic needle.

In the following three blocks A, B, and C we list our main
results.

(A) We have demonstrated that the quantitative predictions of Sec.
\ref{SP} for the rich structure of the orientation-dependent
interaction of non-spherical particles with a ``small mesoscopic
size'' can actually be observed in a lattice model. This is a
nontrivial result in view of a twofold size condition: the
particle size being small compared with other geometric features
and being large on the scale of the lattice constant. Our Monte
Carlo simulations with needle lengths of about 20 lattice
constants~\cite{micromeso} have the potential to closely approach
the asymptotic regime of the ``small mesoscopic needle'', leading
to results for $\Delta F$ which agree quite well with the analytic
predictions, without adjusting any parameter. At the same time,
the parameter range in which there is good agreement provides a
(conservative) estimate for the range of validity of the ``small
needle approximation'' (``sna''), which is a truncated form of the
systematic expansion for the universal quasi-torque $\Delta F$ in
terms of the needle size.

\begin{figure}
\includegraphics[width=0.49\textwidth]{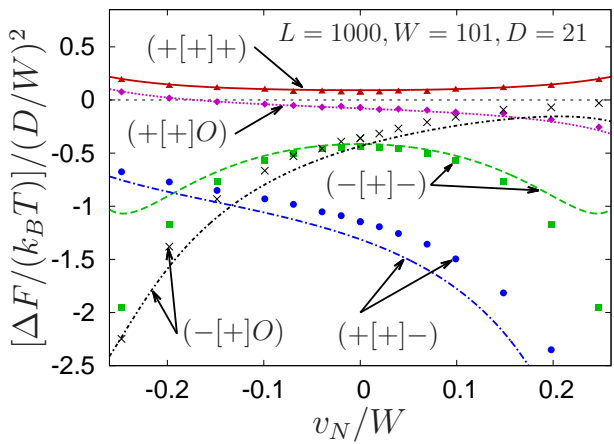}
\caption{Quasi-torque $\Delta F$ acting on a small mesoscopic
``normal'' needle $+$ in a long strip in which one or both
boundaries break the Ising symmetry. We show the dependences of
$\Delta F$ on the position $v_{N}$ of the needle in the strip and
on the boundary conditions $(i,j)$ of the strip. There is
reasonable agreement of the simulation data for $W/L=0.1$ with the
analytic predictions of the ``small needle approximation'' for
$W/L=0$ which are discussed in the first paragraph of
Subsec.~\ref{compnorm}. For the case $(-[+]-)$ the ``sna'' can be
trusted only for $|v_{N}|/W$ smaller than $\approx 0.15$ (see the
discussion in the last but one paragraph of
Subsec.~\ref{compnorm}). The statistical error of the simulation
data is comparable with the symbol sizes.} \label{FIG_V_7}
\end{figure}
\begin{itemize}
\item[A1.] For a strip with double periodic boundary conditions,
for which the aforementioned effect (i) is absent, our data for
$\Delta F$ reproduce, for the larger needle lengths $D=20, \, 21$,
quite well the predicted dependence of $\Delta F$ on the aspect
ratio $W/L$ of the strip and the predicted independence of $\Delta
F$ of the universality classes $h=O$ and $h=+$ of the needle, in
leading order (see Fig.~\ref{FIG_V_1} and the discussion in
Subsec.~\ref{comperper}).

\item[A2.] In strips with actual boundaries, effects of both type
(i) and (ii) are present and the former ones lead to a dependence
of $F_{\parallel}, \, F_{\perp}$, and $\Delta F$ on the position
$v_{N}$ of the needle center within the strip. Here the analytic
predictions hold for strips of infinite length $L= \infty$ and
Figs.~\ref{FIG_V_3},~\ref{FIG_V_4}, and~\ref{FIG_V_6} show the
dependence of the simulation data on the aspect ratio $W/L$ of the
strip.

\item[A2a.] For the five possible and relevant combinations
$(i[h]j)=(O[O]O), \, (+[O]+)$, $(-[O]+), \, (O[O]+)$, and
$(O[+]O)$ of universality classes of the two boundaries of the
strip and of the embedded needle, for which the expansion in Eqs.
(\ref{expand})-(\ref{expand''}) of $\Delta F$ in terms of powers
of $D$ is appropriate, the dependence on $v_{N}$ of our data for
$\Delta F$ is discussed in Subsec.~\ref{compord}. In
Figs.~\ref{FIG_V_2} and~\ref{FIG_V_5} this dependence is compared
with the approximate analytic predictions. There is fair agreement
for the case $(-[O]+)$ while in the four other cases the agreement
is very good.

In the case $(O[O]+)$, due to effect (i) one has $\Delta
F(v_{N})>0$ and $\Delta F(v_{N})<0$ near the $O$ and $+$ boundary,
respectively, so that the needle prefers the perpendicular and
parallel orientation, respectively. Since effect (ii) for unequal
strip boundaries turns the needle parallel to the strip, the range
of $v_{N}$ values allowing for parallel orientation is the larger
part of the accessible values of $v_{N}$, i.e., the point $v_{N}$
where $\Delta F$ changes sign is closer to the $O$ boundary.

In the strip of infinite length $L= \infty$ duality arguments~\cite{duality,dual}
predict for the cases $(+[O]+)$ and $(O[+]O)$
the same expression for $\Delta F (v_{N})$ which within the
``sna'' (Eqs.~(\ref{expand})-(\ref{expand''})) is visualized as
the full line in Fig.~\ref{FIG_V_2} and as the dashed line in
Fig.~\ref{FIG_V_5}. It predicts that in this case the needle
prefers the orientation perpendicular to the strip for small
$|v_{N}|$ where effect (ii) prevails so that $\Delta F>0$, while
for larger $|v_{N}|$ where effect (i) prevails it prefers the
parallel orientation with $\Delta F<0$. The good agreement with
the data in Figs.~\ref{FIG_V_2} and~\ref{FIG_V_5} tells again
that within our lattice model one can access the scaling region
and thus capture the corresponding universal small particle
behavior.

\item[A2b.] Unlike the combinations considered above (in A2a.) the
five combinations $(i[h]j)=(+[+]+), (-[+]-), (+[+]-), (+[+]O),
(-[+]O)$ discussed in Subsec.~\ref{compnorm} involve a needle
which at $T_{c}$ is subject to the order parameter profile induced
by the strip boundaries $(i,j)$  via its nonvanishing amplitude
${\cal A}_{\phi}^{(h)}$ (see Eqs.~(\ref{Fparal})-(\ref{zetaA})).
This leads to a contribution to the partition function $\propto
D^{1/8}$. For the comparison with the simulation data for $\Delta
F$, in these five cases the corresponding ``small needle
approximation'' should not be implemented by directly expanding
$\Delta F$ in terms of powers of $D$ but by expanding the
corresponding partition functions (see the remarks above
Eq.~(\ref{expand}) and in the first paragraph of
Subsec.~\ref{compnorm}). The comparison of the analytical
predictions for these cases with the corresponding MC data is
shown in Fig.~\ref{FIG_V_7}. The orientation preferred by the
needle (i.e., the sign of $\Delta F$) is in conformance
with~\cite{(+[+]-)} the qualitative observation (i); also
quantitatively the MC data agree quite well with the predictions.
According to Fig.~\ref{FIG_V_7}, in the case $(-[+]-)$, however,
the good agreement for small values of $|v_{N}|/W$ does not extend
beyond $|v_{N}|/W \approx 0.15$. In this case  the expected
breakdown of the ``small needle approximation'' near strip
boundaries starts already rather close to the strip center
producing an unphysical minimum of $\Delta F (v_{N})$ at
$|v_{N}|/W \approx 0.25$.
\end{itemize}

(B) We have investigated the effective interaction between the
needle and the boundaries also for needles of {\it arbitrary}
mesoscopic length $D$, i.e., beyond the regime of the ``small
needle expansion''.

\begin{itemize}

\item[B1.] These exact results have been derived in Sec.~\ref{ARB}
and Appendix~\ref{apparb} for the free energy of a needle in the
half plane with an orientation perpendicular to its boundary $i$
and of a needle embedded along the midline of a strip of infinite
length and with the two boundaries being members of the same
surface universality class ($i=j$). These geometries are related
to the $W \times L$ strip without a needle
(Appendix~\ref{appstripbound}) via conformal transformations of
the Schwarz-Christoffel type \cite{SC}. For various combinations
$(h,i)$ of the surface universality classes these effective
interactions are shown in Figs.~\ref{FIG_III_1}-\ref{FIG_III_3}
and display the crossover between the small needle regime and the
limiting behavior for which the needle approaches and nearly
touches the boundary or becomes much longer than the width of the
strip.

\item[B2.] Apart from the importance in their own right these
universal results allow us to better understand the limitations of
the ``small needle approximation''. For example, in
Fig.~\ref{FIG_III_1}(a), for $(i[h])=(-[+])$ this approximation
reproduces the exact result for $F_{\perp}^{(-[+])}$ very well
down to a distance of the closer end of the perpendicular needle
from the boundary corresponding to half its length
($a_{<}=D/2=a_{N}/2$ so that $\vartheta=D/(2a_{N})=0.5$). For even
smaller distances the approximation develops an unphysical
maximum, which is the counterpart of the minimum produced  by the
approximation for $\Delta F$ in the case $(-[+]-)$ (see
Fig.~\ref{FIG_V_7}). In contrast, the exact result crosses over to
a logarithmic increase which diverges when the needle ``touches''
the boundary. We recall that  the validity of the ``small needle
approximation'' requires that the mesoscopic length $D$ of the
needle is small compared with the distance $a_{<}$ between the
closer end of the needle and the boundary, whereas the exact
result is valid if only the microscopic lengths are sufficiently
small compared with $a_{<}$. Concerning the free energy
$F_{\parallel}^{(-[+]-)}$ for embedding a $+$ needle extending
along the midline of a $(-,-)$ strip, Fig.~\ref{FIG_III_2}
explicitly demonstrates the ensuing improvement if one expands the
corresponding partition function rather than the free energy in
terms of the needle length (see the remarks above Eq.
(\ref{expand}) and in the first and second paragraph of Subsec.
\ref{compnorm}).

\end{itemize}

(C) Our results allow us to conclude which of the features studied
here are of a more general character and thus can be expected to
show up also in spatial dimension $d=3$ and which ones are
specific for $d=2$.

\begin{itemize}
\item[C1.] In dimensions $d > 2$ nonspherical mesoscopic particles
such as uniaxial ellipsoids \cite{kondrat,e} or dumbbells of two
touching or interpenetrating spheres of equal size \cite{e} in a
half space and interacting with its planar boundary wall have been
considered for various universality classes $(i[h])$. Both for
particle sizes small \cite{e,d3lead} and comparable \cite{kondrat}
with respect to the distance from the wall, these particles
prefer, at {\it fixed particle center}, orientations perpendicular
(parallel) to the boundary of the half space if $i=h$ ($i \neq
h$). These are the same preferences as described in paragraph (i)
above for our small needles in the half plane. In the mean-field
treatment of Ref. [\onlinecite{kondrat}] it was pointed out that
at {\it fixed closest surface-to-surface distance} of particle and
wall the preferred orientations display the opposite trend, i.e.,
being parallel (perpendicular) to the boundary of the half space
if $i=h$ ($i \neq h$). These latter preferences are in agreement
with our finding in $d=2$ that the free energy
$F_{\parallel}^{(i[h])} - F_{\perp}^{(i[h])}$, required to turn a
long needle in the half plane about its closer end from the
perpendicular to the parallel orientation, is dominated by the
first term, $F_{\parallel}^{(i[h])}=k_{B} T \Delta_{i,h} D/a_{<}$
(note the signs of $\Delta_{i,h}$ given below Eq.
(\ref{stressscale'})). The reason is that the second term, given
by Eq.~(\ref{long'}), depends only logarithmically on the large
ratio $D/a_{<}$ of the needle length $D$ and the distance $a_{<}$
of the closer end of the needle from the boundary and thus can be
neglected relative to the linear increase exhibited by
$F_{\parallel}^{(i[h])}$.

\item[C2.] However, there are also effects which are specific to
two dimensions, due to the symmetries based on the duality
transformation and the much wider class of conformal mappings. For
example, the aforementioned equality of the particle insertion
free energies as function of $v_{N}$ for the cases $(O[+]O)$ and
$(+[O]+)$ with infinite strip length - based on the duality
symmetry \cite{duality,dual} of the $d=2$ Ising model at the bulk
critical point - has no correspondence for a particle between
parallel walls in $d>2$. Likewise, the independence of $h$ of the
leading small particle contribution $\Delta F_{l}$ to the
quasi-torque $\Delta F$ in the case $(O[h]O)$ (see Eq.
(\ref{expand'})) is valid in $d=2$ but not in $d>2$. The
derivation of this result in $d=2$ hinges on using a conformal
transformation as described in Ref. [\onlinecite{parthalf}] which
is not available for a non-spherical particle in $d>2$.

\item[C3.] There is an interesting difference between ``ordinary''
needles in $d=2$ and $d=3$ which arises from the dependence on
their {\it width} ${\cal W}$. Here we have considered needles with
a mesoscopic width $\cal W$ which is much smaller than the needle
length $D$. For both ``ordinary'' and ``normal'' needles in $d=2$
the effects they induce in the embedding critical system (such as
the density profiles), and thus their effective interactions with
the boundaries leading to force and torque, only depend on their
length $D$ but not \cite{micaround} on their width ${\cal W}$.
This applies also \cite{h+d1999} to a ``normal'' needle in $d=3$.
However, for an ``ordinary'' needle in $d=3$ with fixed length $D$
the strength of these effects decreases \cite{e2006} upon
decreasing the width ${\cal W}$. For example, within the small
needle expansion the prefactor of the energy density ${\cal
O}=\epsilon$ is proportional to $D{\cal W}^{x_{\epsilon}-1}=D
{\cal W}^{0.42}$ in $d=3$ while in $d=2$ it is proportional to
$D^{x_{\epsilon}}=D$ and independent of ${\cal W}$ (see Eq.
(\ref{SPE'}) and Ref. [\onlinecite{e'}]).

\end{itemize}

Extending our detailed investigations to non-spherical particles
in {\it three} spatial dimensions and for the whole neighborhood
\cite{offcrit} of the critical point is desirable but beyond the
scope of the present study. For Monte Carlo simulations in $d=3$,
studies of (square shaped) disks and of ``normal'' needles with
their properties being independent of their width look most
promising \cite{micaround}. Corresponding quantitative analytical
predictions remain a challenge. The ``small particle expansion''
can in principle be extended to, e.g., a circular disk. However,
the amplitudes and averages of the corresponding operators can be
obtained only approximately. A mean field treatment corresponding
to $d=4$ is given in Ref.~[\onlinecite{e}] but already one-loop
field theoretic calculations, corresponding to the first order
contribution in an expansion in terms of $4-d$, look quite
demanding.

\begin{acknowledgments}
We are grateful to T.W. Burkhardt for communicating his
unpublished results~\cite{burkunpub} concerning Eqs.~(\ref{Tnneedles}) and
(\ref{TEO}) and for useful discussions.
\end{acknowledgments}
%

%
\appendix
\section{STRIP WITHOUT NEEDLE} \label{appwithout}
\subsection{Double periodic boundary conditions} \label{appperper}
The  behavior near bulk criticality of finite Ising strips with
periodic boundary conditions in both Cartesian directions has
found a long lasting theoretical interest, starting to the best of
our knowledge in 1969 with the seminal paper~\cite{FF} by
Ferdinand and Fisher. Therein the aspect ratio dependence of the
free energy is given explicitly and was re-derived by other
methods in later studies. The quantity $-\Phi_{\rm ST}^{(P)}$
addressed below Eqs.~(\ref{expand''}) and~(\ref{expandP}) is given
by the scale free part of $\ln Z_{nm}$ in Eq.~(3.37) of
Ref.~[\onlinecite{FF}]. It is equal to $\ln Z^{I}$ in
Ref.~[\onlinecite{FSZ}] and to $\ln Z_{PP}$ for the Ising model in
Ref. [\onlinecite{Cardybc}]. Adopting Cardy's notation (see
Eqs.~(2.4)-(2.6) and Table 1 in Ref.~[\onlinecite{Cardybc}]) the
relation
for the Casimir (or stress tensor) amplitude $\Delta_{P}$ given
below our Eq.~(\ref{expand''}) yields the result
\begin{eqnarray} \label{DeltaP}
&&\Delta_{P}(W/L) \equiv \Delta_{P}(1/\delta) = - \pi/12  -(d/d
\delta) \ln \{[\chi_{11}(\delta)]^2 \nonumber \\
&& \qquad \qquad \qquad \qquad \qquad + [\chi_{21}(\delta)]^2 +
[\chi_{22}(\delta)]^2 \}
\end{eqnarray}
where for the Ising model one has the conformal charge $c=1/2$ and
the parameter $m=3$ in the functions $\chi_{p q}$ introduced by
Cardy.

The energy density $\langle \epsilon \rangle_{\rm
ST}=f_{\epsilon}^{(P)}(W/L)/W$ defined below Eqs.~(\ref{SPE''})
and~(\ref{expandP}) is of course independent of $u$ and $v$ and
turns out to vanish \cite{FF,zeroepsPP} for  strips of infinite
length $L$, i.e., $f_{\epsilon}^{(P)}(0)=0$. The aspect ratio
dependence has been determined in Refs.~[\onlinecite{FF}]
and~[\onlinecite{FSZ}] and in our notation reads
\begin{eqnarray} \label{epsP}
f_{\epsilon}^{(P)}(W/L) \equiv f_{\epsilon}^{(P)}(1/\delta) =- \pi
E(\delta)/Z(\delta)
\end{eqnarray}
where
\begin{eqnarray} \label{epsP'}
E(\delta)= [U^{1/24} \Pi_{n=1}^{\infty} (1-U^n)]^2
\end{eqnarray}
with $U = \exp (-2 \pi \delta)$ and
\begin{eqnarray} \label{epsP''}
&&Z(\delta)=(1/2)\{ 2E(2\delta)/E(\delta) +
[E(\delta)]^2/[E(2\delta)E(\delta/2)] \nonumber \\
&&  \qquad \qquad \qquad \qquad \qquad \qquad \quad +
E(\delta/2)/E(\delta) \} \, .
\end{eqnarray}
Our quantities $f_{\epsilon}^{(P)}$, $\delta$, $E$, and $Z$
correspond  to the quantities $-\langle \epsilon \rangle^{I}$,
$-i\tau$, $|\eta^{2}|$, and $Z^{I}$, respectively, in
Ref.~[\onlinecite{FSZ}] (see the introductory remarks in
Sec.~\ref{ARB} and Eqs.~(\ref{arbperp+O}),~(\ref{iJ}),
and~(\ref{arbord+O}) therein).

\subsection{Strip with boundaries (i, j)} \label{appstripbound}
For our strip (ST) with the aspect ratio $1/\delta=W/L$ and
boundaries
\begin{eqnarray} \label{Deltaijrect'}
(i,j)=[(O,O),(+,+),(+,-),(+,O)]
\end{eqnarray}
Cardy's results \cite{Cardybc} for the partition functions and our
Eqs.~(\ref{stressscale}) and~(\ref{stressscale'}) yield
\begin{eqnarray} \label{Deltaijrect}
&&\Delta_{i,j}(1/\delta)=-\pi/48 -
(d/d\delta)\ln[\chi_{11}(\delta/2)+\chi_{21}(\delta/2), \nonumber
\\
&& \qquad \qquad \qquad \quad \chi_{11}(\delta/2), \,
\chi_{21}(\delta/2), \, \chi_{22}(\delta/2)]
\end{eqnarray}
for the universal stress tensor amplitudes in terms of the
functions $\chi_{pq}$ introduced in Ref.~[\onlinecite{Cardybc}].
For later reference we rewrite Cardy's expressions in the form
\begin{eqnarray} \label{Deltaijclose}
&&\Delta_{i,j}(1/\delta)=(\pi/48)\Biggl( -1+12 \kappa(d/d\kappa)
\ln \Bigl\{ [\Sigma_{11}+\Sigma_{21}, \nonumber \\
&&\qquad \qquad \quad \Sigma_{11}, \, \Sigma_{21}, \,
\Sigma_{22}]\Pi_{n=1}^{\infty}(1-\kappa^{4n})^{-1} \Bigr\} \Biggr)
\end{eqnarray}
and alternatively as
\begin{eqnarray} \label{Deltaijdistant}
&&\Delta_{i,j}(1/\delta)=(\pi \delta^{-2}/12) \Biggl(1-24
\sigma(d/d\sigma)\ln\Bigl\{ \Bigl(1+  \nonumber \\
&&\quad + [({\cal S}_{11}+{\cal S}_{21})/2, \, {\cal S}_{11}, \,
{\cal S}_{21}, \, {\cal S}_{22}] \Bigr)
\prod_{n=1}^{\infty}(1-\sigma^{2n})^{-1}\Bigr\}\Biggr) \nonumber
\\
\end{eqnarray}
which converge rapidly for long strips with large $\delta$
(corresponding to extended, closely spaced boundaries) and for
short strips with small $\delta$ (i.e., short, widely separated
boundaries), respectively. In Eqs.~(\ref{Deltaijclose})
and~(\ref{Deltaijdistant}) one has
\begin{eqnarray} \label{Deltaclose'}
&&\kappa=e^{-\pi\delta/4} \, , \quad \Sigma_{pq} \equiv
\Sigma_{pq}(\kappa)= \nonumber \\
&& \quad = \sum \limits_{l=- \infty}^{\infty}
\Biggl(\kappa^{[(24l+4p-3q)^{2}-1]/12} -(q \to -q) \Biggr)
\end{eqnarray}
and
\begin{eqnarray} \label{Deltadistant'}
&&\sigma=e^{-2 \pi/\delta} \, , \quad
{\cal S}_{pq} \equiv {\cal S}_{pq}(\sigma)=\sum \limits_{r=2}^{\infty} \sigma^{(r^{2}-1)/24} \times \nonumber \\
&&\qquad \quad \times \sin{\pi p r \over 3}\sin{\pi q r \over 4}
\Big/ \Bigl(\sin{\pi p \over 3}\sin{\pi q \over 4}\Bigr),
\end{eqnarray}
respectively. The three functions $\Sigma_{pq}(\kappa)$ in
Eq.~(\ref{Deltaijclose}) can be written as
\begin{eqnarray} \label{Deltaclose''}
\Sigma_{p q}(\kappa) = \kappa^{[(4p-3q)^{2}-1]/12} \Bigl( 1+\sum
\limits_{n=1}^{\infty} a_{n} \kappa^{4n} \Bigr)
\end{eqnarray}
where the coefficients $a_{n}$ take the integer values 0, $\pm 1$
as determined by (\ref{Deltaclose'}). For small $\sigma$ the three
functions $S_{pq}(\sigma)$  have the following explicit forms:
\begin{eqnarray} \label{Deltadistant''}
{\cal S}_{11}&=&2^{1/2} \sigma^{1/8} + \sigma - \sigma^{2} -
2^{1/2} \sigma^{4+1/8} - \sigma^{5} + ... \,
\nonumber \\
{\cal S}_{21}&=& - 2^{1/2} \sigma^{1/8} + \sigma - \sigma^{2} +
2^{1/2} \sigma^{4+1/8}
- \sigma^{5} + ... \,  \nonumber \\
{\cal S}_{22}&=&-\sigma - \sigma^{2} + \sigma^{5} + ... \;\; .
\end{eqnarray}

For closely spaced boundaries with $i \neq j$ the derivative with
respect to $\kappa$ in Eq.~(\ref{Deltaijclose}) contributes due to
the prefactor in Eq.~(\ref{Deltaclose''}) even in leading order
$W/L \ll 1$ so that $\Delta_{i,j}(0)$ depends on $(i,j)$ as given
below Eq.~(\ref{stressscale'}). For widely spaced boundaries
$(i,j)$ the derivative with respect to $\sigma$ in
Eq.~(\ref{Deltaijdistant}) does not contribute to the leading
behavior and
\begin{equation} \label{Deldist}
\Delta_{i,j}(1/\delta) \to \pi \delta^{-2}/12  = -
\Delta_{P}(0)/\delta^{2} , \quad 1/\delta \to \infty \, ,
\end{equation}
is independent of $(i,j)$. In this case of $W/L \gg 1$ the stress
tensor averages of the strip ST are dominated by the periodic
boundary condition in $u$ direction and determined by $\langle
T_{\parallel \, \parallel} \rangle_{\rm ST} \equiv \langle T_{u \,
u} \rangle_{\rm ST} \to \Delta_{P}(0)/L^{2}$ with $\Delta_{P}(0)$
given below Eq.~(\ref{expand''}).

The above expressions in
Eqs.~(\ref{Deltaijrect'})-(\ref{Deltadistant''}) do not only serve
to provide the aspect ratio dependence of the leading contribution
$\Delta F_{l}$ in Eq.~(\ref{expand'}) of the free energy required
to rotate the small needle in the strip, but also to calculate the
effective interactions for certain particles of {\it arbitrary}
size: in the first entry of Ref.~[\onlinecite{ber}] for two
particles of circular shape as well as in
Appendix~\ref{apparbbreak} below for certain configurations of two
needles in the unbounded plane, of one needle in the half plane,
and of one needle in a strip.

Now we present the explicit forms of the scaling functions
$f_{\cal O}^{(i,j)} (v_{N}/W,0)$ of the density profiles in
Eq.~(\ref{densscale}) in a strip of infinite length $L=\infty$.
These can be inferred from, e.g., Ref.~[\onlinecite{BX}]. In our
notation they  are given by
\begin{eqnarray} \label{f}
f_{\phi}^{(O,O)}=0&,& \, f_{\epsilon}^{(O,O)}=C/2 \nonumber \\
f_{\phi}^{(+,+)}=(2C)^{1/8}&,& \, f_{\epsilon}^{(+,+)}=-C/2 \nonumber \\
f_{\phi}^{(+,-)}=-(2C)^{1/8} s&,& \, f_{\epsilon}^{(+,-)}=(C/2) [3-4s^{2}] \nonumber \\
f_{\phi}^{(+,O)}=(C/2)^{1/8} [1-s]^{1/4}&,& \, f_{\epsilon}^{(+,O)}=(C/2) s \nonumber \\
\end{eqnarray}
where
\begin{eqnarray} \label{CS}
C \equiv \pi/\cos(\pi v_{N}/W), \quad s \equiv \sin(\pi v_{N}/W) \, .
\end{eqnarray}
One can easily check that near the boundaries the corresponding
profiles $\langle{\cal O}({\bf r}_{N})\rangle$ reduce to the half
plane limits $\langle{\cal O}({\bf r}_{N})\rangle_{\rm half \,
plane}$ with the amplitudes provided in Eq.~(\ref{halfamp}). In
particular, in the strip with $i=+$ the three profiles
$f_{\phi}^{(+,j)}$  given in Eq.~(\ref{f}) exhibit the behavior
$f_{\phi}^{(+,j)}(v_{N}/W \to -1/2) \to 2^{1/8}
((v_{N}/W)+(1/2))^{-1/8}=(2W/a_{N})^{1/8}$ corresponding to the
half plane with boundary~$+$.

\section{NEEDLES OF ARBITRARY LENGTH} \label{apparb}
\subsection{Symmetry preserving cases} \label{apparbord}
{\it Needle and both boundaries belonging to $O$}

Here we  establish Eqs.~(\ref{arbperp}) and~(\ref{arbord}) for the
free energy associated with the insertion of a needle in the case
that both the needle and the boundaries are  of the symmetry
preserving ``ordinary'' type. We start with Burkhardt's result
\cite{burkunpub} for the thermal average of the stress tensor
induced by $n$ nonoverlapping ``ordinary'' needles embedded in the
$x$ axis of the unbounded $(x,y)$ plane. If the $n$ needles extend
from $x_{1<}$ to $x_{1>}$, from $x_{2<}$ to $x_{2>}$, ... , and
from $x_{n<}$ to $x_{n>}$, respectively, with arbitrary real
numbers $x_{1<} \leq x_{1>} \leq x_{2<} \leq x_{2>} \leq ... \leq
x_{n<} \leq x_{n>}$, the stress tensor averages $\langle T_{kl}
(x,y) \rangle$ at a point $(x,y)$ follow from the analytic
function
\begin{eqnarray} \label{Tnneedles}
&&\langle T(z) \rangle^{([O][O] ... [O])} = 2^{-6} \Biggl({1 \over
z-x_{1<}}-{1 \over z-x_{1>}} \nonumber \\
&&+{1 \over z-x_{2<}}-{1 \over z-x_{2>}}+ ... +{1 \over
z-x_{n<}}-{1 \over z-x_{n>}} \Biggr)^2 \nonumber \\
\end{eqnarray}
of the complex variable $z=x+iy$ via \cite{TklTz} the relations
$\langle T_{xx} (x,y) \rangle = -\langle T_{yy} (x,y) \rangle = -
{\rm Re} \langle T(z) \rangle /\pi$ and $\langle T_{xy} (x,y)
\rangle = \langle T_{yx} (x,y) \rangle = {\rm Im} \langle T(z)
\rangle /\pi$ where Re and Im denote real and imaginary parts,
respectively. We point out the consistency of
Eq.~(\ref{Tnneedles}) in the special cases $x_{m<}\to x_{m>}$, in
which  the needle $m$ disappears, and for $x_{m>} \to x_{m+1<}$,
in which two consecutive needles $m$ and $m+1$ merge into a single
one. For $n \, = \, 1$ this reproduces the expression $\langle
T(z) \rangle^{([h])}$ which is independent~\cite{parthalf} of $h$
and follows from mapping the half plane onto the entire plane
outside a single needle (cp. Eqs.~(A8) and~(A9) in
Ref.~[\onlinecite{e'}].

In order to derive Eq.~(\ref{arbperp}) we
use~Eq.~(\ref{Tnneedles}) for $n=2$, put $(x_{1<} ,
x_{1>})=(-\infty,0)$, so that the first needle fills the negative
$x$-axis, and denote $(x_{2<},x_{2>})$ by $(x_{<},x_{>})$. The
conformal transformation $z=c^2 /l$ with $c=a+ib$ maps the $(x,y)$
plane with the two ``ordinary'' needles onto the half plane
$(a>0,b)$ with the ``ordinary'' boundary line $a=0$ and a single
embedded ``ordinary'' needle extending from $(a=\sqrt{lx_{<}}
\equiv a_{<}, b=0)$ to $(a=\sqrt{lx_{>}} \equiv a_{>}, b=0)$. This
leads to a stress tensor function
\begin{eqnarray} \label{Thalfperp}
\langle T(c) \rangle = -A/8 + c^{2} A^{2} /16
\end{eqnarray}
where
\begin{eqnarray} \label{Thalfperp'}
A = {1 \over c^{2}-a_{<}^{2}}-{1 \over c^{2}-a_{>}^{2}} \, .
\end{eqnarray}
In order to obtain this result one uses the transformation formula
for the stress tensor which includes the Schwartz derivative
\cite{TklTz}. The arbitrary length $l$, introduced for dimensional
reasons, does not appear in the relation between $\langle T(c)
\rangle$ and $a_{<}, a_{>}$. Shifting the needle away from the
boundary line, i.e., increasing the distance
$a_{N}=(a_{>}+a_{<})/2$ of its midpoint from the boundary while
keeping its length $a_{>}-a_{<}$ and orientation fixed, leads to a
change in the free energy $F_{\perp}$ determined
by~\cite{Cardy,ber}
\begin{eqnarray} \label{shifthalfperp}
{\partial \over \partial a_{N}} {F_{\perp} \over k_{B}T} = -
\int_{- \infty}^{\infty} d b \, \langle T_{\perp \perp} (a,b)
\rangle \,
\end{eqnarray}
where $\langle T_{\perp \perp} (a,b) \rangle = - {\rm Re} \langle
T(c) \rangle /\pi$ is the diagonal stress tensor component
perpendicular to the boundary line $a=0$. The integration path
must extend between the boundary and the needle, i.e., $0 \leq a <
a_{<}$ in Eq.~(\ref{shifthalfperp}). In this region the integral
over $b$ is independent of $a$ and is carried out most easily for
$a=0$. Integrating the result with respect to $a_{N}$ and denoting
the needle length $a_{>}-a_{<}$ by $D$ leads to the result in
Eq.~(\ref{arbperp}) for $F_{\perp} \equiv F_{\perp}^{(O[O])}$.

In order to establish Eq.~(\ref{arbord}) one uses the
transformation $c/l=\exp(\pi w/W)$, $w=u+iv$, in order to map the
half plane $(a>0,b)$ with its embedded needle onto a needle in
strip geometry as described in the context of Eq.~(\ref{arbord}).
With the ends of the needle at $u= \pm D/2$, this way one finds
the stress tensor function
\begin{eqnarray} \label{Tstripparall}
\langle T(w) \rangle = (\pi/W)^{2} (-1/48 -B/8 + B^{2} /16)
\end{eqnarray}
where
\begin{eqnarray} \label{Tstripparall'}
B = {\sinh \theta \over \cosh \theta -\cosh (2\pi w/W)}
\end{eqnarray}
with $\theta$ defined in Eq.~(\ref{theta}). The free energy change
$(d/dW)F_{\parallel}^{(O[O]O)}/(k_{B}T)$ upon widening the strip
follows from the right hand side of Eq.~(\ref{shifthalfperp}) by
replacing $(a,b)$ by $(u,v)$ and identifying $\langle T_{\perp
\perp} (u,v) \rangle$ with ${\rm Re} \langle T(w) \rangle/\pi$.
Here and below Eq.~(\ref{shifthalfperp}) the real parts of
$\langle T(w) \rangle$ and $\langle T(c) \rangle$ enter with a
plus and minus sign, respectively, because in the complex $w$ and
$c$ planes the directions $\perp$ perpendicular to the boundaries
point along the imaginary and the real axis, respectively.
Performing the integral and integrating with respect to $W$ leads
to the result for $ F_{\parallel}^{(O[O]O)}$ given in
Eq.~(\ref{arbord}).

{\it Needle in a strip with periodic boundary condition}

Now we consider the geometry corresponding to Eq.
(\ref{arbperiod}). For the infinitely long strip with periodic
boundary condition containing a needle in parallel direction the
stress tensor function $\langle T(w) \rangle$ follows from
Eq.~(\ref{Tnneedles}) with $n=1$ and the transformation
$z/l=\exp(2 \pi w/W)$ yielding
\begin{eqnarray} \label{Tperstrip}
\langle T(w) \rangle = (\pi/W)^{2} (-1/12 + B^{2} /16)
\end{eqnarray}
with $B$ given in Eq.~(\ref{Tstripparall'}). Proceeding as for the
$(O,O)$ strip above, Eq.~(\ref{Tperstrip}) leads to the result for
$F_{\parallel}$ given in Eq.~(\ref{arbperiod}).

{\it Ordinary needle and $+$ boundaries}

Finally, we derive Eqs.~(\ref{arbperp+O}) and~(\ref{arbord+O})
from the expression \cite{burkunpub}
\begin{eqnarray} \label{TEO}
&&\langle T(z) \rangle^{([E][O])} = 2^{-6} \times \nonumber \\
&&\times \Biggl({1 \over z-x_{1<}}-{1 \over z-x_{1>}}-{1 \over
z-x_{2<}}+{1 \over z-x_{2>}} \Biggr)^{2} \nonumber \\
\end{eqnarray}
for the stress tensor function induced by an ``extraordinary''
needle $E$ extending from $x_{1<}$ to $x_{1>}$ and an ``ordinary''
needle from $x_{2<}$ to $x_{2>}$. An ``extraordinary'' needle
preserves the $(+ \leftrightarrow -)$ symmetry and can be realized
in a lattice model as a line of spins with infinitely strong
nearest neighbor ferromagnetic couplings between them so that they
all point either in the $+$ or all in the $-$
direction~\cite{duality,dual}. The partition function
$Z^{([E][O])} \equiv Z^{([+][O])}+Z^{([-][O])}$ in the presence of
the two needles $E$ and $O$ differs by only a factor of 2 from the
two identical partition functions $Z^{([+][O])}=Z^{([-][O])}$ in
the presence of two needles $+$ and $O$ or $-$ and $O$. (This
argument holds if the partition functions are finite. This can be
achieved by enclosing the whole system in a large box with
``ordinary'' boundaries.) Thus the corresponding free energies
differ by an additive constant $-k_{B}T \ln 2$ which drops out
from the free energy difference upon changing the needle geometry
as well as from the stress tensor so that
\begin{eqnarray} \label{TEOT+O}
\langle T(z) \rangle^{([E][O])} = \langle T(z) \rangle^{([+][O])}
= \langle T(z) \rangle^{([-][O])} \, .
\end{eqnarray}
Note the different sign sequences in $\langle T(z)
\rangle^{([E][O])}$ on the rhs of Eq.~(\ref{TEO}) and in $\langle
T(z) \rangle^{([O][O])}$ on the rhs of Eq.~(\ref{Tnneedles}) with
$n=2$. This implies that for obtaining the $(+[O])$ case
Eqs.~(\ref{Thalfperp}) and~(\ref{arbperp}) have to be modified by
replacing $A \to -A$ and $\vartheta \to -\vartheta$, respectively,
which leads to Eq.~(\ref{arbperp+O}). Similarly, for the $(+[O]+)$
case Eqs.~(\ref{Tstripparall}) and~(\ref{arbord}) have to be
modified by replacing $B \to -B$ and $\theta \to -\theta$,
respectively, in order to obtain Eq.~(\ref{arbord+O}).

\subsection{Cases with genuinely broken symmetry} \label{apparbbreak}

In this subsection
 we derive the general expressions in Eqs.~(\ref{iJ})
 and~(\ref{arbiJ}) for the Casimir forces which encompass also the
cases $(+[+])$, $(-[+])$ and $(+[+]+)$, $(-[+]-)$ which cannot be
reduced to cases with symmetry preserving boundaries. We start by
considering two needles 1 and 2 on the $x$ axis, similar as in
Appendix~\ref{apparbord}, but with {\it arbitrary} universality
classes $[i]$ and $[j]$, respectively. The corresponding stress
tensor $\langle T(z) \rangle^{([i][j])}$ can be inferred from the
difference
\begin{eqnarray} \label{Tbreak}
&&\langle T(z) \rangle^{([i][j])} -  \langle T(z)
\rangle^{([O][O])}= \nonumber \\
&&= {1
\over (z-x_{1<})(z-x_{1>})(z-x_{2<})(z-x_{2>})} \times \nonumber \\
&&\quad \times {(x_{1>}-x_{1<})(x_{2>}-x_{2<}) \over (1-k)^2}
\tau_{i,j}(k)
\end{eqnarray}
where
\begin{eqnarray} \label{Tbreak'}
\tau_{i,j}(k) \, = \, {1 \over 48} (-1+6k-k^2) -{\pi
\Delta_{i,j}(1/\delta) \over (2 K(k))^2}
\end{eqnarray}
and $\langle T(z) \rangle^{([O][O])}$ is the stress tensor for two
``ordinary'' needles given by Eq.~(\ref{Tnneedles}) with $n=2$.
The quantity $0\leq k \leq1$ is related to cross ratios of the four
needle end points via one of the two following equivalent relations:
\begin{eqnarray} \label{cross}
{(x_{1>}-x_{1<})(x_{2>}-x_{2<}) \over
(x_{2<}-x_{1>})(x_{2>}-x_{1<})}={(1-k)^2 \over 4k}
\end{eqnarray}
and
\begin{eqnarray} \label{cross'}
{(x_{2<}-x_{1<})(x_{2>}-x_{1>}) \over
(x_{2<}-x_{1>})(x_{2>}-x_{1<})}={(1+k)^2 \over 4k} \,.
\end{eqnarray}
The amplitude functions  $\Delta_{i,j}(1/\delta)$ are given by
Eqs.~(\ref{Deltaijrect})~-~(\ref{Deltadistant'}) with the argument
$1/\delta$ related to $k$ via
\begin{eqnarray} \label{Ellipt}
1/\delta=K(k)/K^{*}(k) \,
\end{eqnarray}
with the complete elliptic integral~\cite{GR} functions $K(k)$ and
$K^{*}(k)=K(\sqrt{1-k^2})$ (see Eq.~(\ref{theta'})).

Note that the pole of second order at $z$ equal to a needle end,
present with equal residues $2^{-6}$ in the stress tensor averages
in Eqs.~(\ref{Tnneedles}) and~(\ref{TEO}), is absent in the
difference of averages in Eq.~(\ref{Tbreak}). This implies that,
e.g., near the needle end at $x_{1<}$ the leading contribution to
the corresponding average of the stress tensor elements $T_{kl}$
(see below Eq.~(\ref{Tnneedles})) is independent not only
\cite{parthalf} of the length and universality class of needle 1
but also of the presence (i.e., distance, length, and universality
class) of another needle 2. Moreover - as shown by Eqs.
(\ref{Thalfperp}),~(\ref{Tstripparall}), and~(\ref{Tperstrip})
above as well as by Eqs.~(\ref{Thalfbreak})
and~(\ref{Tstripbreak}) below - this contribution is also independent
of the presence of the concomitant boundaries of the half space
and the strip.

Equations~(\ref{Tbreak})~-~(\ref{Ellipt}) follow from the
Schwarz-Christoffel transformation~\cite{SC, SC'} which
conformally maps the $z=x+iy$ plane with the two needles $[i]$ and
$[j]$ embedded in the $x$ axis onto the rectangle or strip ST with
boundaries $(i,j)$ at $v= \pm W/2$, periodic boundary condition in
$u$ direction, and Casimir amplitudes $\Delta_{i,j}$, as
introduced in the paragraph containing
Eqs.~(\ref{densscale})-(\ref{stressscale'}). The Schwartz
derivative in the corresponding transformation law of stress
tensors\cite{Cardy,TklTz} drops out from the stress tensor
difference given by Eq.~(\ref{Tbreak}).

The vanishing of $\tau_{i,j}$ in Eq.~(\ref{Tbreak'}) for
$(i,j)=(O,O)$ provides, together with Eq.~(\ref{Ellipt}), another
expression for $\Delta_{O,O}(1/\delta)$ besides (but equivalent
to) the ones in Eqs.~(\ref{Deltaijrect}),~(\ref{Deltaijclose}),
and~(\ref{Deltaijdistant}).
 For $\Delta_{+,O}(1/\delta)$ the corresponding
other expression follows  from Eqs.~(\ref{Tbreak}) and
(\ref{Tbreak'}) with $(i,j)=(+,O)$ when combined with
Eqs.~(\ref{Tnneedles}),~(\ref{TEO}), and~(\ref{TEOT+O}).

The difference of the  stress tensors in Eq.~(\ref{Tbreak}) must vanish
in the limit of distant needles with no correlation between them
because the stress tensor for a single needle in unbounded space
is independent of its
universality class~\cite{parthalf}. For the same reason it must
also vanish if two needles of the same universality class
 $i=j=+$ (or $-$) come
close so that they merge, i.e., $x_{1>}=x_{2<}$, and form a single $+$
(or $-$) needle. These expectations are in agreement with the
behaviors
\begin{eqnarray} \label{taudistant}
{\rm lim}_{k \to 1} \, \tau_{i,j}(k) = 0 \,
\end{eqnarray}
and
\begin{eqnarray} \label{tauclose}
{\rm lim}_{k \to 0} \, \tau_{i,j}(k) =
[\Delta_{i,i}(0)-\Delta_{i,j}(0)]/\pi
\end{eqnarray}
of $\tau_{i,j}$ for distant and close needles with $\delta
\searrow 0$, $k \nearrow 1$ and $\delta \nearrow \infty$, $k
\searrow 0$, respectively. Equation~(\ref{taudistant}) follows from
Eqs.~(\ref{Deldist}) and~(\ref{Ellipt}).

Proceeding as described above Eq.~(\ref{Thalfperp}), the
difference of the stress tensors for the needle in the half plane
is obtained as
\begin{eqnarray} \label{Thalfbreak}
\langle T(c) \rangle^{(i[j])} -  \langle T(c) \rangle^{(O[O])} = -
\frac{4 A}{ (1-k)^2} \tau_{i,j}(k)
\end{eqnarray}
with $\langle T(c) \rangle^{(O[O])} \equiv \langle T(c) \rangle$
from Eq.~(\ref{Thalfperp}) and the function $A$ from
Eq.~(\ref{Thalfperp'}). Via
\begin{eqnarray} \label{Thalfbreak'}
&&a_{<} / a_{>} = 2 k^{1/2} /(1+k) , \nonumber \\
&&k = (1-\vartheta^{1/2})^{2} / (1+\vartheta^{1/2})^{2}
\end{eqnarray}
$k$ is related to the needle parameters $a_{<}/a_{>}$  and
$\vartheta=(a_{>}-a_{<})/(a_{>}+a_{<})$ introduced above Eq.
(\ref{Thalfperp}) and in Eq.~(\ref{vartheta}). Using the relation
between $k$ and $\vartheta$ in Eq.~(\ref{Thalfbreak'}) together
with the suitable functional relations
\begin{eqnarray} \label{K+K}
&&K(k)=(1+\vartheta^{1/2})^{2} \, K^{*}(\vartheta)/4 \, , \nonumber \\
&&K^{*}(k)=(1+\vartheta^{1/2})^{2} \, K(\vartheta) \,
\end{eqnarray}
between elliptic integrals (see Eqs. 8.126.1 and 8.126.3 in Ref.
[\onlinecite{GR}]) yields together with Eq.~(\ref{Tbreak'}) the
more convenient expression
\begin{eqnarray} \label{conven}
{4 \over (1-k)^2} \tau_{i,j}(k) = {1 \over 4 \vartheta}
\tilde{\tau}_{i,j}(\vartheta)
\end{eqnarray}
for the amplitude in Eq.~(\ref{Thalfbreak}) where
$\tilde{\tau}_{i,j}$ is taken from
Eqs.~(\ref{sigmatilde})-(\ref{K+K^{*}}).
Equation~(\ref{shifthalfperp}) finally yields the expression
\begin{eqnarray} \label{forcehalfbreak}
-{\partial \over \partial a_{N}} {F_{\perp}^{(i[j])} -
F_{\perp}^{(O[O])} \over k_{B} T} &=& - {1 \over 4 \vartheta}
\tilde{\tau}_{i,j}(\vartheta) \Biggl( {1 \over a_{<}} - {1 \over
a_{>}} \Biggr) \nonumber \\
\end{eqnarray}
for the difference of the Casimir forces acting on the needle in
the cases $(i,j)$ and $(O,O)$ which implies Eq.~(\ref{iJ}) upon
renaming the dummy index $j$ as $h$.

For the special combinations $(+,O)$, $(O,+)$, $(-,O)$, and
$(O,-)$ the expression in Eq.~(\ref{conven}) becomes independent
of $k$ and $\vartheta$ and equals $-1/4$. This follows from
comparing Eq.~(\ref{Tbreak}) with the simple expressions in
Eqs.~(\ref{Tnneedles}) and~(\ref{TEO}) and by using Eq.~(\ref{TEOT+O}).
For combinations $(i,j)$ with both $i$ and $j$ being ``normal''
universality classes the expression in Eq.~(\ref{conven}) displays
a nontrivial dependence on $k$ and $\vartheta$.

In Eq.~(\ref{sigma}) the dependence on $\vartheta$ of
$\tilde{\tau}_{i,h}$ and $\rho_{i,h}$, with $(i,h)$ being
arbitrary, can be readily calculated for a needle nearly touching
the boundary ($\bar \vartheta \, = \, \sqrt{1-\vartheta^{2}}
\searrow 0, \, \delta \nearrow \infty$) as well as for a distant
or small needle ($\vartheta \searrow 0, \, \delta \searrow 0$) by
combining the first and second expression in Eq.~(\ref{sigma})
with Eqs.~(\ref{Deltaijclose}) and~(\ref{Deltaclose''}) and with
Eqs.~(\ref{Deltaijdistant}),~(\ref{Deltadistant''}),
 and~(\ref{Deldist}), respectively.
 In obtaining these limiting behaviors  one
uses the fact that Eqs.~(\ref{Deltaclose'})
and~(\ref{Deltadistant'}) and Eqs.~(\ref{K+K^{*}})
and~(\ref{theta'}) imply the simple dependences
\begin{eqnarray} \label{phi+gamma}
\kappa=[(\bar \vartheta/4)P( \bar \vartheta^{2})]^{2} \, , \quad
\sigma=(\vartheta/4)P(\vartheta^{2})
\end{eqnarray}
of $\kappa$ and $\sigma$ in Eqs.~(\ref{Deltaijclose})
and~(\ref{Deltaijdistant}) on $\bar \vartheta$ and $\vartheta$,
where $P(x^{2})$ can be expressed in terms of a power series in
$x^{2}$ given by
\begin{eqnarray} \label{P}
P(x^{2})=\exp \{-Q(x^{2})/R(x^{2})\}=1+x^{2}/4+{\cal O}(x^{4}) \,
, \nonumber \\
\end{eqnarray}
with
\begin{eqnarray} \label{P'}
R(x^{2})=(2/\pi)K(x)=1+x^{2}/4+{\cal O}(x^{4}) \,
\end{eqnarray}
and (see Ref.~[\onlinecite{GR}])
\begin{eqnarray} \label{P''}
&&Q(x^{2})=K(\sqrt{1-x^2})-R(x^{2})\ln(4/x) \nonumber \\
&& = -2 \Biggl\{  \Biggl( {1 \over 2}\Biggr)^{2} {1 \over 1 \times
2}x^{2} + \nonumber \\
&&\; \, + \Biggl( {1 \times 3 \over 2 \times 4 } \Biggr)^{2}
\Biggl[ {1 \over 1 \times 2}+{1 \over 3 \times 4} \Biggr]x^{4}
+... \Biggr\} \, .
\end{eqnarray}

We use these relations to show that for small $\vartheta$ the
exact expressions in Eq.~(\ref{iJ}) for the Casimir forces are
consistent with the ``small needle expansion''. To this end we
first note that expanding the expressions in Eq.~(\ref{iJ}) yields
\begin{eqnarray} \label{+-expand}
- a_{N} {\partial \over \partial a_{N}}{F_{\perp}^{(i[h])} \over
k_{B} T} =- \vartheta {d \over d \vartheta} \ln \{  \} + {\cal
O}(\vartheta^{4})
\end{eqnarray}
with the curly bracket being identical to the curly bracket in the
expression for $\Delta_{i,h}$  in Eq.~(\ref{Deltaijdistant}). This
follows from inserting Eqs.~(\ref{sigma})
and~(\ref{Deltaijdistant}) into the second expression in
Eq.~(\ref{iJ}) and by using Eq.~(\ref{P'}) for $K(\vartheta)$ as
well as the relation $\sigma \, d/d\sigma =(1-\vartheta^{2}/2
+{\cal O}(\vartheta^{4}))\vartheta \, d/d \vartheta$ due to
Eq.~(\ref{phi+gamma}). Furthermore, apart from terms of order
$\vartheta^{4}$ the curly bracket in Eq.~(\ref{+-expand}) is equal
to $1+\zeta_{I}+\zeta_{A}$ with the  expressions $\zeta_{I}$ and
$\zeta_{A}$ from Eq.~(\ref{halfzeta}) for a small needle in a half
plane. For example, in the case $(i[h])=(-[+])$, with the small
needle approximation~\cite{ellipses}
\begin{eqnarray} \label{+-predict}
&&- a_{N} {\partial \over \partial a_{N}}{F_{\perp}^{(-[+])} \over
k_{B} T} = -\vartheta {d \over d \vartheta} \ln
(1+\zeta_{I}+\zeta_{A}) \nonumber \\
&&= - \vartheta {d \over d \vartheta} \ln
[1-2^{1/4}\vartheta^{1/8}+\vartheta/4 \nonumber
\\
&& \qquad \qquad \qquad
-2^{1/4}(3/32)\vartheta^{2+1/8}+(5/4^{3})\vartheta^{3}] \, ,
\nonumber \\
\end{eqnarray}
the aforementioned relation between the curly bracket and
$1+\zeta_{I}+\zeta_{A}$ follows from the second part of
Eq.~(\ref{Deltadistant''}) and from Eqs.~(\ref{phi+gamma})
and~(\ref{P}). Accordingly, the difference between the exact
expression for the critical Casimir forces in Eq.~(\ref{iJ})
 and their small needle approximations
is of the order $\vartheta^{4}$.

Proceeding as in the paragraph containing
Eqs.~(\ref{Tstripparall}) and~(\ref{Tstripparall'}), for the
geometry of a needle of class $j$ extending along the midline of
an $(i,i)$ strip, as discussed in paragraph (ii) in Sec.
\ref{ARB}, one obtains
\begin{eqnarray} \label{Tstripbreak}
\langle T(w) \rangle^{(i[j]i)} -  \langle T(w) \rangle^{(O[O]O)} =
- B \times  (\pi/W)^{2} \, \tilde{\tau}_{i,j}(t)/(4t) \nonumber \\
\end{eqnarray}
with the function $B$ given by Eq.~(\ref{Tstripparall'}), $\langle
T(w) \rangle^{(O[O]O)} \equiv \langle T(w) \rangle$ from
Eq.~(\ref{Tstripparall}), $\tilde{\tau}_{i,j}$ as in
Eq.~(\ref{sigmatilde}), and $t$ related to $\theta \equiv \pi D/W$
as stated in Eq.~(\ref{arbiJ}). The force in Eq.~(\ref{arbiJ})
follows from the stress tensor difference in
Eq.~(\ref{Tstripbreak}) upon replacing $[j]$ by $[h]$ and by using
a so-called shift equation as in Eq.~(\ref{shifthalfperp}).

Similar as in the paragraph containing Eq.~(\ref{+-expand}), the
expanded expression
\begin{eqnarray} \label{expandpar}
&&- W {\partial \over \partial W}{F_{\parallel}^{(i[h]i)} \over
k_{B} T} = -2 \hat{e} \theta^{2} - \theta {d \over d \theta} \ln
\{1+\hat{a} \theta^{1/8}+\hat{b} \theta + \nonumber \\
&&+\hat{c} \theta^{2+1/8}+\hat{d} \theta^{3} \}+{\cal
O}(\theta^{4}) \, , \qquad \hat{e} = -(1/3)2^{-7} \, ,
\end{eqnarray}
of the exact result in Eq.~(\ref{arbiJ}) with coefficients
$\hat{a}-\hat{e}$ follows from Eqs.~(\ref{sigma}),
(\ref{Deltaijdistant}), (\ref{Deltadistant''}), (\ref{phi+gamma}),
and~(\ref{P}) and is related to the ``small needle approximation''
\cite{ellipses}
\begin{eqnarray} \label{approxpar}
&&- W {\partial \over \partial W}{F_{\parallel}^{(i[h]i)} \over
k_{B} T} = -\theta {d \over d \theta} \ln
(1+\zeta_{I}-\zeta_{A}) \nonumber \\
&&=- \theta {d \over d \theta} \ln (1+\hat{a} \theta^{1/8}+\hat{b}
\theta+ \hat{e} \theta^{2}+\hat{f} \theta^{2+1/8}+\hat{g}
\theta^{3}) \nonumber \\
\end{eqnarray}
where $\zeta_{I}$ and $\zeta_{A}$ are given by Eqs.~(\ref{zetaI})
and~(\ref{zetaA}) with needle center $v_{N}=0$ at the midline of
the $(i,i)$ strip. Note that the term $\hat{e} \theta^{2}$ with
$\hat{e}$ given in Eq.~(\ref{expandpar}) equals the contribution
$(\pi/2)(D/2W)^{2}\Delta_{i,i}$ with the critical Casimir
amplitudes $\Delta_{i,i}=-\pi/48$ for  $(i,i)$ strips which enter
into $\zeta_{A}$ in Eq.~(\ref{zetaA}) and Eq.~(\ref{approxpar}).
Unlike $\hat{e}$, the other coefficients $\hat{a}, \, \hat{b}, \,
\hat{c}, \, \hat{d}, \, \hat{f}$, and $\hat{g}$ depend on the
boundary and needle universality classes $(i,h)$. The deviation of
the exact expression $-W(d/d W)F_{\parallel}^{(i[h]i)} /(k_{B}T)$
from its ``small needle approximation'' is given by the difference
of Eq.~(\ref{expandpar}) and Eq.~(\ref{approxpar}) and is of the
order $\theta^{4}$ because
\begin{eqnarray} \label{since}
\hat{a} \hat{e}+\hat{c}-\hat{f}=\hat{b} \hat{e}+\hat{d}-\hat{g}=0
\, .
\end{eqnarray}
For example, in  the case $(i[h]i)=(-[+]-)$, in which the
expression for the argument of the logarithm in
Eq.~(\ref{approxpar}) equals $Z_{\parallel}$ as given by
Eq.~(\ref{full}), one has
\begin{eqnarray} \label{since'}
\hat{a}&=&-2^{1/8}, \, \hat{b}=1/8, \, \hat{c}=-(5/3)2^{-7+1/8}, \nonumber \\
\hat{d}&=&-(1/3)2^{-9}, \, \hat{f}=-(1/3)2^{-5+1/8}, \,
\hat{g}=-2^{-10} \, , \nonumber \\
\end{eqnarray}
which satisfy Eq.~(\ref{since}).

For the convenience of the reader we provide the explicit
expressions
\begin{eqnarray} \label{SNE}
\Bigl(f_{O[O]}^{\perp}(\vartheta)\Bigr)_{\rm sna}&=&-\ln \{1+
(\vartheta/4)[1+(5/16)\vartheta^{2}]\} \, , \nonumber \\
\Bigl(f_{+[+]}^{\perp}(\vartheta)\Bigr)_{\rm sna}&=&-\ln
\{1+2^{1/4}\vartheta^{1/8}[1+(3/32) \vartheta^{2}] \nonumber \\
&& \qquad + (\vartheta/4)[1+(5/16)\vartheta^{2}] \} \, , \nonumber
\\
\Bigl(f_{-[+]}^{\perp}(\vartheta)\Bigr)_{\rm sna}&=&-\ln
\{1-2^{1/4}\vartheta^{1/8}[1+(3/32) \vartheta^{2}] \nonumber \\
&& \qquad + (\vartheta/4)[1+(5/16)\vartheta^{2}] \} \,
\end{eqnarray}
and
\begin{eqnarray} \label{barfpar}
&&\Bigl(f_{O[O]O}^{\parallel}(\theta)\Bigr)_{\rm sna}=-\ln \Bigl
\{ 1-(1/3)2^{-7} \theta^{2} \nonumber \\
&& \qquad \qquad \qquad \qquad \qquad \qquad +
(\theta/8)[1-2^{-7} \theta^{2}] \Bigr \} \, , \nonumber \\
&&\Bigl(f_{+[+]+}^{\parallel}(\theta)\Bigr)_{\rm sna}=-\ln \Bigl
\{ 1-(1/3)2^{-7} \theta^{2} \nonumber \\
&& \quad +(2\theta)^{1/8}[1+ (1/3) 2^{-5} \theta^{2}] +
(\theta/8)[1-2^{-7} \theta^{2}]
\Bigr \} \, , \nonumber \\
&&\Bigl(f_{-[+]-}^{\parallel}(\theta)\Bigr)_{\rm sna}=-\ln \Bigl
\{ 1-(1/3)2^{-7} \theta^{2} \nonumber \\
&& \quad -(2\theta)^{1/8}[1+ (1/3) 2^{-5} \theta^{2}] +
(\theta/8)[1-2^{-7} \theta^{2}]\Bigr \} \, \nonumber \\
\end{eqnarray}
for the~\cite{ellipses} ``small needle approximations''
$(f_{i[h]}^{\perp})_{\rm sna}$ and $(f_{i[h]i}^{\parallel})_{\rm
sna}$ of $f_{i[h]}^{\perp} \equiv F_{\perp}^{(i[h])}/(k_{B}T)$ and
$f_{i[h]i}^{\parallel} \equiv F_{\parallel}^{(i[h]i)}/(k_{B}T)$.
For the cases $(+[O])$ and $(+[O]+)$ the expressions follow from
those for $(O[O])$ and $(O[O]O)$ by the replacements $\vartheta
\to -\vartheta$ and $\theta \to -\theta$, respectively.
%
%
\section{GLOSSARY} \label{appnotations}
In order to ease the reading of the text, here we compile the
symbols and notations used and explain their meanings.
\vspace{0.4cm}

$A$ : Expression given by Eq.~(\ref{Thalfperp'}) which serves to
display the dependence on $c$ of the analytic functions $\langle
T(c) \rangle^{(i[h])}$ (see Eqs.~(\ref{Thalfperp}) and
(\ref{Thalfbreak}) as well as the text below Eq.~(\ref{TEOT+O}))
and thus to display the position dependence of the stress tensor
averages for the geometries of a needle of boundary universality
class $h$ in the half plane with orientation perpendicular to the
boundary of boundary universality class $i$.
\vspace{0.4cm}

${\cal A}_{\cal O}^{(h)}$ : Universal amplitudes of the profiles
$\langle {\cal O}(\bf r) \rangle_{\rm half \, plane}$ in the half
plane with boundary universality class $h$ where ${\cal O}=\phi$
and ${\cal O}=\epsilon$ are the normalized operators (see Eq.
(\ref{norm})) of the order parameter density and of the deviation
of the energy density from its bulk value, respectively (see Eq.
(\ref{halfamp})).
\vspace{0.4cm}

$a$ and $b$ : Half plane coordinates perpendicular and parallel to
the boundary $a=0$ (see above Eqs.~(\ref{halfamp}) and
(\ref{Thalfperp})).
\vspace{0.4cm}

$a_{N}$ : Distance of the needle center from the boundary of the
half plane (or from the lower boundary of the strip) (see Fig.~\ref{FIG_I_2}).
\vspace{0.4cm}

$a_{<}$ and $a_{>}$ : Distance of the closer and farther needle
end from the boundary for a needle in the half plane with
perpendicular orientation (see Fig.~\ref{FIG_I_2}(b) and the text
above Eqs.~(\ref{asmaller}) and~(\ref{Thalfperp})); $a_{>}= a_{<}
+ D$.
\vspace{0.4cm}

$B$ : Expression given by Eq.~(\ref{Tstripparall'}) which serves
to display the dependence on $w$ of the analytic functions
$\langle T(w) \rangle^{(i[h]i)}$ (see Eqs.~(\ref{Tstripparall})
and~(\ref{Tstripbreak}) as well as the text below Eq.
(\ref{TEOT+O})) and thus to display the position dependence of the
stress tensor averages for the geometries of a needle of class $h$
embedded in the midline of a strip with $(i,i)$ boundaries.
\vspace{0.4cm}

$c=a+ib$ : Complex variable specifying the position vector $(a,b)$
in the half plane (see above Eq.~(\ref{Thalfperp})).
\vspace{0.4cm}

$c_{i,j} \equiv [\Delta F_{nl}/(k_{B}T)]/(D/W)^{2}$ : Normalized
next-to-leading contribution to the quasi-torque acting on a small
``ordinary'' needle $O$ in an $(i,j)$ strip (see the paragraph
following Eqs.~(\ref{nlexplicit}) and~(\ref{nlexplicit'}) and the
caption of Fig.~\ref{FIG_V_2}).
\vspace{0.4cm}

$D$ : The number of missing bonds (fixed spins) in the lattice
description of the ``ordinary'' (``normal'') needle in Secs.
\ref{intro} and~\ref{SIM} (see Fig.~\ref{FIG_I_1} as well as
Figs.~\ref{FIG_IV_1} and~\ref{FIG_IV_2}) or~\cite{micromeso} the length
of the needle in the continuum descriptions used in
Secs.~\ref{intro},~\ref{SP},~\ref{ARB}, and Appendix~\ref{apparb} (see
Fig.~\ref{FIG_I_2}).
\vspace{0.4cm}

$E$ : ``Extraordinary'' boundary with infinitely strong
ferromagnetic nearest neighbor couplings between surface spins
(see Ref. [\onlinecite{duality}]).
\vspace{0.4cm}

$F_{\parallel}$ and $F_{\perp}$ : Free energy cost to transfer the
needle from the bulk into the strip (or into the half plane) with
its orientation parallel and perpendicular to the boundaries,
respectively (see Eqs.~(\ref{Fparal})-(\ref{zetaA}) and the text
above Eq.~(\ref{theta}) or below Eq.~(\ref{asmaller})). In
particular, $F_{\parallel} \equiv F_{\parallel}^{(i[h]i)} = k_{B}T
f_{i[h]i}^{\parallel}(\theta)$ and $F_{\perp} \equiv
F_{\perp}^{(i[h])} = k_{B}T f_{i[h]}^{\perp}(\vartheta)$ for
needles $h$ of arbitrary length embedded in the midline of an
$(i,i)$ strip and perpendicular to the boundary $i$ of the half
plane, respectively.
\vspace{0.4cm}

$\Delta F \equiv F_{\parallel}-F_{\perp} \equiv F_{1}-F_{0}$ :
Free energy required to turn the needle about its center from the
perpendicular to the parallel orientation (see Eq.
(\ref{DeltaF'})).
\vspace{0.4cm}

$\Delta F_{l}$ and $\Delta F_{nl}$ : Leading and next-to-leading
contributions, respectively, to $\Delta F$ for the cases $(i[O]j)$
and $(O[+/-]O)$ of a small needle (see Eqs.
(\ref{expand})-(\ref{expand''}) as well as the corresponding half
plane relations in Eq.~(\ref{halfDelF})).
\vspace{0.4cm}

$F_{\rm ST}^{(i,j)}$ : Free energy of the strip ST without needle
and with boundaries $(i,j)$ (see Ref.
[\onlinecite{shapefreeenergy}]).
\vspace{0.4cm}

$F_{\rm cr}(\lambda)$ : The free energy belonging to the lattice
crossover Hamiltonian (see the paragraph containing Eqs.
(\ref{eq:hcr})-(\ref{eq:alg})).
\vspace{0.4cm}

$f_{i[h]i}^{\parallel}(\theta) \equiv F_{\parallel}^{(i[h]i)}
/(k_{B}T)$ : See Eq.~(\ref{longiJ}) as well as
Figs.~\ref{FIG_III_2} and~\ref{FIG_III_3}.
\vspace{0.4cm}

$f_{i[h]}^{\perp}(\vartheta) \equiv F_{\perp}^{(i[h])} /(k_{B}T)$
: See Fig.~\ref{FIG_III_1} and below Eq.~(\ref{asmaller}).
\vspace{0.4cm}

$f_{\rm b}$ and $f_{\rm s}$ : Bulk free energy per area and
surface free energy per length (see Ref.
[\onlinecite{shapefreeenergy}]).
\vspace{0.4cm}

$f_{\cal O}^{(i,j)}$ : Universal scaling functions of the profiles
$\langle {\cal O}(\bf r) \rangle_{\rm ST}$ for ${\cal O}=\phi$,
$\epsilon$ in the needle-free strip ST with boundaries $(i,j)$
(see Eqs.~(\ref{densscale}) and~(\ref{f})).
\vspace{0.4cm}

$f_{\epsilon}^{(P)}$ : Position-independent universal scaling
function of $\langle \epsilon \rangle_{\rm ST}$ in the needle-free
strip ST with double periodic boundary conditions (see below Eq.
(\ref{expand''}) and Appendix~\ref{appperper}).
\vspace{0.4cm}

$g_{i[h]}^{\perp}(\vartheta)$ and $g_{i[h]i}^{\parallel}(\theta)$
: Scaling functions for the effective force acting on a needle $h$
perpendicular to the boundary $i$ of the half plane and for the
disjoining force induced in an $(i,i)$ strip by a needle $h$
embedded in its midline (see the text below Eq.~(\ref{asmaller})
or above Eq.~(\ref{theta})).
\vspace{0.4cm}

${\cal H}_{\rm ST}$ : Lattice Hamiltonian for strips without a
needle (see Eq.~(\ref{eq:hini})).
\vspace{0.4cm}

${\cal H}_{\rm ST}+{\cal H}_{\perp}^{(h)} \equiv  {\cal H}_{0}$
and ${\cal H}_{\rm ST}+{\cal H}_{\parallel}^{(h)} \equiv {\cal
H}_{1}$ : Lattice Hamiltonians for strips containing an embedded
needle with orientation perpendicular and parallel to the
boundaries, respectively (see Eqs.~(\ref{Hperp}) and
(\ref{Hparallel})).
\vspace{0.4cm}

$\Delta {\cal H} \equiv {\cal H}_{1}-{\cal H}_{0} \equiv {\cal
H}_{\parallel}^{(h)} - {\cal H}_{\perp}^{(h)}$ : Difference of
lattice Hamiltonians for parallel and perpendicular orientation of
the needle $h$ with a fixed center (see Eqs.
(\ref{eq:hcr})-(\ref{eq:hcr'})).
\vspace{0.4cm}

${\cal H}_{\rm cr}^{(O)}(\lambda)$ and ${\cal H}_{\rm
cr}^{(+)}(\lambda)$ : Crossover Hamiltonian for the needle of
broken bonds and of fixed spins, respectively
(see Eqs.~(\ref{eq:HOcr}) and~(\ref{+cross}) as well as
Figs.~\ref{FIG_IV_1}(c) and~\ref{FIG_IV_3}(c)).
\vspace{0.4cm}

$\tilde{\cal H}^{(O)}$ and $\tilde{\cal H}^{(+)}$ : Lattice
Hamiltonian for a strip containing a cross-shaped hole with the
two bars corresponding to the two orientations of the needle of
broken bonds and of fixed spins, respectively (see below Eqs.~(\ref{eq:HOcr})
and~(\ref{+cross})).
\vspace{0.4cm}

$h$ : Characterizes the universality class of the needle surface.
\vspace{0.4cm}

$(i,j)$ : Characterizes the surface universality classes of the
(lower, upper) boundary of the strip.
\vspace{0.4cm}

$(i[h])$ : Needle of class $h$ in the half plane with boundary of
class $i$.
\vspace{0.4cm}

$(i[h]j)$ : Needle of class $h$ in the strip with boundaries of
classes $i$ and $j$.
\vspace{0.4cm}

${\cal J} > 0$ : Ferromagnetic coupling strength between nearest
neighbor Ising spins on the square lattice implying the bulk
critical temperature $T_{c}=[2/\ln (\sqrt{2}+1)]({\cal J}/k_{B})$
(see Eq.~(\ref{eq:hini}) and the paragraph above Eq.
(\ref{Hperp})).
\vspace{0.4cm}

$J_{u,v;u',v'}  {\cal J}$ : Coupling between nearest neighbors
$(u,v)$ and $(u',v')$ in a strip containing a needle of broken
bonds (see Fig.~\ref{FIG_IV_1}(a) and~\ref{FIG_IV_1}(b)). $J$
equals $0$ and $1$ for broken and unbroken bonds, respectively.
\vspace{0.4cm}

$k$ : Parameter characterizing the configuration of two needles
embedded in the $x$ axis of the unbounded $(x,y)$ plane via the
cross ratio of their endpoints (see Eqs.~(\ref{cross}) and
(\ref{cross'})).
\vspace{0.4cm}

$L$ : The number of columns in the lattice model for the strips
considered in Secs.~\ref{intro} and~\ref{SIM} (see
Fig.~\ref{FIG_I_1} and Figs.~\ref{FIG_IV_1}-\ref{FIG_IV_3}) or
\cite{micromeso} the length of the strip in the continuum
descriptions used in Secs.~\ref{intro},~\ref{SP},~\ref{ARB}, and
Appendix~\ref{appwithout} (see Fig.~\ref{FIG_I_2}).
\vspace{0.4cm}

$l$ : Arbitrary length in the conformal transformations given in
the text above Eqs.~(\ref{Thalfperp}), (\ref{Tstripparall}),
and~(\ref{Tperstrip}) which relate different geometries. It is
introduced for dimensional reasons only and, due to dilatation
invariance, drops out from equations relating quantities which
belong to the same geometry.
\vspace{0.4cm}

${\bf n}$ : Unit vector describing the orientation of the needle
(see Eq.~(\ref{SPE''})). In the strip, ${\bf
n}=(n_{\parallel},n_{\perp})$ (see Eq.~(\ref{uvnparnperp})).
\vspace{0.4cm}

$O$ : The ``ordinary'' surface universality class. Corresponding
boundaries or needles induce disorder in the system of Ising
spins, i.e., in their vicinity the probability to find parallel
nearest neighbor spins is smaller than in the bulk.
\vspace{0.4cm}

$+$ and $-$ : The two ``normal'' surface universality classes with
the tendency to order the Ising spins in the $+$ and $-$
directions, respectively.
\vspace{0.4cm}

${\bf r}_{N}$ : Position vector of the center of the needle (see
Eqs.~(\ref{SPE'}) and~(\ref{SPE''})). In the strip, ${\bf
r}_{N}=(u_{N}, v_{N})$ (see Eq.~(\ref{uvnparnperp})).
\vspace{0.4cm}

$S_{I}$ and $S_{A}$ : Operator contribution to the normalized
Boltzmann weight of the small needle which is isotropic and
anisotropic, respectively, with respect to the orientation of the
needle (see Eqs.~(\ref{SPE})-(\ref{SPE''})).
\vspace{0.4cm}

ST : Denotes the strip in the absence of the needle.
\vspace{0.4cm}

``sna'' (``small needle approximation''): Truncated form of the
``small needle expansion'' (see the paragraph containing Eqs.
(\ref{expand})-(\ref{expand''}) and the first paragraph of Subsec.
\ref{compnorm}). For explicit expressions see Eqs.~(\ref{full}),
(\ref{SNE}), and~(\ref{barfpar}).
\vspace{0.4cm}

$s_{0}$ : Exterior spin fixed to the value $+1$ (see the first
paragraph in Sec.~\ref{simB3} as well as Fig.~\ref{FIG_IV_3}(c)).
\vspace{0.4cm}

$T_{kl}$ : Stress-tensor operator (see Eq.~(\ref{SPE''})) with its
elements in the strip denoted by $T_{\parallel \,
\parallel}, \, T_{\parallel \,
\perp}, \, T_{\perp \,
\parallel}$, and $T_{\perp \,
\perp}$ (see Eq.~(\ref{stressscale})).
\vspace{0.4cm}

$\langle T(z) \rangle$, $\langle T(c) \rangle$, and $\langle T(w)
\rangle$ : Analytic functions in the unbounded plane with needles,
in the half plane with a needle, and in the strip with a needle,
respectively, which determine the corresponding stress tensor
averages as explained in Appendix~\ref{apparbord} (see Eqs.~(\ref{Tnneedles}),
(\ref{Thalfperp}), (\ref{Tstripparall}),
(\ref{Tperstrip})-(\ref{Tbreak}), (\ref{Thalfbreak}), and
(\ref{Tstripbreak}) as well as Ref.~[\onlinecite{TklTz}]).
\vspace{0.4cm}

$t \equiv \tanh(\theta/2)$ : Useful short notation according to
Eqs.~(\ref{arbiJ}) and~(\ref{Tstripbreak}).
\vspace{0.4cm}

$u$ and $v$ : Strip coordinates \cite{micromeso} parallel and
perpendicular, respectively, to the boundaries of the strip. In
the continuum description the boundaries of the strip are at $v=
\pm W/2$ (see Fig.~\ref{FIG_I_2}). In the lattice description of a
``normal'' (``ordinary'') needle the coordinates $u$ and $v$ of
the lattice vertices have integer (half odd integer) values (see
Fig.~\ref{FIG_I_1} and Figs.~\ref{FIG_IV_1}-\ref{FIG_IV_3}).
\vspace{0.4cm}

$u_{N}$ and  $v_{N}$ : Coordinates parallel and perpendicular,
respectively, to the strip of the position vector ${\bf
r}_{N}=(u_{N},v_{N})$ of the center of the needle. As explained in
Sec.~\ref{intro} and Ref. [\onlinecite{micromeso}], $u_{N}$ and
$v_{N}$ are lengths in the continuum description used in Sec.
\ref{SP} and in Eqs.~(\ref{f}) and~(\ref{CS}) of Appendix
\ref{appwithout}, while in the lattice description used in Sec.
\ref{SIM} they are measured in units of the lattice constant and
have integer values for both ``ordinary'' and ``normal'' needles.
\vspace{0.4cm}

$W$ : The number of rows in the lattice model for the strip or
\cite{micromeso} the width of the strip in the continuum
description.
\vspace{0.4cm}

$w=u+iv$ : Complex variable specifying the position vector $(u,v)$
in the strip (see above Eq.~(\ref{Tstripparall})).
\vspace{0.4cm}

$x_{\phi}=1/8$ and $x_{\epsilon}=1$ : Scaling dimensions of the
order parameter and energy densities, respectively (see below Eq.
(\ref{norm})).
\vspace{0.4cm}

$Z_{\parallel}$ and $Z_{\perp}$ : Partition functions
corresponding to $F_{\parallel}$ and $F_{\perp}$ (see
Eqs.~(\ref{Fparal}) and~(\ref{Fperp})).
\vspace{0.4cm}

$Z_{\rm ST}^{(i,j)}$ : Partition function of a $W \times L$ strip
ST without needle and with boundaries $(i,j)$ (see Refs.
[\onlinecite{shapefreeenergy}] and [\onlinecite{dual}]).
\vspace{0.4cm}

$Z^{([h_{1}][h_{2}])}$ : Partition function of a large system
containing two needles $h_{1}$ and $h_{2}$ (see below Eq.
(\ref{TEO})).
\vspace{0.4cm}

$z=x+iy$ : Complex variable specifying the position vector $(x,y)$
in the unbounded plane (see below Eq.~(\ref{Tnneedles})).
\vspace{0.4cm}

$\Delta_{i,j}(1/\delta)$ : Casimir amplitude describing the
universal contribution $-L^{-1} \partial \Phi_{\rm ST}^{(i,j)}
(\delta) / \partial W = \Delta_{i,j}(1/\delta) /W^{2}$ to the
disjoining pressure per $k_{B}T$ of an $(i,j)$ strip without
needle (see Eqs.~(\ref{stressscale}),~(\ref{stressscale'}) and
Appendix~\ref{appstripbound}).
\vspace{0.4cm}

$\Delta_{i,j} \equiv \Delta_{i,j}(0)$ : Casimir amplitude for an
$(i,j)$ strip of infinite length $L=\infty$ and without needle
(see below Eq.~(\ref{stressscale'})).
\vspace{0.4cm}

$\Delta_{P}(1/\delta)$ : Casimir amplitude for the double periodic
strip without needle (see below Eq.~(\ref{expand''}) and Appendix
\ref{appperper}).
\vspace{0.4cm}

$\Delta_{P} \equiv \Delta_{P}(0) = -\pi/12$.
\vspace{0.4cm}

$\delta \equiv L/W$ : Characterizes the shape of the strip (see
Eq.~(\ref{stressscale'})). We call $W/L \equiv 1/\delta$ the
aspect ratio of the strip.
\vspace{0.4cm}

$\epsilon({\bf r})$ : Energy-density operator with its average in
the unbounded plane (bulk) at bulk criticality subtracted (see
below Eq.~(\ref{SPE''})) and normalized according to Eq.
(\ref{norm}).
\vspace{0.4cm}

$\zeta_{I}$ and $\zeta_{A}$ : Contributions to $Z_{\parallel}$ and
$Z_{\perp}$ which arise via the ``small needle expansion'' from
the operator contributions $S_{I}$ and $S_{A}$ to the normalized
Boltzmann weight of the needle; they are isotropic and anisotropic
with respect to the needle orientation, respectively (see
Eqs.~(\ref{SPE})-(\ref{SPE''}), (\ref{zetaI}), and~(\ref{zetaA})).
\vspace{0.4cm}

$\vartheta \equiv  D/(2 a_{N})$ : Characterizes the size versus
the distance to the boundary for a needle in the half plane with
its orientation perpendicular to the boundary line (see Eq.
(\ref{asmaller})).
\vspace{0.4cm}

$\bar{\vartheta} \equiv \sqrt{1-\vartheta^{2}} =
\sqrt{a_{<}a_{>}}/a_{N}$ : Approaches zero if the closer end of
the needle approaches the boundary.
\vspace{0.4cm}

$\vartheta_{0}$ : Threshold value of $\vartheta$ above which the
interaction $f_{-[+]}^{\perp}(\vartheta)$ between a $-$ boundary
of the half plane and a $+$ needle perpendicular to it deviates
significantly from the corresponding ``small needle
approximation'', i.e., from the last equation in Eq.~(\ref{SNE})
(see the discussion of Fig.~\ref{FIG_III_1}(a) in the paragraph
between Eqs.~(\ref{long'}) and~(\ref{theta})).
\vspace{0.4cm}

$\theta \equiv \pi D/W$ : Characterizes the size versus the
distance to the boundaries for a needle embedded within the
midline of the strip (see Eq.~(\ref{theta})).
\vspace{0.4cm}

$\kappa \equiv \exp (-\pi \delta/4)$ : Variable used for the
aspect ratio dependence of $\Delta_{i,j}$ in the case of a long
strip with $L/W \equiv \delta \gg 1$ and close boundaries (see
Eqs.~(\ref{Deltaijclose}) and~(\ref{Deltaclose'})).
\vspace{0.4cm}

$\Lambda_{i}^{(1)}$ and $\Lambda_{j}^{(1)}$ : Strengths of the
coupling to the lower and upper additional outside row,
respectively, of fixed spins generating strip boundaries $i$ and
$j$ of ``ordinary'' or ``normal'' character in the lattice model
(see Eq.~(\ref{eq:hini}) and Figs.~\ref{FIG_I_1}
and~\ref{FIG_IV_1}-\ref{FIG_IV_3}).
\vspace{0.4cm}

$\lambda$ : Parameter within the lattice model describing the
crossover of the needle orientation from perpendicular
($\lambda=0$) to parallel ($\lambda=1$) orientation with respect
to the boundaries of the strip (see Eq.~(\ref{eq:hcr'}) as well as
Figs.~\ref{FIG_IV_1}(c) and~\ref{FIG_IV_3}(c)).
\vspace{0.4cm}

$\rho_{i,h}(\vartheta)$ : Auxiliary function determining the
Casimir force on a needle $h$ in the half plane with boundary $i$
and determining the disjoining force induced in an $(i,i)$ strip
of infinite length upon inserting a needle $h$ (see
Eqs.~(\ref{iJ}) and~(\ref{arbiJ}), respectively). Via
Eqs.~(\ref{sigma}) and~(\ref{K+K^{*}}) the function
$\rho_{i,h}(\vartheta)$ is related to the dependence
$\Delta_{i,h}(1/\delta)$ on the aspect ratio of the Casimir
amplitude of the needle-free strip with boundaries $(i,h)$.
\vspace{0.4cm}

$\sum \limits_{\langle {\rm inc.}\rangle }$ and $\sum
\limits_{\langle {\rm decr.}\rangle }$ : Sum of products of those
nearest neighbor spins in the crossover Hamiltonian ${\cal H}_{\rm
cr}^{(O)}(\lambda)$ the ferromagnetic coupling strength of which
increases and decreases, respectively, upon increasing $\lambda$
(see Eqs.~(\ref{eq:HOperp})-(\ref{eq:HOcr}) and Fig.~\ref{FIG_IV_1}(c)).
\vspace{0.4cm}

$\sum \limits_{\langle {\rm one}\rangle}^{(+)}$ : Sum of those
four spins which are coupled to an external spin (and carry a
northeast arrow) for {\it both} needle orientations shown in
Figs.~\ref{FIG_IV_3}(a)
and~\ref{FIG_IV_3}(b) (see Eq.~(\ref{ext1})).
\vspace{0.4cm}

$\sum \limits_{\langle {\rm zero}\rangle}$ : Sum of products of
the center spin of the ``normal'' $(+)$ needle and of its four
nearest neighbor spins (see Eq.~(\ref{nn1})). These four products
are missing in the crossover Hamiltonian ${\cal H}_{\rm cr}^{(+)}$
(see Eq.~(\ref{+cross})) and the four corresponding couplings are
absent in Figs.~\ref{FIG_IV_3}(a),~\ref{FIG_IV_3}(b),
and~\ref{FIG_IV_3}(c).
\vspace{0.4cm}

$\widehat{\sum \limits_{\langle {\rm inc.} \rangle}}$ and
$\widehat{\sum \limits_{\langle {\rm decr.} \rangle}}$ :
Sum of products of those nearest neighbor spins in the crossover
Hamiltonian ${\cal H}_{\rm cr}^{(+)}(\lambda)$ the ferromagnetic
coupling strength of which increases and decreases, respectively,
upon increasing $\lambda$ (see Eqs.~(\ref{nn2}), (\ref{nn3}),
and~(\ref{+cross}) as well as Fig.~\ref{FIG_IV_3}(c)).
\vspace{0.4cm}

$\sum \limits_{\langle {\rm inc.}\rangle}^{(+)}$ and $\sum
\limits_{\langle {\rm decr.}\rangle}^{(+)}$ : Sum of those spins
which in the crossover Hamiltonian ${\cal H}_{\rm
cr}^{(+)}(\lambda)$ are coupled with increasing and decreasing
strength, respectively, to the external spin $s_{0}=1$ (see Eqs.
(\ref{ext2})-(\ref{+cross}) and Fig.~\ref{FIG_IV_3}(c)).
\vspace{0.4cm}

$\sigma \equiv \exp (-2 \pi / \delta)$ : Variable used for the
aspect ratio dependence of $\Delta_{i,j}$ in the case of a short
strip with $L/W \equiv \delta \ll 1$ and distant boundaries (see
Eqs.~(\ref{Deltaijdistant}) and~(\ref{Deltadistant'})).
\vspace{0.4cm}

$\tilde{\tau}_{i,h}(\vartheta)$ : Auxiliary function with the same
character as (and simply related to) $\rho_{i,h}(\vartheta)$ (see
Eq.~(\ref{sigmatilde})).
\vspace{0.4cm}

$\Phi_{\rm ST}^{(i,j)} (\delta \, = \, L/W)$ : Universal shape
dependent and scale free contribution to the free energy per
$k_{B}T$ of $L \times W$ strips with boundaries $(i,j)$ but
without needle (see Eq.~(\ref{stressscale}) and Ref.
[\onlinecite{shapefreeenergy}]). $\Phi_{\rm ST}^{(P)} (\delta)$
denotes the corresponding contribution for the double periodic
strip.
\vspace{0.4cm}

$\phi({\bf r})$ : Order-parameter-density operator, normalized
according to Eq.~(\ref{norm}).
\vspace{0.4cm}

$\chi_{pq}(\delta)$ or $\chi_{pq}(\delta/2)$ : Auxiliary functions
for the Casimir amplitudes of $W \times L$ strips without a needle
and with double periodic boundary conditions or with $(i,j)$
boundaries (see Eqs.~(\ref{DeltaP}) or (\ref{Deltaijrect}) as well
as Ref.~[\onlinecite{Cardybc}]).
\vspace{0.4cm}

$\langle ... \rangle$ : Thermal average which may be specified by
means of subscripts and superscripts such as
\vspace{0.4cm}

$\langle ... \rangle_{\rm bulk}$ : for the unbounded plane without
embedded particles (see Eq.~(\ref{norm})).
\vspace{0.4cm}

$\langle ... \rangle_{\rm half \, plane}$ : for the half plane
without embedded particles (see above Eq.~(\ref{halfamp})).
\vspace{0.4cm}

$\langle ... \rangle_{\rm ST}$ and $\langle ... \rangle_{\rm
ST}^{(i,j)}$ : for a strip without embedded particles and with
boundaries $(i,j)$ (see Eqs.~(\ref{densscale}) and
(\ref{stressscale})).
\vspace{0.4cm}

$\langle ... \rangle^{(i[h])}$ : for the half space with boundary
$i$ and an embedded needle $h$ (see Eq.~(\ref{Thalfbreak})).
\vspace{0.4cm}

$\langle ... \rangle^{(i[h]j)}$ : for the strip with boundaries
$(i,j)$ and an embedded needle $h$ (see Eq.~(\ref{Tstripbreak})).
\vspace{0.4cm}

$\langle ... \rangle^{([h_{1}][h_{2}]..[h_{n}])}$ : for the
unbounded plane with $n$ embedded needles $h_{1}$, $h_{2}$, ..,
$h_{n}$ (see Eqs.~(\ref{Tnneedles}) and
(\ref{TEO})-(\ref{Tbreak})).
\vspace{0.4cm}

$\langle ... \rangle_{\rm cr}$ : Thermal average based on the
lattice crossover Hamiltonian (see Eqs.~(\ref{eq:df}) and
(\ref{eq:alg})).
\vspace{0.4cm}

\newpage


\begin{thebibliography}{99}
\bibitem{opera}The ``operators'' in the present systems of
classical statistical mechanics are classical objects fluctuating
with the basic statistical variables (which in the Ising lattice
model are the Ising spins). In the operator expansions one uses
local operators in coarse grained models (such as the densities of
the order parameter and of the energy), see Refs.
[\onlinecite{e}], [\onlinecite{e'}], and
[\onlinecite{coarsegrain}] as well as the entries in Refs.
[\onlinecite{ber}] and [\onlinecite{WK}].
\bibitem{bulkatcritpoint}Our study is exclusively about
properties for which the bulk of the $d=2$ Ising model is at the
critical point. Thus for all expressions shown we correspondingly
assume that the temperature is at bulk $T_{c}$ and that there is
no bulk field coupled to the order parameter.
\bibitem{normzd}With the normalization used in Eq.~(\ref{norm}) the inverse
length dimensions of $\phi$ and $\epsilon$ are equal to their
scaling dimensions and the half plane amplitudes in Eq.
(\ref{halfamp}) are universal.
\bibitem{parthalf}This property is special for two-dimensional conformal models
in which the space outside a (simply connected) particle and its
surface, embedded in the unbounded plane, can be conformally
mapped onto the half plane $a>0$ and its straight line boundary
$a=0$, respectively. Since the thermal average of the stress
tensor in the half plane vanishes for all surface universality
classes $h$, its average outside the particle is solely determined
by the Schwartz derivative of the mapping \cite{Cardy} and thus
depends only on the shape of the particle and not on its surface
class $h$. This should be compared with the density profiles
$\langle {\cal O} \rangle_{\rm half \, plane} = {\cal A}_{\cal
O}^{(h)} a^{-x_{\cal O}}$ in the half plane (see Eq.
(\ref{halfamp})) and the ones outside the particle, which both do
depend on $h$.
\bibitem{shapefreeenergy}For the strip ST with boundaries
$(i,j)$, free energy $F_{\rm ST} \equiv F_{\rm ST}^{(i,j)} =
-k_{B} T \ln Z_{\rm ST}^{(i,j)}$, and partition function $Z_{\rm
ST}^{(i,j)}$ the universal contribution $\Phi_{\rm
ST}^{(i,j)}(L/W)$ introduced above Eq.~(\ref{stressscale}) is the
limit of $[F_{\rm ST}^{(i,j)}-WLf_{\rm b}-L(f_{\rm s}^{(i)}+f_{\rm
s}^{(j)})]/k_{B} T$ as $L$ and $W$ tend to infinity with $L/W$
fixed. Here $f_{\rm b}$ is the bulk free energy per area while
$f_{\rm s}^{(i)}$ and $f_{\rm s}^{(j)}$ are the surface free
energies per length corresponding to the surface classes $i$ and
$j$, respectively.
\bibitem{useless}Unlike $Z_{\parallel}, \, Z_{\perp}$ an expansion
of $F_{\parallel}, \, F_{\perp}$ would involve arbitrary integer
powers of $(D/W)^{1/8}$ for a ``normal'' needle in a strip with at
least one ``normal'' boundary.
\bibitem{symmfinL} There are two consequences of duality for the energy
density scaling functions. The property, that
$f_{\epsilon}^{(+,O)}$ is an odd function of $v_{N}/W$, is not
only valid for infinite strips $W/L=0$ (see Eq.~(\ref{f})) but
even for {\it arbitrary} values of $W/L$. However, the identity
$f_{\epsilon}^{(O,O)}=-f_{\epsilon}^{(+,+)}$  only holds for
infinite strips (compare the corresponding arguments for
$\Delta_{O,O}=\Delta_{+,+}$ in the present Ref.
[\onlinecite{dual}]).
\bibitem{integration}For the needle perpendicular to the boundary
of the half plane the dimensionless and universal scaling function
$f_{i[h]}^{\perp}(\vartheta)=F_{\perp}^{(i[h])}/(k_{B}T)$ of the
free energy is related to the scaling function
$g_{i[h]}^{\perp}(\vartheta)=-a_{N}(\partial/\partial
a_{N})F_{\perp}^{(i[h])}/(k_{B}T)$ of the force by
$f_{i[h]}^{\perp}(\vartheta)= \int_{0}^{\vartheta} {\rm d}
\hat{\vartheta} g_{i[h]}^{\perp}(\hat{\vartheta})/
\hat{\vartheta}$ because $f_{i[h]}^{\perp}(\vartheta=0)=0$. The
corresponding relations for the needle with parallel orientation
and at the center line of a symmetric strip follow upon the
replacements $F_{\perp}^{(i[h])} \to F_{\parallel}^{(i[h]i)}$,
$a_{N} \to W$, $(\vartheta,\hat{\vartheta}) \to
(\theta,\hat{\theta})$, and $f_{i[h]}^{\perp} \to
f_{i[h]i}^{\parallel}$, $g_{i[h]}^{\perp} \to
g_{i[h]i}^{\parallel}$, where $\theta=\pi D/W$.
\bibitem{Maghrebi}The geometry underlying Eq.~(\ref{long}) of an
infinitely long needle perpendicular to a boundary line is
reminiscent of an (idealized) AFM geometry of a cone near a wall
(mimicking the surface of a colloidal particle), albeit in two
spatial dimensions and for a vanishing opening angle of the cone.
It would be interesting to use AFM in order to investigate
critical Casimir forces for such a system immersed in a critical
binary liquid mixture. The somewhat related critical behavior of
the cone-plate interaction induced by a long flexible polymer
chain attached to the cone tip has been discussed in Ref.
[\onlinecite{Magh}]. Still another example for two interacting
scale-free objects is provided in Ref. [\onlinecite{gies}]
analyzing the quantum-electrodynamical Casimir effect between a
plane and a semi-infinite plate. In all these cases only a single
length appears and the force between such objects is related, via
simple dimensional considerations, to a negative integer power of
this length times a universal constant, like in the original
geometry of two infinitely extended parallel plates.
\bibitem{ii}The effect (ii) described in the Summary prevails
for cases in which $\Delta F$ can be expanded in terms of $D$ and
is dominated by its leading order $\propto D^{2}$. This applies to
needle positions near the strip center for the cases $(i[h]j)$
considered in Subsec.~\ref{compord}. However, for the cases
considered in Subsec.~\ref{compnorm} in which the needle is
subject to the order parameter profile induced by the boundaries
$(i,j)$ there appear terms $\propto (D/W)^{1/8}$ inside the
logarithm determining $\Delta F$ which cannot be made sufficiently
small in simulations with mesoscopic $D$ (in order not to exceed a
tractable lattice size $W \times L$) and prevent a useful
expansion (see the remarks above Eq.~(\ref{expand})). Actually,
for the case $(-[+]-)$ the naive application of (ii) would predict
the wrong orientation of the needle near the strip center (compare
the discussion in the paragraph containing Eq.~(\ref{full})).
\bibitem{(+[+]-)}For the case $(+[+]-)$ the simulation data and
the analytic prediction indicate that $\Delta F$ becomes {\it
positive}, in agreement with the observation (i) in the Summary,
for values $v_{N}/W$ which are closer to the $+$ boundary (i.e.,
which are more negative) than the ones shown in
Fig.~\ref{FIG_V_7}.
\bibitem{TklTz}The relations given below Eq.~(\ref{Tnneedles})
between $\langle T_{kl}(x,y) \rangle$ and $\langle T(z) \rangle$
are based on the normalization of $T_{kl}$ as specified by Eq.
(\ref{stressscale}) and consistent with Eq.~(\ref{shifthalfperp}).
Here $T(z)$ is the usual \cite{Cardy} complex stress tensor which
within thermal averages is an analytic function of $z$ and its
normalization is specified by its transformation law. For two
geometries $G$ and $\tilde{G}$ related by the conformal mapping
$z=z(\tilde{z})$, the corresponding analytic functions $\langle
T(z) \rangle_{G}$ and $\langle T(\tilde{z}) \rangle_{\tilde{G}}$
are related by $\langle T(\tilde{z})
\rangle_{\tilde{G}}=(dz/d\tilde{z})^{2} \langle T(z) \rangle_{G}
+S/24$ with the Schwartz derivative
$S=(z'''/z')-(3/2)(z''/z')^{2}$ where the primes denote
derivatives with respect to $\tilde{z}$. For example, for the
mapping $z=z(c)=c^{2} /l$ considered above Eq.~(\ref{Thalfperp})
this relation allows one to obtain $\langle T(c)
\rangle_{\tilde{G}} \equiv \langle T(c) \rangle$ in Eq.
(\ref{Thalfperp}) from $\langle T(z) \rangle_{G} \equiv \langle
T(z) \rangle^{([O][O])}$ according to Eq.~(\ref{Tnneedles}).
\bibitem{SC'}For the simple needle configuration $x_{1<}=-1/k, \, x_{1>}=-1, \,
x_{2<}=1, \, x_{2>}=1/k$ with $0 \le k \le1$ the mapping is
provided by $u+iv=i(W/2)[\int_{0}^{z} dt/\sqrt{(1-t^2)(1-k^2
t^2)}]/K(k)$. This maps the upper and lower halves of the $z$
plane with $y>0$ and $y<0$ onto the left and right halves of the
rectangle or strip ST with $u<0$ and $u>0$, respectively. Crossing
the $x$ axis outside of the interval where the needles are, i.e.,
at $|x|>1/k$ corresponds to jumping within the rectangle from
$(u=-L/2, v)$ to $(u=L/2, v)$ which amounts to the periodic
boundary condition in $u$ direction. The more general needle
configuration corresponding to Eq.~(\ref{Tbreak}) is related to
the simple one by a special conformal transformation which
preserves the real axis.
\end{thebibliography}
\end{document}